\documentclass{aa}  
\usepackage{graphicx}
\usepackage{txfonts}
\usepackage{newtxtext,newtxmath}
\usepackage[T1]{fontenc}
\usepackage{ae,aecompl}
\usepackage{graphicx}   
\usepackage{amsmath}    
\usepackage{hyperref}
\usepackage{multirow}
\usepackage{threeparttable}
\usepackage[export]{adjustbox}
\usepackage{fontawesome}\usepackage[dvipsnames]{xcolor}

\begin{document}

   \title{Physical properties of circumnuclear ionising clusters\\ II. NGC\,7469}


   \author{S. Zamora
          \inst{1} \inst{2}\fnmsep\thanks{PhD fellow of Ministerio de Educación y Ciencia, Spain, BES-2017-080509, CEAL-AL/2017-02}
          \and
          A. I. Díaz \inst{1}\inst{2}
          }

   \institute{Departamento de Física Teórica, Universidad Autónoma de Madrid, 28049 Madrid, Spain
         \and
             CIAFF, Universidad Autónoma de Madrid, 28049 Madrid, Spain
             }

   \date{ }

 
  \abstract
   {
   Circumnuclear star-forming regions (CNSFRs) found close to galactic nuclei are ionised by massive clusters. These entities give us an excellent opportunity to study star formation in environments with high metallicity, and to relate it with active galactic nuclei.
   Our principal aim is to derive the physical properties and dynamical masses of the CNSFRs in the two rings of the spiral NGC~7469, categorised as a luminous infrared galaxy (LIRG) and hosting a Seyfert 1 nucleus. 
   We used archival data obtained with the MUSE spectrograph.
   The galaxy shows two prominent star-forming rings, one of them very close to their active galactic nucleus, within 1.5 arcsec from the galaxy centre. We   constructed 2D flux maps of the  different emission lines and two continuum bands. A map of the EW(H$\alpha$) emission shows the circumnuclear regions within the rings having EW(H$\alpha$) > 50 \AA, consistent with the presence of recent star formation. All emission lines appear to have at least two kinematical components. We   ascribe the most intense and narrow component to the emission lines originated by the ionising star-forming complexes (SFC) since they follow the radial velocity of the galaxy disc. 
   For each HII region, we  derived the number rate of Lyman continuum photons; the gas electron density; the ionisation parameter; the filling factor; and the mass of ionised hydrogen. We  used sulphur as a tracer for chemical abundances with the temperature-sensitive [SIII]$\lambda$ 6312 \AA\ emission line having been measured in $\sim$ 50 \% of the total, allowing the derivation of abundances by the direct method. The evolutionary state of the SFC was inferred with the help of population synthesis models yielding mean ages of 5.7 Ma, agreeing  with the presence of the broad Wolf--Rayet (WR) carbon feature at $\lambda$ ~5800 \AA\ detected in all the regions. Ionising (lower limits) and photometric SFC masses were estimated from the number of Lyman continuum photons and absolute r-magnitudes using stellar population synthesis techniques, and give median values of 2.3 $\times$ 10$^6$ and 6.9 $\times$ 10$^6$, respectively. The dynamical masses (upper limit) were derived from the measured absorption CaT velocity dispersion and the sizes of each cluster were measured on continuum light, assuming virialisation, and yield a median value of 6.7 $\times$ 10$^8$.
   Regions in the studied galaxy show sizes larger than implied by simple photo-ionisation models, which can be explained by the stellar winds produced by WR stars. The inner ring regions seem to be more compact than the outer ones. The young stellar population of the clusters has contributions of ionising populations with ages around 5 Ma, and its masses constitute less than  1\% of the total dynamical mass of each SFC. Finally, the comparison between the characteristics of the inner and outer ring ionising clusters, together with their derived dynamical masses, point to circumnuclear regions close to the active galactic nucleus being more compact and having higher gas density.}


   \maketitle
%

\section{Introduction}
\label{sec:introduction}

This is the second paper in a series to study the peculiar conditions of star formation in circumnuclear regions of early-type spiral galaxies, in particular the kinematics of the connected stars and gas using archival data obtained with the MUSE spectrograph attached to one of the ESO VLT telescopes.

Circumnuclear star-forming regions (CNSFRs) represent a common mode of star formation found close to galactic nuclei. Some of these regions, being a few hundred parsecs in size and showing integrated H$\alpha$ luminosities which overlap with those of HII galaxies (typically higher than 10$^{39}$ erg s$^{-1}$), seem to be composed of several HII regions ionised by luminous compact stellar clusters whose sizes, as measured from high spatial resolution HST images, are seen to be of only a few parsecs. These regions are young (age < 10 Ma ) and massive (up to 2 $\times$ 10$^8$ M$_\odot$) \citep{hagele2007,Hagele2013}. In the UV and B wavebands, they contribute substantially to the emission of the entire nuclear region, even in the presence of an active nucleus \citep[see e.g.][]{2002ApJ...579..545C}.  In some nearby galaxies presenting circumnuclear star-forming rings, this is the strongest organised source of far-UV (FUV) emission, and 30\% of the total observed FUV emission is produced within a radius of 10”. At redshifts of z ~ 2–3, this structure would be confined to a region 0.2” in diameter for $\Omega$ = 1 and would appear point-like in low-resolution observations. Consequently, in the absence of diagnostic spectroscopy, some of these objects could be mistaken for an active galactic nucleus (AGN). 
It is currently generally accepted that some connection exists between star formation and activity in galactic nuclei, and young stars appear as one component of the unified model of AGN giving rise to the blue featureless continuum that is observed in Seyfert 2 galaxies where the broad line region is obscured \citep[see][and references therein]{1998ApJ...505..174G}.

NGC 7469 gives us an excellent opportunity to study these phenomena in detail. 
It is one of the brightest blue galaxies first listed by Seyfert (1943); it is  included  in Arp's Atlas for Peculiar Galaxies \citep{1966ApJS...14....1A} as number 298. It is relatively nearby (z=0.01627), has been classified as an SABa(rs), and categorised as a luminous infrared galaxy (ULIRG). The galaxy has a close companion, IC~538; together they  form an isolated interacting pair first catalogued by \citet{ 1966ApJS...14....1A}. The companion is located at $\sim$ 22 kpc \citep{1963ApJ...137.1022B}. The pair interaction was studied in \citet{1994AJ....108...90M}, and \citet{1995ApJ...444..129G} suggested that this interaction may have taken place more than 150 Ma ago, triggering the powerful starburst found in the central 3 arcsec of the galaxy that is responsible for 60 \% of the bolometric luminosity of the whole galaxy. The stellar population of their young stellar clusters has been studied in detail by \citet{2007ApJ...661..149D}.

In Sect. 2 we describe the observations and the selection of the sample objects. Our results are presented in Sect. 3 and discussed in Section 4. Finally, Sect. 5 summarises this work and our conclusions.

\section{Observations and sample selection}
\label{observations}

For this work we analysed the circumnuclear environment of the almost face-on galaxy NGC~7469, which shows two prominent star-forming rings, using publicly available observations obtained by the IFS MUSE \citep{MUSE}.  Some characteristics of this galaxy are given in Table \ref{tab:galaxy characteristics}.
The galaxy is a well-studied early spiral (SABa) hosting a Seyfert 1 nucleus and an actively star-forming ring very close to it, within ~1.5 arcsec from the galaxy centre, with circular appearance. In addition, it also shows, further out, a second incomplete ring of elliptical appearance, and with dimensions of its major and minor axes of 21 and 13.2 arcsec, respectively \citep[see][]{ 1993AJ....105.1344B}, at the limit of what can be considered circumnuclear according to the definition given in \citet{2015MNRAS.451.3173A}. These two described structures can be easily identified in moderate resolution images of the galaxy. All throughout this paper we  refer to them respectively as the inner ring and the outer ring. \footnote{In Buta and Crocker's notation these structures are referred to as the nuclear ring and the inner ring, respectively.}

\begin{table}
\centering
\caption{NGC~7469 global properties.}
\label{tab:galaxy characteristics}
\begin{tabular}{cc}
\hline
Galaxy & \href{https://ned.ipac.caltech.edu/byname?objname=NGC7469&hconst=67.8&omegam=0.308&omegav=0.692&wmap=4&corr_z=1}{NGC7469} \\ \hline
RA J2000 (deg)$^a$ & 345.815095\\               
Dec J2000 (deg)$^a$ & 8.873997\\ 
Morphological type & (R')SAB(rs)a\\     
Nuclear type & Sy 1 \\
z & 0.01627\\ 
Distance (Mpc)$^b$ & 66.47 \\
Scale (pc/arcsec)$^c$ & 316\\   
\hline
\end{tabular}
\begin{tablenotes}
\centering
\item $^a$ \citet{10.1093/mnras/197.4.829}.\\
\item $^b$ \citet{1988cng..book.....T}.\\
\item $^c$ Cosmology-corrected scale.
\end{tablenotes}
\end{table}

NGC~7469 was observed as part of the first MUSE Science Verification run in 2014 August 19 under ESO Programme 60.A-9339(A). The observing time was split into four exposures of 600s with an offset of 1 arcsec in declination and different rotations among observations. Offset sky observations were taken after the target observations for adequate sky subtraction. The median seeing was 1.6 arcsec.
The reduction of the data was performed by the Quality Control Group at ESO in an automated process applying version 0.18.5 of the MUSE pipeline \citep{MUSEpipeline} including all the steps listed in \citet{ngc7742paper}. 

We also used additional data from \textit{Hubble} Space Telescope (HST) in the F336W and F606W filters. The UV data were acquired on 2018 October 29 with the Wide Field and Planetary Camera 3 (WFPC3) as part of the programme \href{https://archive.stsci.edu/proposal_search.php?mission=hst&id=15472}{GO/15472} providing high spatial resolution images ($\simeq$ 0.1 arcsec pixel$^{-1}$) and a FoV of 150 arcsec$^2$. These data were retrieved from the \textit{Hubble} Legacy Archive and organised in 3 exposures of 820 s each. The optical data were acquired on 1994 June 10 with the Wide Field and Planetary Camera 2 (WFPC2) as part of the programme \href{https://archive.stsci.edu/proposal_search.php?mission=hst&id=5479}{ID/5479} and the data has an exposure time of 500s. Their reduction was performed by the Space Telescope Science Institute (STScI) using available calibration files taken for this observation and taking into account different dithering positions.

\section{Results}
\label{sec:results}
\subsection{Ionised gas}
\label{gas}
The data presented here have been analysed following the methodology already used and tested in \cite{ngc7742paper}. The analysis is based on: (i) performing 2D maps of different emission lines and continuum bands; (ii) selecting HII regions from the H$\alpha$ emission line map; (iii) extracting each region spectrum and measuring the available emission lines; (iv) calculating the integrated SDSS magnitudes in the r and i SDSS filters; and (v) deriving chemical abundances for each of the CNSFRs. In this section only specific details introduced in this analysis due to the particular characteristics of NGC~7469 are explained.

\subsubsection{Emission line and continuum maps}
\label{emmision maps}
\begin{figure*}
\includegraphics[width=\textwidth]{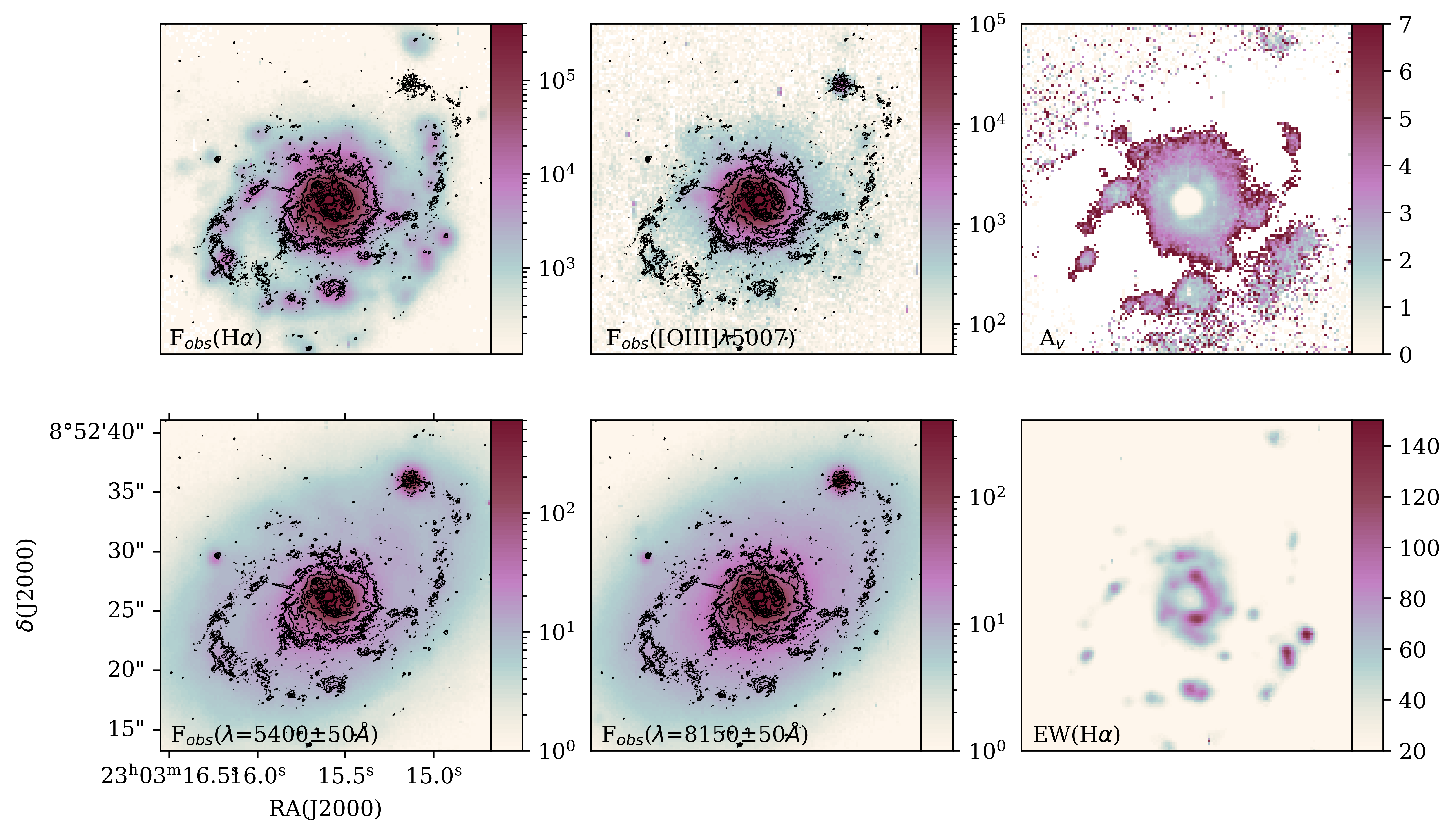}
\caption{Emission line and continuum maps. From left to right and top to bottom: Maps of the observed H$\alpha$ and [OIII]$\lambda$5007 \r{A} emission line fluxes (in units of 10$^{-20}$ erg/s/cm$^2$ and in logarithmic scale); A$_V$ extinction (in magnitudes); observed continuum fluxes in the blue and red parts of the spectrum (5400 \AA\ and 8150\AA, respectively; in units of 10$^{-17}$ erg/s/cm$^2$ and in logarithmic scale); and EW ($H\alpha$) in \r{A}. The top and bottom left and centre images show superimposed the contours of the HST-UV image described in the text. Orientation is north up, east to the left.}
\label{fig:Ha_OIII_map}
\end{figure*}

From the observed data cubes we have constructed 2D maps of different emission lines and two continuum bands. The top left panel of Fig. \ref{fig:Ha_OIII_map} shows the spatial distribution of the observed H$\alpha$ flux with contours of HST images from the WFC3 camera in the F336W filter superimposed. In this filter young star clusters should be most clearly visible and hence this comparison provides information about the spatial resolution of our MUSE instrumental configuration. The agreement between the HST contours and the MUSE maps ensures the existence of young ionising stellar populations in the observed clusters. On this map the emission of the nucleus and the inner ring of the galaxy are also clearly distinguished. The top central panel of this figure shows the [OIII]$\lambda $5007 \AA\ emission map. The emission from the active nucleus is predominant in this line and it seems to blur along the galaxy disc.

The H$\alpha$ and H$\beta$ maps have been combined to produce an extinction map which is shown in the top right panel of the figure. It has been calculated by adopting the Galactic extinction law of \citet{reddening}, with a specific attenuation of R$_V$ = 2.97 and the theoretical ratio H$\alpha $/H$\beta $ = 2.87 from \citet{Osterbrock2006} (n$_e$ = 100 cm$^{-3}$, T$_e$ = 10$^4$ K, case B recombination). In this map the inner ring is clearly visible with the whole of it showing a similar extinction ($\sim$ 2 mag). At the galaxy nucleus itself, A$_V$ $\sim$ 0, suggesting some kind of observational problem with the H$\alpha$ line emission. Actually, the existence of 30 pixels in this area of the galaxy with a H$\alpha$/H$\beta$ ratio < 2.7 has been confirmed. The H$\alpha$ line seems to be saturated in these pixels as probably it is also in other pixels around where number counts are close to the non-linear regime of the detector. On the other hand, there are apparently higher extinction values at the edges of the HII regions which could be due to the low S/N ratio in the H$\beta$ emission line.

\begin{figure}
\centering
\includegraphics[width=\columnwidth]{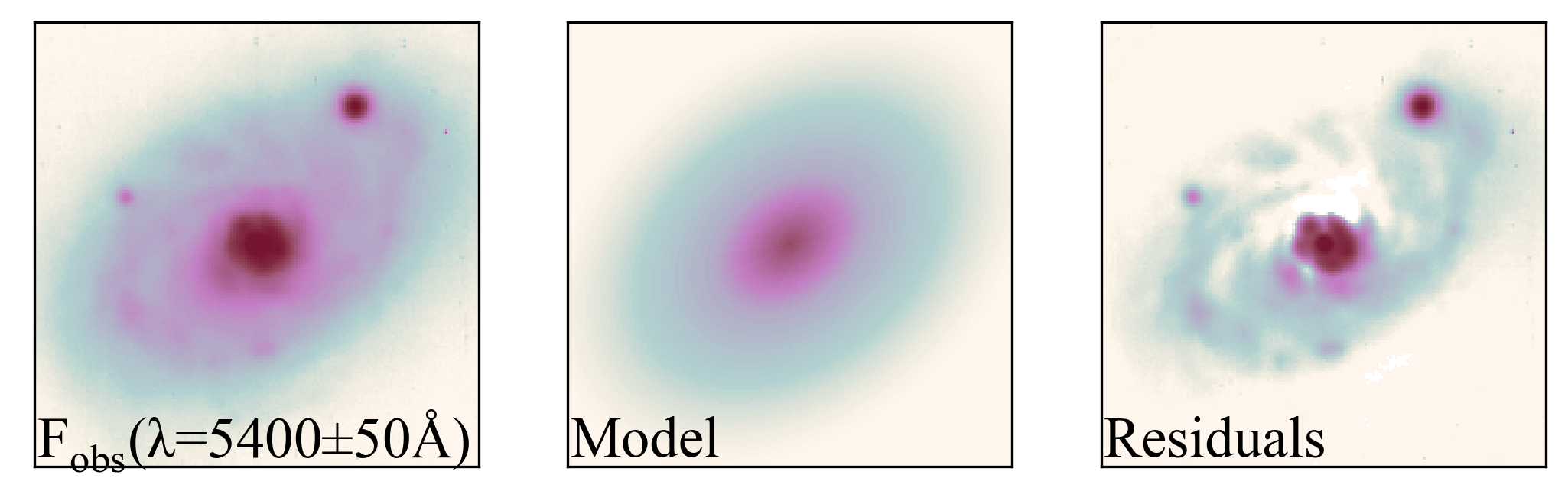}
\includegraphics[width=\columnwidth]{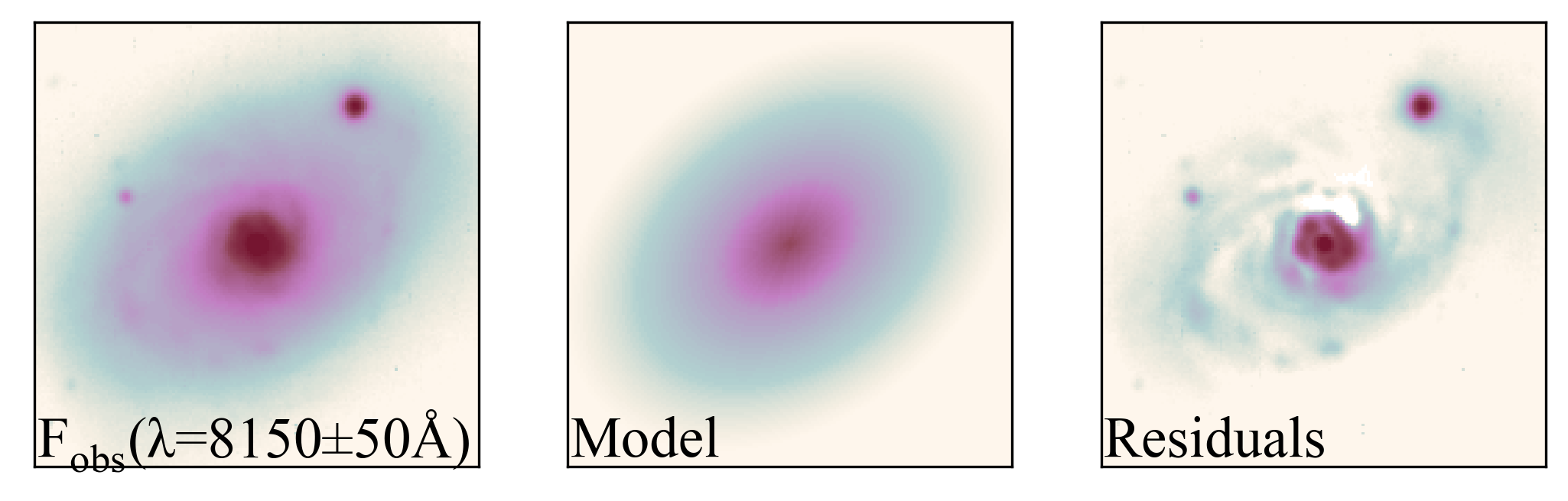}
\caption{Maps of observed continua. Left panels: Maps of the observed continua in the blue and red parts of the spectrum (5400 \AA\ and 8150\AA, respectively) in logarithmic scale). Centre panels: Fitted models to the galaxy disc profile (see text for details). Right panels: Residuals between the continuum maps and the disc fitted profile. Orientation is north up,  east to the left. The panel sizes are 32 arcsec $\times$ 32 arcsec.}
 \label{fig:sersic-fit_cont}
\end{figure}

The two bottom left panels of Fig. \ref{fig:Ha_OIII_map} show maps of the observed continuum fluxes at blue and red wavelengths, 5400 \AA\ and 8150 \AA\ respectively. Superimposed are the contours of HST-WFC3 data in the F336W filter. In both maps, the dominant continuum emission is shown to proceed from the galaxy disc. 
A Sersic fit to the stellar surface brightness has been done in order to better understand the behaviour of the continuum in our HII regions. Three different components with different scale-lengths have been fitted taking the geometrical values of this galaxy into account \citep[PA = 128º, i = 45º][]{2004ApJ...602..148D}. Figure \ref{fig:sersic-fit_cont} shows from left to right the original map, the fitted model and the residuals for the blue (upper panels) and red (lower panels) continuum maps. Only a few HII regions within the outer ring stand out of the galaxy profile while all inner-ring regions show high fluxes in these bands. Also, there is a diffuse excess in both maps that follow the area of the outer ring and seems to fall to the central part of the galaxy.
Finally, in the bottom right panel of Fig. \ref{fig:Ha_OIII_map} we can see the map of the equivalent width (EW) of $H\alpha$ (in \AA ). All circumnuclear regions have values EW(H$\alpha$) > 50 \AA\ , consistent with the presence of recent star formation, not older than 10 Ma.

\subsubsection{HII region selection}
\label{sec:segmentation}

Our HII region selection method is described in detail in \cite{ngc7742paper}. It is based on an iterative procedure that works on a line emission map and detects high-intensity clumps. It requires several input parameters: the maximum size of the regions, the diffuse gas emission level, the relative flux intensity of each of the regions with respect to the emission of its centre, and the maximum and minimum extent of regions according to their typical projected size and the point spread function (PSF) of observations. 

For regions within the outer ring we have used the observed H$\alpha$ flux map to select HII regions in the same way as was done for the NGC~7742 galaxy. However, for the inner-ring regions we have decided to use the observed HeI$\lambda$ 6678 \AA\ flux map due to the already mentioned saturation effects in the H$\alpha$ emission line in the central parts of the galaxy (see Sect. \ref{emmision maps}). Thus, we have constructed an observed HeI$\lambda$ 6678 \AA\ map (see Sect. \ref{emmision maps}) assuming a linear behaviour of the continuum emission and choosing side-bands around the line of a given width ($\lambda_c$ = 6678 \AA, $\Delta \lambda$ = 3 \AA, $\Delta \lambda_{left}$ = 6650 \AA\ and $\Delta \lambda_{right}$ = 6695 \AA, all wavelengths in rest frame). 

\begin{figure}
\centering
\includegraphics[width=\columnwidth]{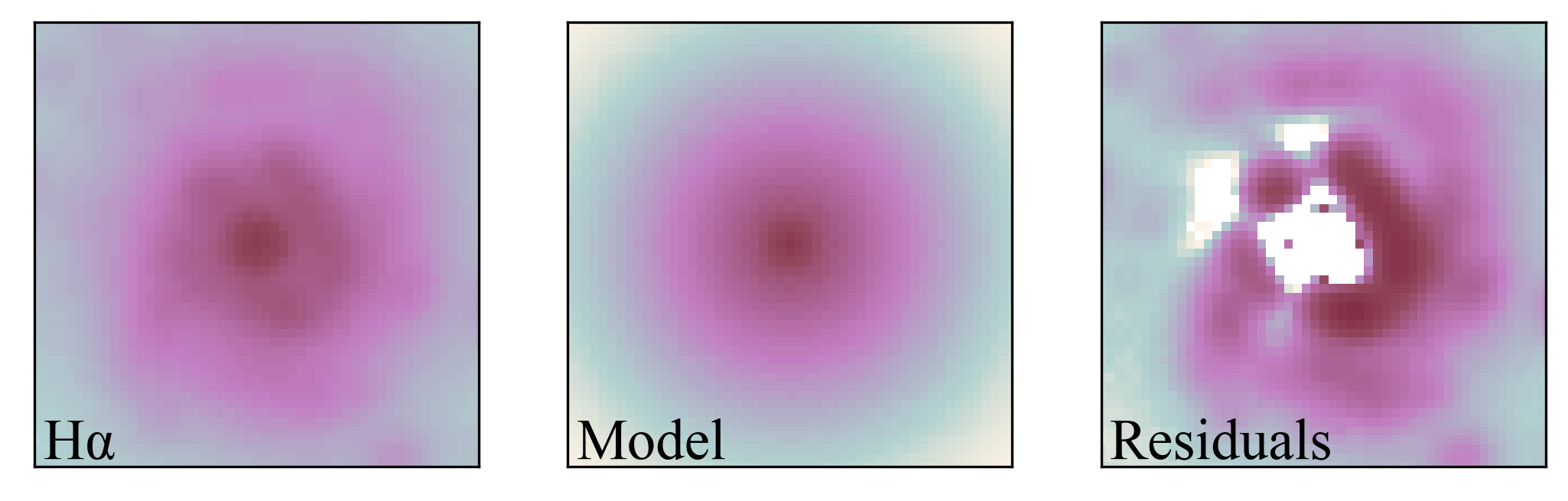}
\includegraphics[width=\columnwidth]{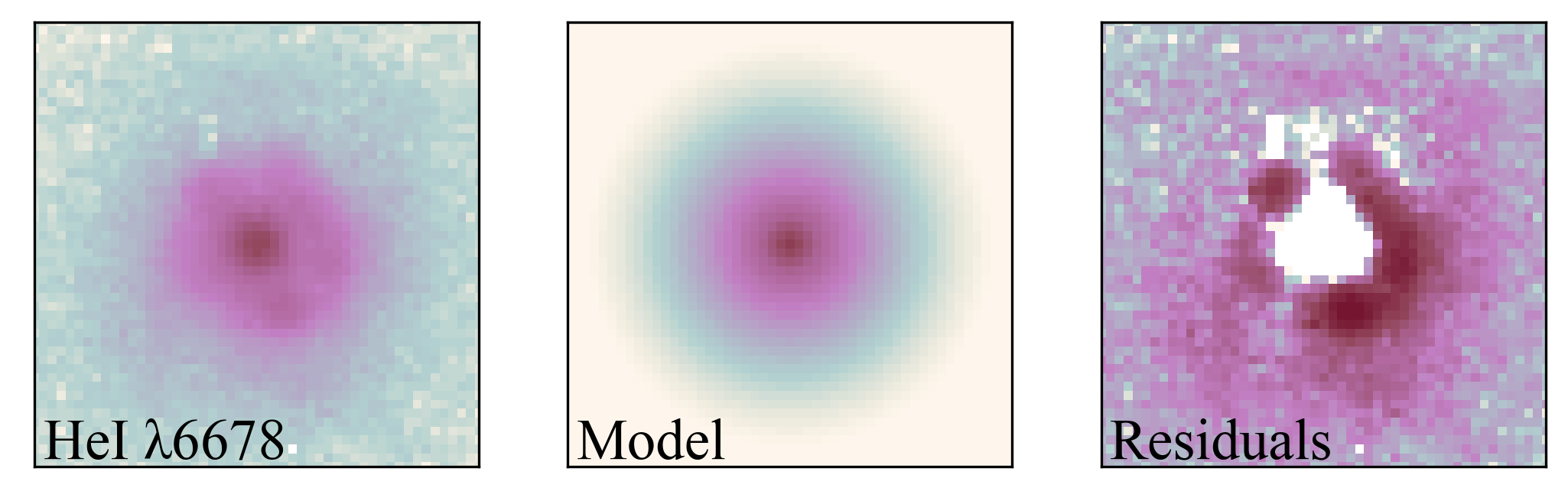}
\caption{Maps of the observed hydrogen and helium emission lines. Upper panels, from left to right: Map of the observed H$\alpha$ emission line, fitted model to the AGN profile (see text), and residuals between the observed map and the AGN fitted profile. Lower panels: Same maps, but of the observed HeI$\lambda$ 6678 \AA\ emission line. Orientation is north up, east to the left. The panel sizes are 10 arcsec $\times$ 10 arcsec.}
 \label{fig:sersic_}
\end{figure}

In order to do that, we have compared  first the spatial distribution of the hydrogen and helium emission. For this purpose, an exponential fit to the AGN brightness has been done. Figure \ref{fig:sersic_} shows from left to right the original map, the fitted model and the residuals. We can see that both maps show the same spatial distribution although the HeI$\lambda$ 6678 \AA\ emission map has a lower S/N. The two procedures are slightly different since the HeI flux intensity is weaker. A longer exposure time would have yielded a higher S/N in this line and also in other weak lines needed for the determination of the physical conditions of the gas \citep[see][]{2022MNRAS.516..749Z}.

Finally, we have imposed the following quality control requirements to the integrated spectra extracted from each selected region to ensure their physical meaning and to be certain that the emission has a star formation origin: EW(H$\alpha$) > 6 \AA\ \citep{Sanchez2015} and 2.7 < H$\alpha$/H$\beta$  < 6.0 \citep[][n$_e$ = 100 cm$^{-3}$, T$_e$ = 10$^4$ K]{Osterbrock2006}.

\begin{figure*}
\centering
 \includegraphics[width=\textwidth]{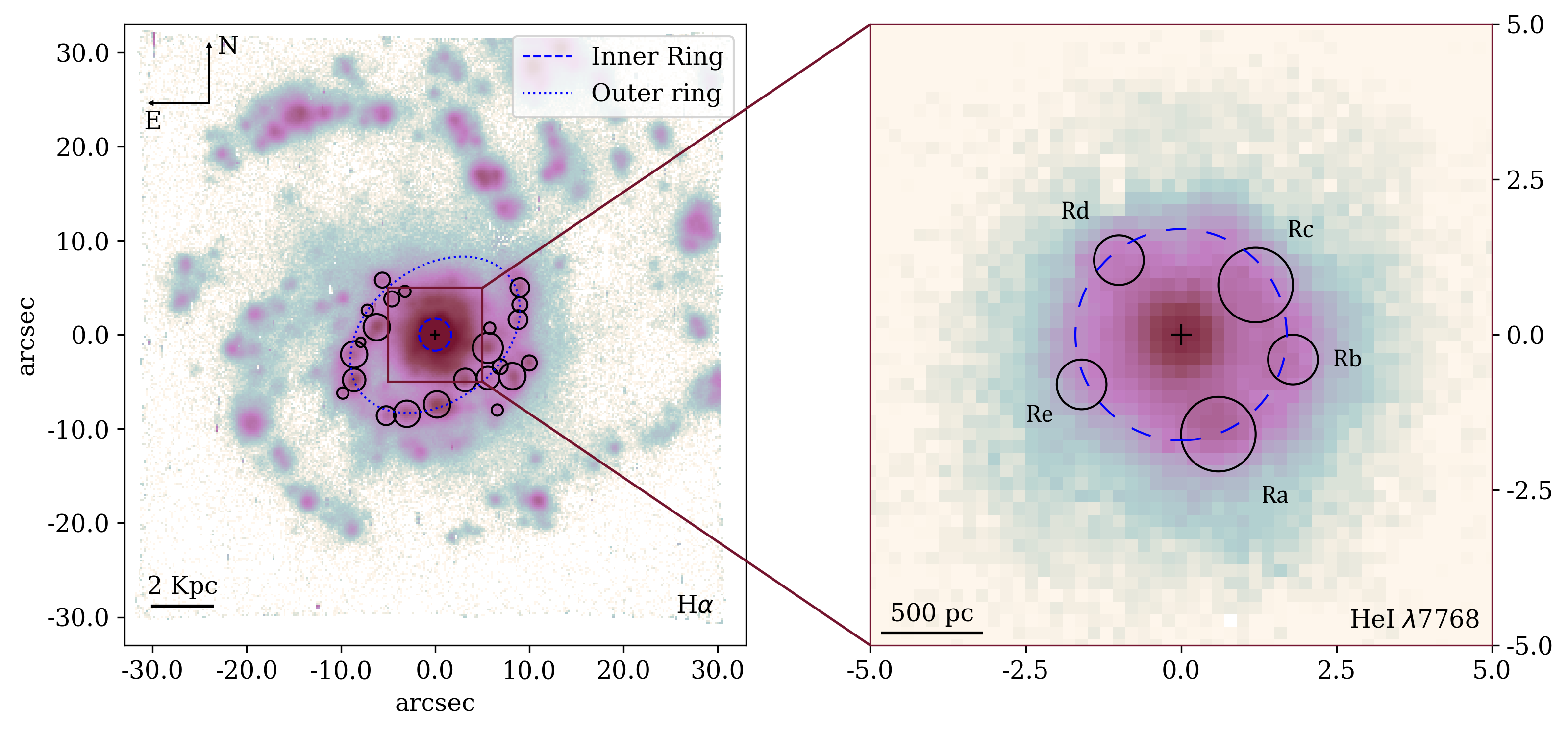}
 \caption{HII region selection. Left panel: HII regions selected using our segregation programme on the H$\alpha$ observed emission line  map. Right panel: HII regions selected with our segregation programme with the HeI $\lambda$ 7768 \AA\ observed emission line map, in logarithmic colour scale. Orientation is north up, east to the left. The physical scale is represented in the bottom left corner of the map. The outer and inner rings are marked with blue ellipses.}
 \label{fig:ring_ha_profile}
\end{figure*}

\begin{table}
\centering
\caption{Selection characteristics for observed CNSFRs.}
\label{tab:seleccion}
\begin{tabular}{ccc}
\hline
Region ID & \begin{tabular}[c]{@{}c@{}} Area \\ (arcsec$^2$)\end{tabular} & \begin{tabular}[c]{@{}c@{}}Offsets from galaxy centre $^a$ \\ (arcsec)\end{tabular}\\ \hline
R1      &       7.72    &       0.2, -7.4       \\
R2      &       7.36    &       -6.2, 0.8       \\
R3      &       5.80    &       -8.6, -4.8      \\
R4      &       8.44    &       5.6, -1.4       \\
R5      &       5.76    &       3.2, -4.8       \\
R6      &       2.76    &       10.0, -3.0      \\
R7      &       7.04    &       8.2, -4.4       \\
R8      &       6.76    &       -3.0, -8.4      \\
R9      &       3.24    &       9.0, 5.0        \\
R10     &       1.48    &       -3.2, 4.6       \\
R11     &       6.80    &       -8.6, -2.1      \\
R12     &       3.56    &       8.8, 1.6        \\
R13     &       2.76    &       -4.6, 3.8       \\
R14     &       3.04    &       -5.2, -8.6      \\
R15     &       1.72    &       6.9, -3.4       \\
R16     &       1.76    &       9.0, 3.2        \\
R17     &       0.88    &       -7.9, -0.8      \\
R18     &       1.40    &       -7.2, 2.6       \\
R19     &       1.36    &       -9.8, -6.2      \\
R20     &       1.56    &       5.8, 0.7        \\
R21     &       2.68    &       -5.6, 5.8       \\
R22     &       4.52    &       5.6, -4.6       \\
R23     &       1.48    &       6.6, -8.0       \\
\hline
Ra      &       1.48    &       0.6, -1.6       \\
Rb      &       1.00    &       1.8, -0.4       \\
Rc      &       1.44    &       1.2, 0.8        \\
Rd      &       0.76    &       -1.0, 1.2       \\
Re      &       0.80    &       -1.6, -0.8      \\
\hline
\end{tabular}
\begin{tablenotes}
\centering
\item $^a$ Offsets from centre of the galaxy to the centre of each individual region.
\end{tablenotes}
\end{table}

At the end of the entire procedure, we have obtained a total of 23 HII regions in  the outer ring and 5 in the inner one. Figure \ref{fig:ring_ha_profile} shows the HII regions selected with the use of the described methodology for the two rings and  Table \ref{tab:seleccion} lists their characteristics: the position of each HII region in the ring with respect to that of the galaxy centre, its size and its observed integrated H$\alpha$ emission flux.

\begin{figure*}
\centering
\includegraphics[width=0.87\textwidth]{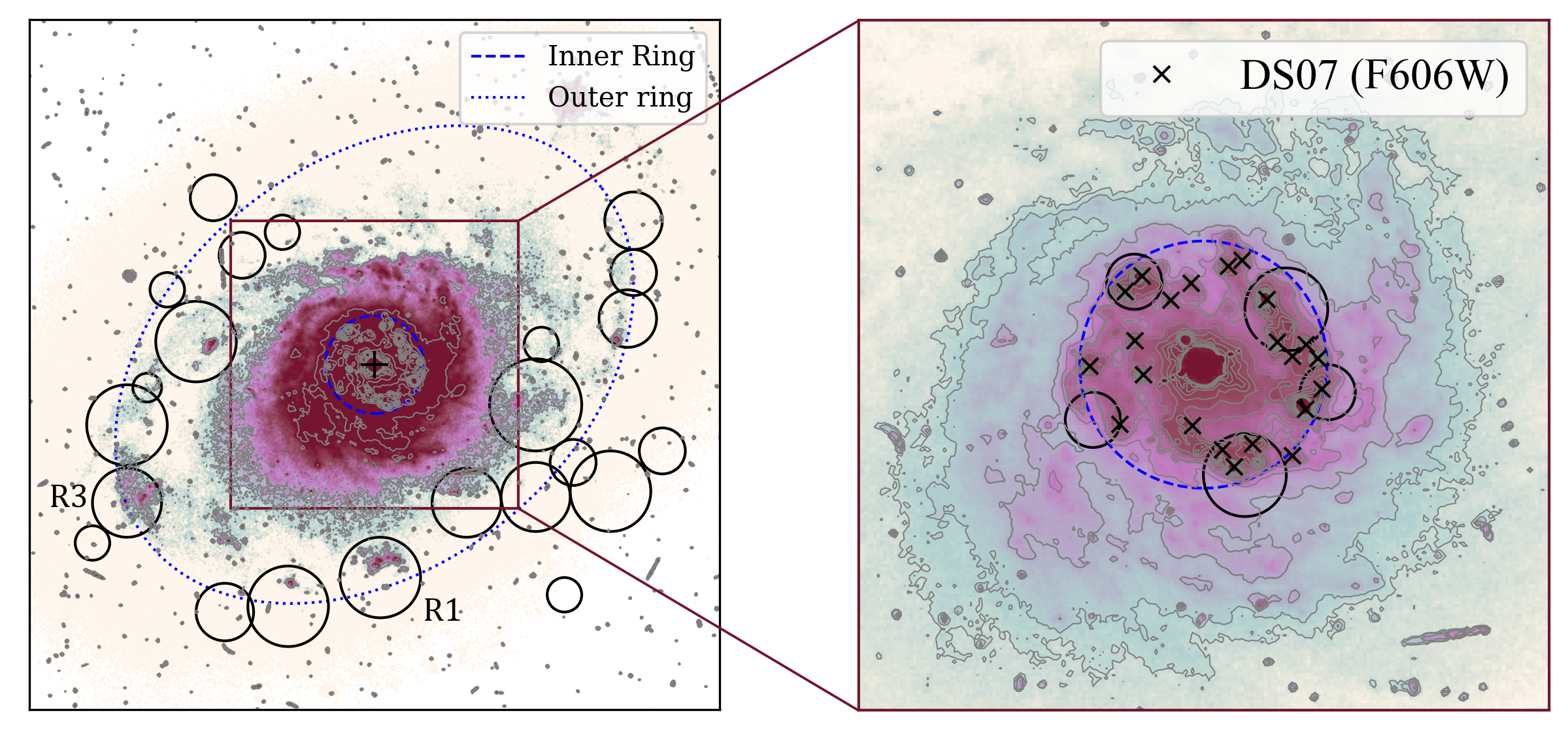}
\caption{Selected HII regions in HST images. Left panel: Outer ring selected HII regions superimposed on the HST WPC3 F606W image. The two regions labelled R1 and R2 seem to be associated with large star-forming complexes. Right panel: Enlargement of the central 5 x 5 arcsec of the galaxy showing the HII regions selected in the inner ring. The young star clusters identified by \citet[][DS07]{2007ApJ...661..149D} are marked with black crosses. Orientation is north up, east to the left.}
\label{fig:HST-606}
\end{figure*}

Next, we have used HST data from the WFC3 camera in the F606W filter in order to determine if our selected regions are associated with single young stellar clusters. Figure \ref{fig:HST-606} shows the emission maps in this band where we have superimposed our selected CNSFR apertures.
The left panel of the figure shows the outer ring of the galaxy where we can identify two regions, R1 and R3, that seem to encompass large star formation complexes in stead of single clusters. Regions R2, R7, and R9 also look like complexes in the HST images and the five mentioned regions show non-symmetric profiles in the H$\alpha$ emission line (see Fig. \ref{fig:Ha_OIII_map}). We will keep these regions within the CNSFRs study sample keeping in mind the possible effects of this result in our analysis.

The right panel of the figure shows enlarged the inner galaxy ring where the star clusters identified by \citep[][DS07]{2007ApJ...661..149D} at this wavelength are marked with crosses. Regions Ra, Rc and Rd seem to contain multiple stellar clusters although Rc only shows one ionising cluster in HST-WFC3 F336W filter (see Fig. \ref{fig:Ha_OIII_map}). On the other hand, region Re does not exhibit flux excess in any of the two filters at 3360 \AA\ or 6600 \AA\ wavelengths.

\subsubsection{Emission line measurements and uncertainties}
\label{sec:line measurements}

\begin{figure*}
\centering
\includegraphics[width=0.8\textwidth]{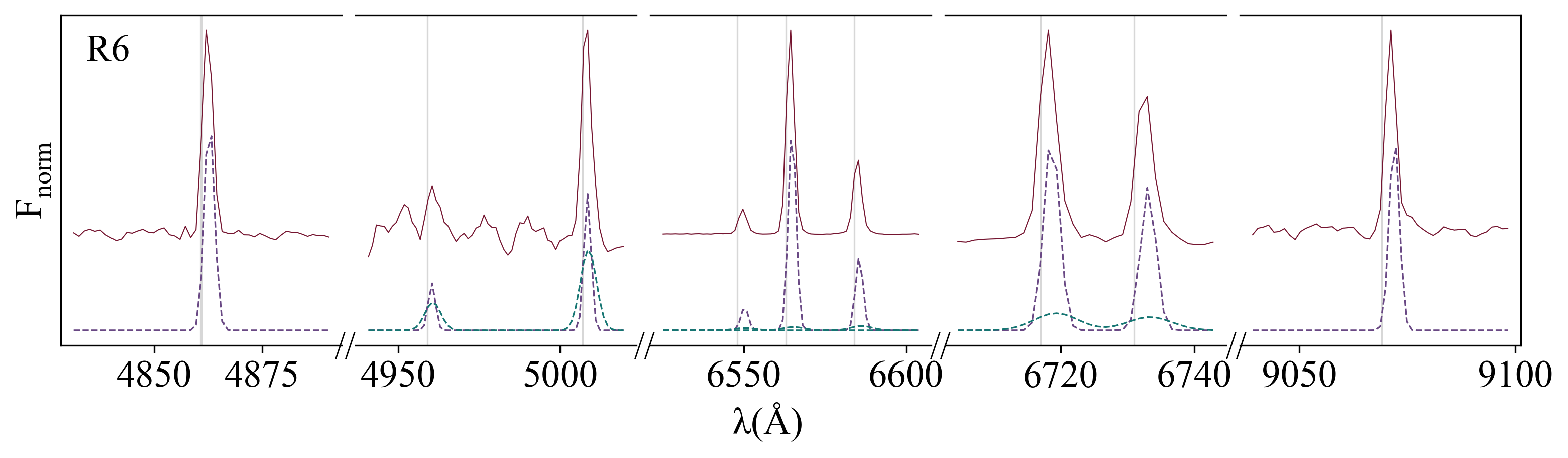}
\includegraphics[width=0.8\textwidth]{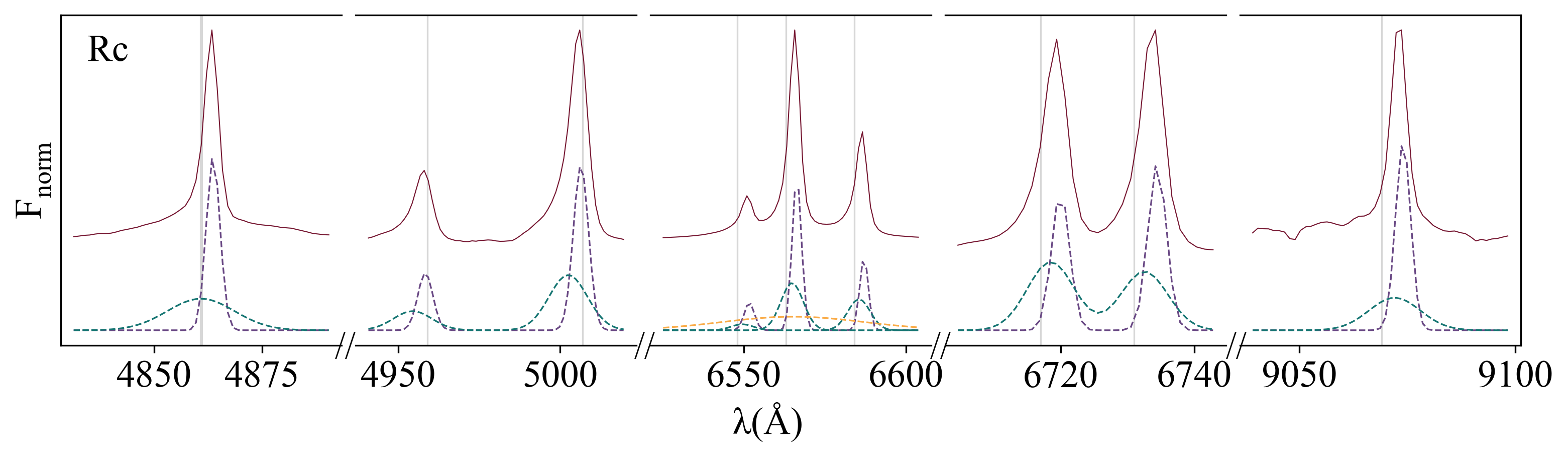}
\caption{Spectra of regions R6 and Rc (top and bottom panels, respectively) showing the kinematical components fitted. From left to right:  H$\beta$, [OIII]$\lambda \lambda$ 4959,5007 \AA , H$\alpha$ and [NII]$\lambda \lambda$ 6548,6584 \AA , [SII]$\lambda \lambda $ 6717,6731 \AA,\ and [SIII]$\lambda$ 9069 \AA\ emission lines are shown. The dotted purple lines show the components associated with the stellar clusters.}
 \label{fig:blend_7469}
\end{figure*}

We   extracted each region spectrum by integrating the flux inside its corresponding aperture. 
All emission lines with sufficient S/N (higher than 2 $\sigma _c$) appear to have at least two kinematical components and four of them: regions R2, R10, R13 and Rd, show three components in the [OIII]$\lambda \lambda$ 4959,5007 \AA\ emission lines. In order to separate the different kinematical components, we have used the code LiMe \citep[LIne MEasuring library][]{2022arXiv221210593F}. We have taken into account only those components that meet the requirement: $A_g >3\sigma_l$, with $A_g$ and $\sigma_l$ being the Gaussian amplitude and the local standard deviation of the residuals of the Gaussian fit in 30 \AA\ around each line centre. With this criteria, we have assigned only one component to the [SIII]$\lambda $ 9069 \AA\ emission line in all the outer ring regions. Figure \ref{fig:blend_7469} shows two de-blended examples: outer ring regions, R6 and R10, showing two and three kinematical components in the [OIII]$\lambda \lambda$ 4959,5007 \AA\ lines, and the inner ring region Rc, showing an extra component in the H$\alpha$ emission line that seems to be associated with high-density and high-velocity gas. For each line, we have ascribed the most intense and narrow component to the emission of the ionising cluster, and we have subtracted the rest of the components of the total spectrum in order to perform our subsequent analysis.

Next, we have measured the intensities of the identified emission lines in our spectra following the procedure descibed in \citet{ngc7742paper}. The intensities of all prominent emission lines with a S/N > 3 have been measured, discarding the most uncertain values. These lines are: H$\beta$ and H$\alpha$ Balmer lines; [OIII]$\lambda\lambda$ 4959,5007 \AA , [NII]$\lambda\lambda$ 6548,84 \AA , [SII]$\lambda\lambda$ 6716,31 \AA , and [SIII]$\lambda$ 9069 \AA\ forbidden lines. We have also measured the weak [SIII]$\lambda$ 6312 \AA\ and HeI$\lambda$ 6678 \AA\ lines detected with S/N > 1 and additionally, in the inner ring regions the HeI$\lambda$ 5875 \AA\ line with the same precision. The [SIII]$\lambda$ 6312 \AA\ and HeI $\lambda$ 6678 \AA\ lines have been measured with in 8 and 10 outer ring regions respectively. For the inner ring regions, there are three regions with [SIII]$\lambda$ 6312 \AA\ measurements and in all the regions both HeI lines have been measured.

\subsubsection{Extinction correction}

\begin{table*}
\centering
\setlength{\tabcolsep}{2pt}
\caption{Reddening-corrected emission line intensities (extract).  The full table is available in Appendix \ref{ap:tables}.}
\label{tab:lines}
\begin{tabular}{ccccccccccc}
\cline{2-11}
& Line & Hb & [OIII] & [OIII] &  [NII] & H$\alpha$ & [NII] & [SII] & [SII] & [SIII]\\
& $\lambda$ & 4861 & 4959 & 5007  & 6548 & 6563 & 6584 & 6717 & 6731 & 9069 \\
& f($\lambda$) & 0.000 & -0.024 & -0.035 & -0.311 & -0.313 & -0.316 & -0.334 & -0.336 & -0.561  \\ \hline
\multicolumn{1}{c}{Region ID} & \multicolumn{1}{c}{c(H$\beta$)} & \multicolumn{1}{c}{I(H$\beta$)$^a$}& \multicolumn{8}{c}{I($\lambda $)$^b$} \\ \hline
R1 & 0.50 $\pm$ 0.02 & 7.53 $\pm$ 0.42 & 42 $\pm$ 10 & 118 $\pm$ 9 & 304 $\pm$ 6 & 2870 $\pm$ 68 & 936 $\pm$ 9 & 469 $\pm$ 7 & 335 $\pm$ 7 & 82 $\pm$ 9\\ 
R2 & 0.78 $\pm$ 0.03 & 8.05 $\pm$ 0.67 & 206 $\pm$ 18 & 553 $\pm$ 18 & 457 $\pm$ 9 & 2870 $\pm$ 101 & 1394 $\pm$ 13 & 595 $\pm$ 13 & 438 $\pm$ 12 & 76 $\pm$ 11\\ 
R3 & 0.57 $\pm$ 0.04 & 4.47 $\pm$ 0.40 & 83 $\pm$ 17 & 229 $\pm$ 17 & 387 $\pm$ 9 & 2870 $\pm$ 108 & 1290 $\pm$ 13 & 551 $\pm$ 9 & 368 $\pm$ 8 & 111 $\pm$ 12\\ 
R4 & 0.54 $\pm$ 0.05 & 4.89 $\pm$ 0.59 & 109 $\pm$ 23 & 303 $\pm$ 23 & 415 $\pm$ 13 & 2870 $\pm$ 146 & 1188 $\pm$ 17 & 610 $\pm$ 15 & 478 $\pm$ 14 & 135 $\pm$ 17\\ 
R5 & 0.46 $\pm$ 0.05 & 2.75 $\pm$ 0.36 & 145 $\pm$ 24 & 407 $\pm$ 24 & 498 $\pm$ 13 & 2870 $\pm$ 158 & 1504 $\pm$ 19 & 720 $\pm$ 16 & 555 $\pm$ 15 & 149 $\pm$ 19\\ 
R6 & 0.88 $\pm$ 0.03 & 3.16 $\pm$ 0.23 & 56 $\pm$ 14 & 148 $\pm$ 13 & 315 $\pm$ 8 & 2870 $\pm$ 90 & 1025 $\pm$ 13 & 385 $\pm$ 10 & 275 $\pm$ 9 & 154 $\pm$ 12\\ 
R7 & 0.81 $\pm$ 0.04 & 5.51 $\pm$ 0.54 & 60 $\pm$ 18 & 159 $\pm$ 17 & 335 $\pm$ 11 & 2870 $\pm$ 120 & 1118 $\pm$ 16 & 521 $\pm$ 12 & 327 $\pm$ 11 & 122 $\pm$ 15\\ 
\hline
\end{tabular}
\begin{tablenotes}
\centering
\item $^a$ In units of 10$^{-15}$ erg/s/cm$^2$.\
\item $^b$ Values normalised to I(H$\beta$) 10$^{-3}$. 
\end{tablenotes}
\end{table*}

For the regions in the outer galaxy ring, the measured line intensities have been corrected using the reddening constant, c(H$\beta$), derived from the ratio of the Balmer H$\alpha$ and H$\beta$ lines assuming a simple screen distribution of the dust and the same extinction for emission lines and the stellar continuum. We have adopted the Galactic extinction law of \citet{reddening}, with a specific attenuation of R$_v$ = 2.97. A theoretical value for the H$\alpha $/H$\beta $ ratio of 2.87 corresponding to n$_e$ = 100 cm$^{-3}$ and T$_e$ = 10$^4$ K for the electron density and temperature respectively has been adopted.  The upper panel of Table \ref{tab:lines} shows, for each selected outer ring HII region, the reddening-corrected emission line intensities of strong lines relative to H$\beta$, and its corresponding reddening constant.

\begin{figure}
\centering
\includegraphics[width=\columnwidth]{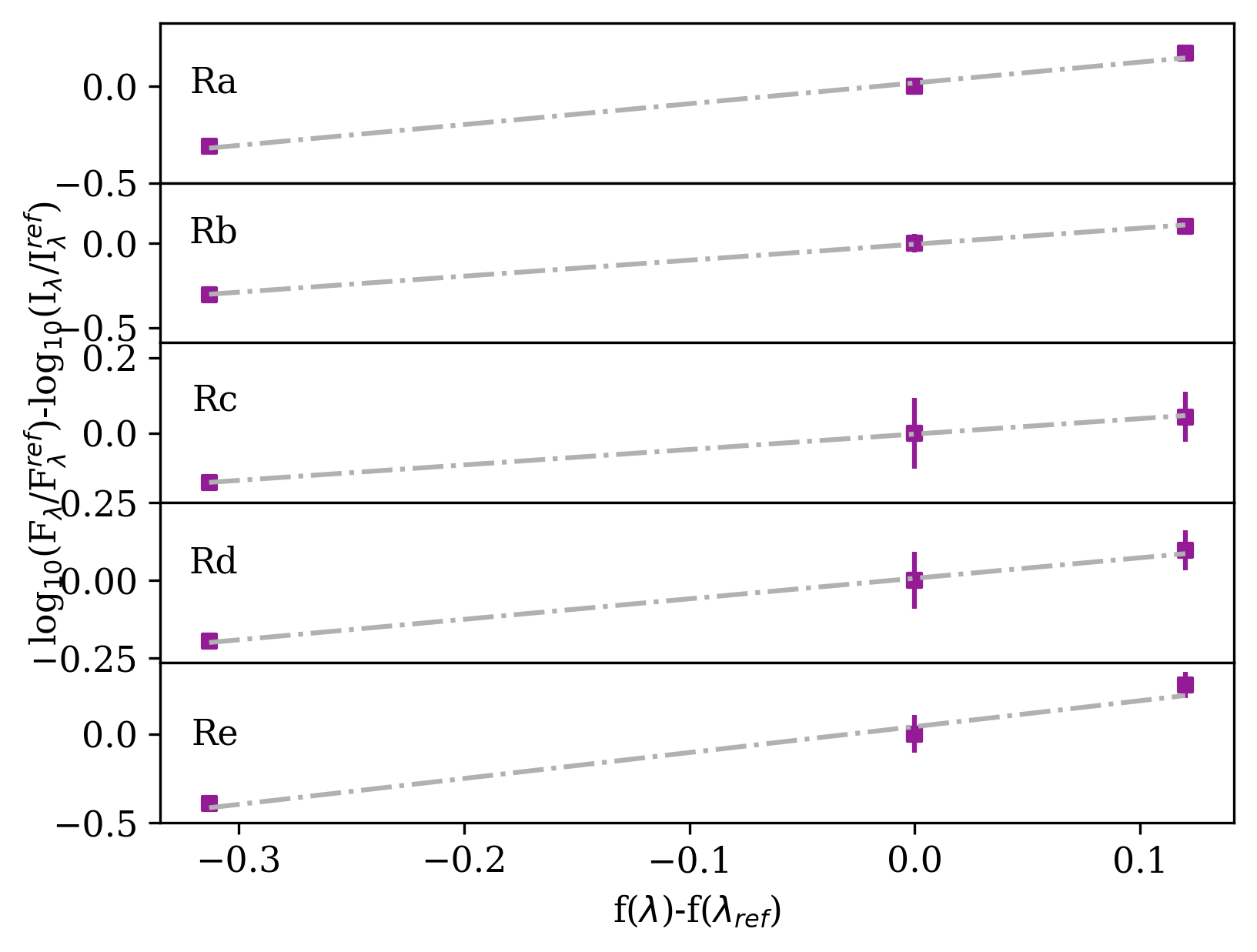}
\caption{Linear regressions of c(H$\beta$) values from hydrogen and helium lines for inner ring regions.}
 \label{fig:redening}
\end{figure}

\begin{table*}
\caption{Measured HeI line intensities and logarithmic extinction coefficients.}
\label{tab:reddening}
\centering
\begin{tabular}{ccccc}
\hline
Region ID & HeI$\lambda$ 5875\AA$^a$ & HeI$\lambda$ 6678\AA$^a$ & c(Hb)$_{H_\alpha}$   & c(Hb)$_{fit}$\\
\hline
Ra        &  6.903 $\pm$ 0.095   & 2.930 $\pm$ 0.070 & 0.99 $\pm$ 0.02 & 1.081 $\pm$ 0.007   \\
Rb        &  5.00  $\pm$ 0.11    & 1.80  $\pm$ 0.16  & 0.97 $\pm$ 0.02 & 0.9506 $\pm$ 0.0004 \\
Rc        &  6.751 $\pm$ 0.061   & 2.12  $\pm$ 0.32  & 0.42 $\pm$ 0.04 & 0.4102 $\pm$ 0.0001 \\
Rd        &  3.382 $\pm$ 0.041   & 1.20  $\pm$ 0.18 & 0.62 $\pm$ 0.05 & 0.661 $\pm$ 0.001   \\
Re        &  2.001 $\pm$ 0.030   & 1.07  $\pm$ 0.18 & 1.25 $\pm$ 0.04 & 1.46 $\pm$ 0.04    \\
\hline
\end{tabular}
\begin{tablenotes}
\centering
\item $^a$ In units of 10$^{-16}$ erg/s/cm$^2$.
\end{tablenotes}
\end{table*}

On the other hand, the study of circumnuclear regions around active galactic nuclei often presents an additional problem due to the high surface brightness at the galaxy centre. If a single exposure image is used for the analysis of outer and inner HII regions simultaneously, it is probable that in the central part of the galaxy some of the strongest nebular emission lines be saturated or the flux values fall within the non-linearity range of the detector. In our particular case, the H$\alpha$ line looks saturated in at least 30 pixels around the galaxy centre (H$\alpha$/H$\beta$ < 2.7; see Sect. \ref{emmision maps}). For this reason, we have used the weaker HeI emission lines at $\lambda$ 5875\AA\ and $\lambda$6678 \AA\  in order to derive the extinction in regions within the inner ring following the methodology proposed in \citet{2022MNRAS.516..749Z}, after checking carefully that the H$\beta$ and HeI $\lambda$ 6678 \AA\ lines show  the same spatial distribution (Fig. \ref{fig:sersic_}) being also similar to that of the continuum emission map at 6060 \AA\ (see Fig. \ref{fig:HST-606}). A theoretical value for the ratio of HeI$\lambda$ 5875 \AA / HeI$\lambda$ 6678 \AA\ ratio of 3.52 \citep[][for n$_e$ = 100 cm$^{-3}$ and T$_e$ = 10$^4$ K]{pyneb} has been assumed. The reddening constant c(H$\beta$) has been derived by performing a linear regression using all the available HI and HeI emission lines ({Fig. \ref{fig:redening}). Table \ref{tab:reddening} shows, for each inner ring HII region, the results obtained with this procedure and lists in columns 1 to 5: (1) the region ID; (2 and 3) the HeI$\lambda$ 5875 \AA\ and HeI$\lambda$ 6678 \AA\ line fluxes respectively; and (4 and 5) the reddening constant calculated using only the H$\alpha$/H$\beta$ ratio and both the HI and HeI lines in the  fit. The results of these two fits using are very similar and particularly, for regions Rb, Rc and Rd the results are fully compatible. Also, the intercept of all regression lines is compatible with zero. The lower panel of Tab. \ref{tab:lines} shows, for the selected inner HII regions, the reddening-corrected emission line intensities of strong lines relative to H$\beta$, and its corresponding reddening constant.

\subsubsection{Chemical abundances}
\label{sec:abundances}

CNSFR metallicities have been traced by their sulphur abundances following the methodology described in \citet{2022MNRAS.511.4377D}, well suited to the use of MUSE data since it is based on red-to-near infrared spectroscopy and presents two interesting advantages:  reddening effects are decresed due to the longer wavelengths involved and  contrary to the case of oxygen, sulphur does not seem to be depleted in diffused clouds \citep{2021A&A...648A.120R}. Additionally, the electron temperature-sensitive line of [SIII] at  $\lambda$ 6312 \AA\ can be detected and measured up to, at least, solar abundances \citep{2007MNRAS.382..251D} as those expected in the central regions of galaxies.

\begin{figure}
 \includegraphics[width=\columnwidth]{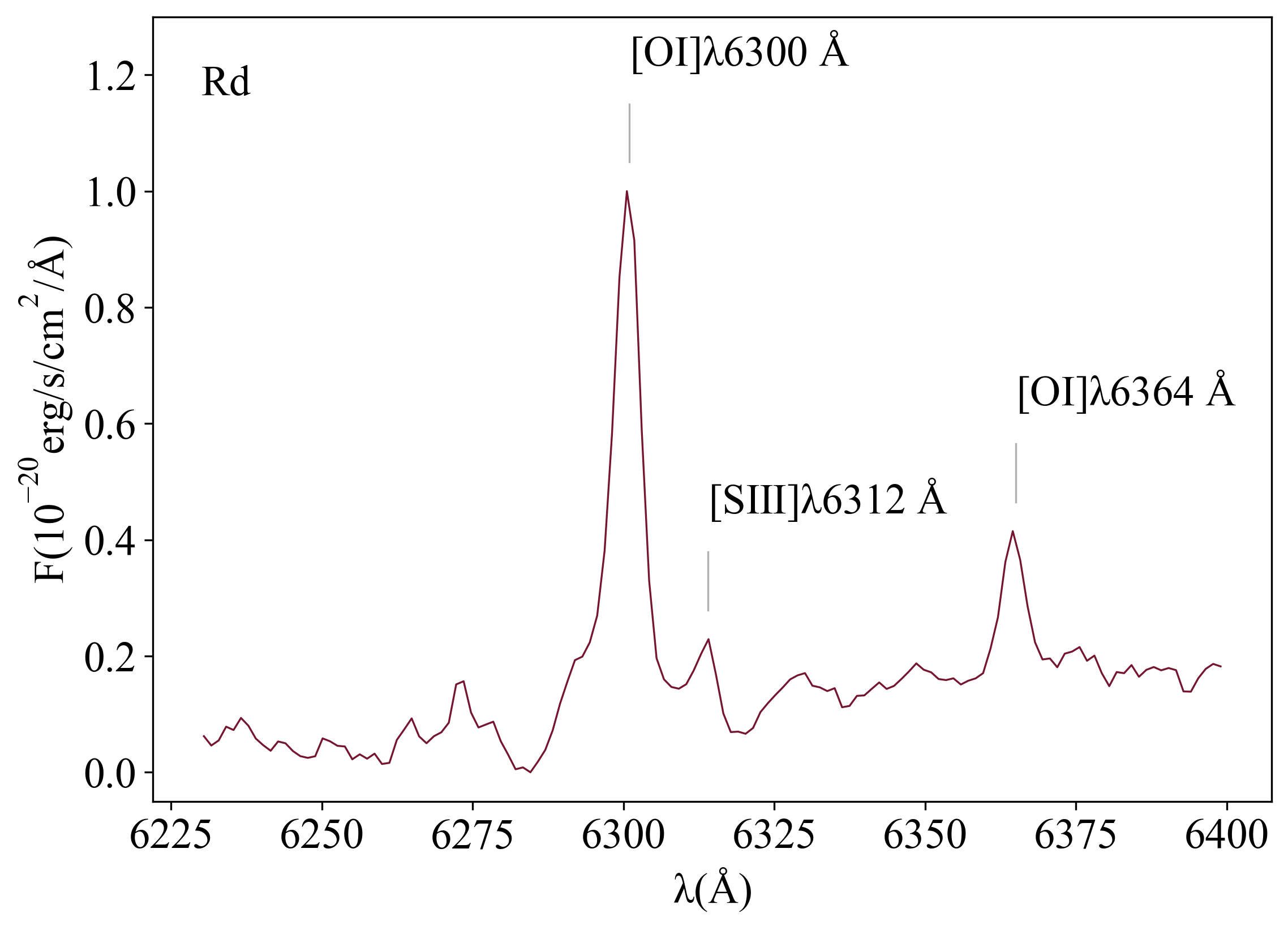}
 \caption{[SIII]$\lambda$ 6312 \AA\ reddening-corrected emission line as detected in region Rd. The positions of the  [OI]$\lambda \lambda $ 6300,6364 \AA\ emission lines are also shown.} 
 \label{fig:6312_Rd}
\end{figure}

\begin{table*}
\centering
\caption{Ionic and total sulphur abundances derived by the direct method for the CNSFRs with measured [SIII]$\lambda$ 6312 \AA\ line intensities.}
\label{tab:sulfur_measurements}
\begin{tabular}{ccccccc}
\hline
Region ID & I([SIII]$\lambda $6312)$^a$& R$_{S3}$& t$_e$([SIII])$^b$ & 12+log(S$^{+}$/H$^{+}$)& 12+log(S$^{++}$/H$^{+}$) & 12+log(S/H)\\ \hline
R1       &      13.4 $\pm$ 0.8   &      158.7 $\pm$ 0.9  &      0.6731 $\pm$ 0.001    &      6.8174 $\pm$ 0.0058      &      6.5331 $\pm$ 0.0459      &       6.999 $\pm$ 0.016\\ 
R2       &      6.9 $\pm$ 0.6    &      306.9 $\pm$ 0.8  &      0.5773 $\pm$ 0.0003   &      7.1755 $\pm$ 0.0075      &      6.6982 $\pm$ 0.0606      &       7.3 $\pm$ 0.016\\ 
R3       &      12.2 $\pm$ 1.1   &      139.8 $\pm$ 0.8  &      0.6958 $\pm$ 0.0011   &      6.8255 $\pm$ 0.0061      &      6.6267 $\pm$ 0.0489      &       7.038 $\pm$ 0.019\\ 
R4       &      19.8 $\pm$ 2.5   &      115.2 $\pm$ 1.1  &      0.735 $\pm$ 0.002    &      6.8197 $\pm$ 0.0089      &      6.6489 $\pm$ 0.0542      &       7.044 $\pm$ 0.023\\ 
R5       &      13.1 $\pm$ 1.7   &      107.3 $\pm$ 0.8  &      0.7507 $\pm$ 0.0018   &      6.8585 $\pm$ 0.0079      &      6.6651 $\pm$ 0.0549      &       7.074 $\pm$ 0.022\\ 
R6       &      20.0 $\pm$ 1.6   &      83.4 $\pm$ 1.2   &      0.8157 $\pm$ 0.004    &      6.4619 $\pm$ 0.0101      &      6.5899 $\pm$ 0.0378      &       6.832 $\pm$ 0.022\\ 
R18      &      3.2 $\pm$ 0.5    &      101.6 $\pm$ 1.3  &      0.7637 $\pm$ 0.0032   &      6.6869 $\pm$ 0.0119      &      6.648 $\pm$ 0.0654       &       6.969 $\pm$ 0.032\\ 
R20      &      2.8 $\pm$ 0.6    &      80.7 $\pm$ 2.3   &      0.8251 $\pm$ 0.0084   &      6.8015 $\pm$ 0.0158      &      6.4714 $\pm$ 0.1094      &       6.968 $\pm$ 0.036\\ 
\hline

Ra       &      167.7 $\pm$ 3.0  &      365.1 $\pm$ 0.7  &      0.555 $\pm$ 0.0002   &      6.7187 $\pm$ 0.0246      &      7.0098 $\pm$ 0.011       &       7.189 $\pm$ 0.011\\ 
Rb       &      49.9 $\pm$ 0.5   &      430.9 $\pm$ 1.2  &      0.5337 $\pm$ 0.0004   &      6.8926 $\pm$ 0.0185      &      6.9548 $\pm$ 0.0174      &       7.226 $\pm$ 0.013\\ 
Rd       &      208.0 $\pm$ 1.5  &      60.8 $\pm$ 1.1   &      0.9209 $\pm$ 0.0068   &      6.2023 $\pm$ 0.0347      &      6.6461 $\pm$ 0.0272      &       6.78 $\pm$ 0.022\\ 
\hline

\end{tabular}
\begin{tablenotes}
\centering
\item $^a$ In units of 10$^{-18}$ erg/s/cm$^2$.\\
\item $^b$ In units of 10$^{4}$ K.\\
\item * Region near SN explosion.
\end{tablenotes}
\end{table*}

This line has been measured with a S/N higher than 1 in $\sim$ 35 \% (8 out of 23) of the HII regions within the outer galaxy ring and in three out of five regions within the inner galaxy ring. Figure \ref{fig:6312_Rd} shows the spectrum of the inner ring region with the highest S/N in this sulphur line as an example. For these regions, total sulphur abundances have been derived by the direct method as described in \citet{ngc7742paper}. Table \ref{tab:sulfur_measurements} lists in columns 1 to 7: (1) the region ID; (2) the measured  [SIII]$\lambda$ 6312 \AA\ emission line intensity; (3) the R$_{S3}$ line ratio; (4) the [SIII] electron temperature; (5 and 6) the ionic abundances of S$^+$ and S$^{++}$ relative to H$^+$; and (7) the total S/H abundance. 

For the rest of the regions we have used the S$_{23}$ parameter and the calibration given in \citet{2022MNRAS.511.4377D} to derive empirical sulphur abundances. The sulphur abundances derived from this calibration for all the objects in our sample are given in Table \ref{tab3}.

\begin{table}
\centering
\caption{Sulphur abundances of the observed CNSFRs derived by empirical methods.}
\label{tab3}
\begin{tabular}{ccc}
\hline
Region ID & S23 &  12+log(S/H)\\ \hline
R1 & 1.086 $\pm$ 0.031 & 6.716 $\pm$ 0.031\\ 
R2 & 1.294 $\pm$ 0.041 & 6.896 $\pm$ 0.035\\ 
R3 & 1.301 $\pm$ 0.044 & 6.902 $\pm$ 0.038\\ 
R4 & 1.554 $\pm$ 0.061 & 7.096 $\pm$ 0.047\\ 
R5 & 1.785 $\pm$ 0.068 & 7.257 $\pm$ 0.048\\ 
R6 & 1.189 $\pm$ 0.044 & 6.807 $\pm$ 0.04\\ 
R7 & 1.268 $\pm$ 0.054 & 6.874 $\pm$ 0.046\\ 
R8 & 1.153 $\pm$ 0.052 & 6.776 $\pm$ 0.047\\ 
R9 & 1.523 $\pm$ 0.054 & 7.074 $\pm$ 0.043\\ 
R10 & 1.578 $\pm$ 0.048 & 7.114 $\pm$ 0.038\\ 
R11 & 1.261 $\pm$ 0.058 & 6.868 $\pm$ 0.05\\ 
R12 & 1.726 $\pm$ 0.063 & 7.217 $\pm$ 0.046\\ 
R13 & 1.891 $\pm$ 0.083 & 7.327 $\pm$ 0.057\\ 
R14 & 1.221 $\pm$ 0.109 & 6.835 $\pm$ 0.093\\ 
R15 & 1.473 $\pm$ 0.072 & 7.036 $\pm$ 0.056\\ 
R16 & 1.543 $\pm$ 0.079 & 7.089 $\pm$ 0.06\\ 
R17 & 1.318 $\pm$ 0.05 & 6.916 $\pm$ 0.043\\ 
R18 & 1.419 $\pm$ 0.079 & 6.995 $\pm$ 0.062\\ 
R19 & 1.303 $\pm$ 0.078 & 6.903 $\pm$ 0.065\\ 
R20 & 1.906 $\pm$ 0.107 & 7.336 $\pm$ 0.071\\ 
R21 & 1.367 $\pm$ 0.079 & 6.954 $\pm$ 0.064\\ 
R22 & 1.469 $\pm$ 0.073 & 7.033 $\pm$ 0.057\\ 
R23 & 1.432 $\pm$ 0.128 & 7.005 $\pm$ 0.1\\ 

\hline
Ra & 0.78 $\pm$ 0.021 & 6.411 $\pm$ 0.026\\ 
Rb & 0.755 $\pm$ 0.022 & 6.383 $\pm$ 0.028\\ 
Rc & 1.126 $\pm$ 0.06 & 6.752 $\pm$ 0.054\\ 
Rd & 1.309 $\pm$ 0.053 & 6.908 $\pm$ 0.045\\ 
Re & 1.032 $\pm$ 0.029 & 6.667 $\pm$ 0.03\\ 
\hline
\end{tabular}
\end{table}

\subsection{Ionising clusters}

\subsubsection{Integrated magnitudes}
\label{sec:int_flux}

\begin{table*}
\centering
\setlength{\tabcolsep}{2pt}
\caption{Colours and magnitudes results (extract). The full table is available in Appendix \ref{ap:tables}.}
\label{tab:magnitudes}
\begin{tabular}{cccccc}
\hline
Region ID & m$_i$ (mag) & m$_r$ (mag) & M$_i$ (mag) & M$_r$ (mag) & r-i ( mag)\\ \hline
R1 & 17.48 $\pm$ (0.35 $ \times$ 10$^{-4}$) & 17.81 $\pm$ (0.30 $ \times$ 10$^{-4}$) & -14.25 $\pm$ (0.35 $ \times$ 10$^{-4}$) & -13.92 $\pm$ (0.46 $ \times$ 10$^{-4}$) & 0.324 $\pm$ (0.460 $ \times$ 10$^{-4}$)\\ 
R2 & 17.36 $\pm$ (0.31 $ \times$ 10$^{-4}$) & 17.78 $\pm$ (0.30 $ \times$ 10$^{-4}$) & -14.37 $\pm$ (0.31 $ \times$ 10$^{-4}$) & -13.95 $\pm$ (0.43 $ \times$ 10$^{-4}$) & 0.427 $\pm$ (0.431 $ \times$ 10$^{-4}$)\\ 
R3 & 17.66 $\pm$ (0.33 $ \times$ 10$^{-4}$) & 17.99 $\pm$ (0.28 $ \times$ 10$^{-4}$) & -14.07 $\pm$ (0.33 $ \times$ 10$^{-4}$) & -13.75 $\pm$ (0.43 $ \times$ 10$^{-4}$) & 0.322 $\pm$ (0.429 $ \times$ 10$^{-4}$)\\ 
R4 & 16.99 $\pm$ (0.27 $ \times$ 10$^{-4}$) & 17.40 $\pm$ (0.25 $ \times$ 10$^{-4}$) & -14.74 $\pm$ (0.27 $ \times$ 10$^{-4}$) & -14.33 $\pm$ (0.37 $ \times$ 10$^{-4}$) & 0.414 $\pm$ (0.367 $ \times$ 10$^{-4}$)\\ 
R5 & 17.52 $\pm$ (0.29 $ \times$ 10$^{-4}$) & 17.90 $\pm$ (0.27 $ \times$ 10$^{-4}$) & -14.21 $\pm$ (0.29 $ \times$ 10$^{-4}$) & -13.83 $\pm$ (0.40 $ \times$ 10$^{-4}$) & 0.385 $\pm$ (0.396 $ \times$ 10$^{-4}$)\\ 
R6 & 19.50 $\pm$ (0.76 $ \times$ 10$^{-4}$) & 19.85 $\pm$ (0.67 $ \times$ 10$^{-4}$) & -12.23 $\pm$ (0.76 $ \times$ 10$^{-4}$) & -11.88 $\pm$ (1.02 $ \times$ 10$^{-4}$) & 0.347 $\pm$ (1.017 $ \times$ 10$^{-4}$)\\ 
R7 & 18.31 $\pm$ (0.64 $ \times$ 10$^{-4}$) & 18.68 $\pm$ (0.57 $ \times$ 10$^{-4}$) & -13.42 $\pm$ (0.64 $ \times$ 10$^{-4}$) & -13.05 $\pm$ (0.86 $ \times$ 10$^{-4}$) & 0.370 $\pm$ (0.856 $ \times$ 10$^{-4}$)\\ 
\hline
\end{tabular}
\end{table*}

\begin{figure}
\centering
\includegraphics[width=\columnwidth]{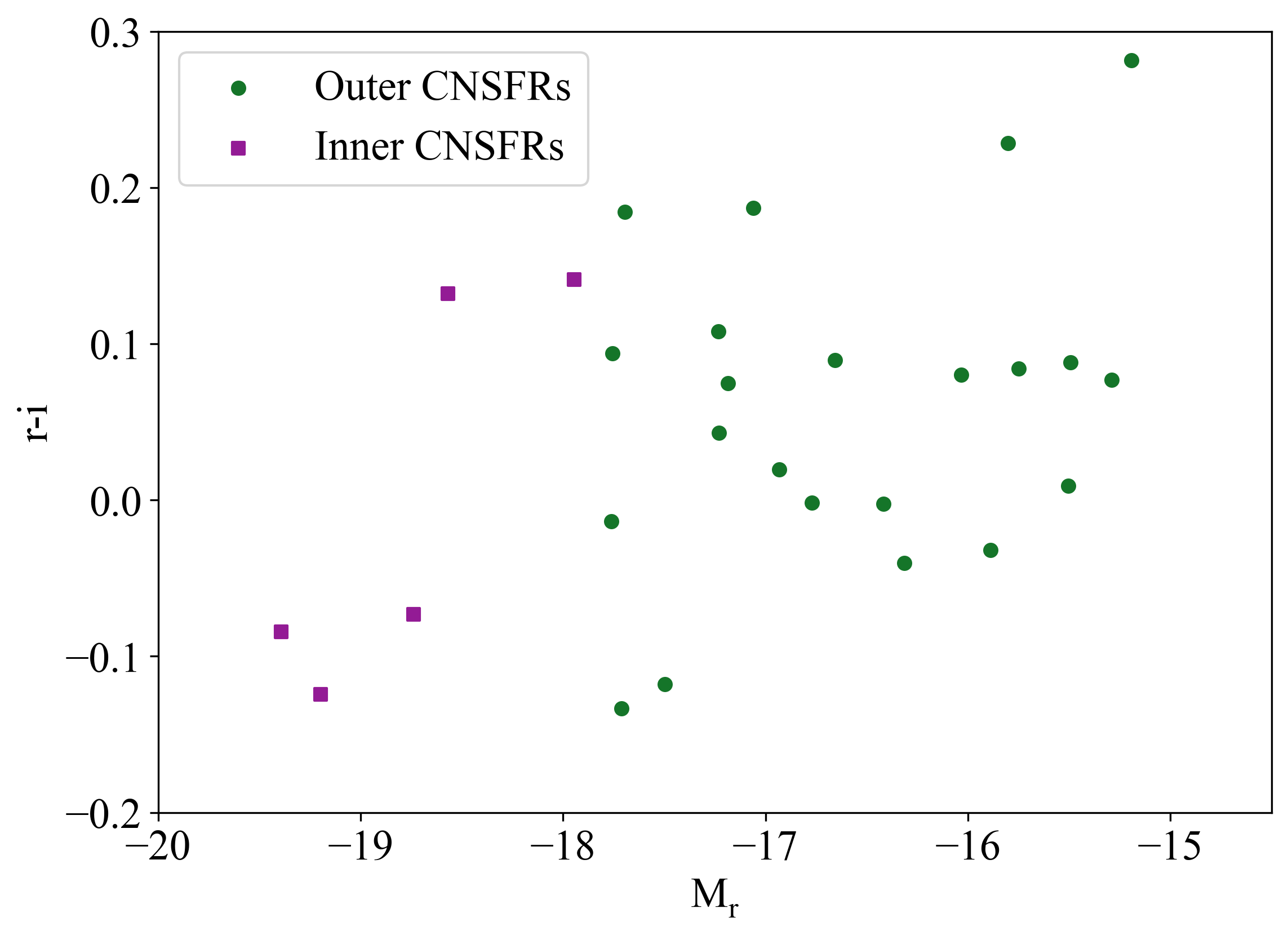}
\caption{Colour-magnitude diagram for outer ring (green dots) and inner ring HII regions (purple squares).}
\label{fig:color-mag}
\end{figure}

For each region of the sample, we have calculated the fluxes inside the \href{http://svo2.cab.inta-csic.es/theory/fps/index.php?mode=browse&gname=SLOAN&asttype=}{Sloan Digital Sky Survey (SDSS) filters} using their reddening-corrected integrated spectrum, previously masking the nebular emission lines, and the expressions shown in \citet{ngc7742paper}. Table \ref{tab:magnitudes} shows these integrated magnitudes and the corresponding derived quantities for each HII region within the inner and outer rings listing in columns 1 to 6: (1) the region ID; (2) the apparent magnitude for the i band; (3) the apparent magnitude for the r band; (4) the absolute magnitude for the i band; (5) the absolute magnitude for the r band; and (6) the r-i colour. Figure \ref{fig:color-mag} shows the colour-magnitude diagram of the studied ionised regions as a first approach to their stellar population properties. Inner ring regions show r-band luminosities larger than the rest. Also, regions Rc and Rd show larger r-i values than the rest. These regions were identified with multiple clusters in the 6060 \AA\ wavelength HST filter in Sect. \ref{sec:segmentation}. In general, HII regions in the outer ring look a factor of up to 40 fainter and somewhat redder than those in the inner ones, an effect that seems to be real given the small reddening correction involved.

\subsubsection{Stellar absorption lines}
\label{EWCaT}

\begin{table*}
\centering
\caption{Extraction parameters for absorption lines.}
\label{tab:line ranges stars}
\begin{tabular}{lcccc}
\hline
Line &  $\lambda_c$ (\r{A})  & $\Delta\lambda$ (\r{A}) & $\Delta\lambda_{left}$ (\r{A}) & $\Delta\lambda_{right}$ (\r{A}) \\ 
\hline
 CaII$\lambda$ 8498 \AA & 6563 & 30& 8467.5 - 8482.5&8702.5 - 8717.5\\
 CaII$\lambda$ 8548 \AA & 4861 & 30& 8467.5 - 8482.5&8702.5 - 8717.5\\
 CaII$\lambda$ 8662 \AA & 5007 & 30& 8467.5 - 8482.5&8702.5 - 8717.5\\
 MgI$\lambda$ 5171 \AA & 6583& 30 & 5128.5 - 5143.5 & 5235.5 - 5250.5\\
\hline
\end{tabular}
\begin{tablenotes}
\centering
\item All wavelengths are in rest frame.
\end{tablenotes}
\end{table*}

In all the studied regions the far red CaII$\lambda \lambda $ 8498, 8542, 8662 \AA\ (CaT) and MgI$\lambda$ 5171 \AA\ stellar lines are clearly detected. We have calculated their equivalent widths (EW) measuring the flux in 30 \AA\ continuum bands at both sides of each of the lines and assuming a linear behaviour of the continuum between them. Table \ref{tab:line ranges stars} gives the identification of each line in column 1, its central wavelength in \AA\ in column 2, its equivalent width in \AA\ in column 3, and the limits of the two continuum side-bands, in \AA, in columns 4 and 5. To calculate the error of these measurements, we have taken into account the standard deviation in the continuum bands propagating them in quadrature.

In principle, absorption line EW measurements in objects with different velocity dispersions have to be corrected for the broadening of the spectral lines. This effect decreases the continuum level and the line fluxes integrated in fixed width apertures, providing lower EW values. In order to evaluate this correction in our data, we have convolved some stellar spectra \citep[giant and supergiant stars][]{2019A&A...629A.100I} with known Gaussian functions of different $\sigma$ (from 1\AA\ to 8 \AA) measuring the CaII lines EW in each broadened spectrum. The correction ($\Delta$EW(CaT)) has been calculated as the average difference between the EW of CaII measurements  with and without broadening \citep[see][]{ 1990MNRAS.242..271T}. Finally, we  fitted a second-order polynomial, which applies to $\sigma$ values higher than 4.5 \AA :
\begin{equation}
\Delta EW(CaT) = 0.02931 \cdot \sigma^2 - 0.197\cdot \sigma + 0.2985.
\end{equation}
Here $\sigma$ is the velocity dispersion in \AA. This correction takes values lower than 0.05 \AA\ for $\sigma$ = 5, which corresponds with a velocity dispersion around 170 km/s. Since the  common values for CNSFRs are lower than this, no correction has been applied.

It is well known that, in the presence of an AGN, inferring stellar population properties from stellar absorption lines can be difficult due to the presence of  a non-thermal extra component to  the continuum coming from the galactic nucleus \citep[see][]{1990MNRAS.242..271T}. But also in regions with high star formation rates there is a non-negligible contribution by the nebular continuum which usually is not taken into account. Both of them can  dilute the starlight weakening and distorting the absorption features. With this in mind we have calculated a dilution factor for each of the lines as the ratio between the observed EWs and a standard value taken as reference, D = EW$_{obs}$ / EW$_{ref}$. For this comparison, we have used representative values of the EW absorption features measured in the spectra of normal spiral galaxy nuclei which correspond to old, metal rich stellar populations \citep[EW$_{ref}$(CaII) = 7.7 $\pm$ 0.5 \AA , EW$_{ref}$(MgI) = 5.18 $\pm$ 0.71 \AA ;][]{1990MNRAS.242..271T,1983ApJ...269..466K}. Figure \ref{fig:Dilution_NGC7469} shows the MgI dilution as a function of CaII dilution for  the studied CNSFR. The red solid line corresponds to the dilution by the nebular continuum associated with a nebula ionised by a young star cluster synthesised using the PopStar code \citep{Popstar} with Salpeter’s IMF \citep[m$_{low}$ = 0.85 M$_\odot$, m$_{up}$ = 120 M$_\odot$][]{Salpeter1955} and 5.5 Ma.

\begin{figure}
\centering
\includegraphics[width=\columnwidth]{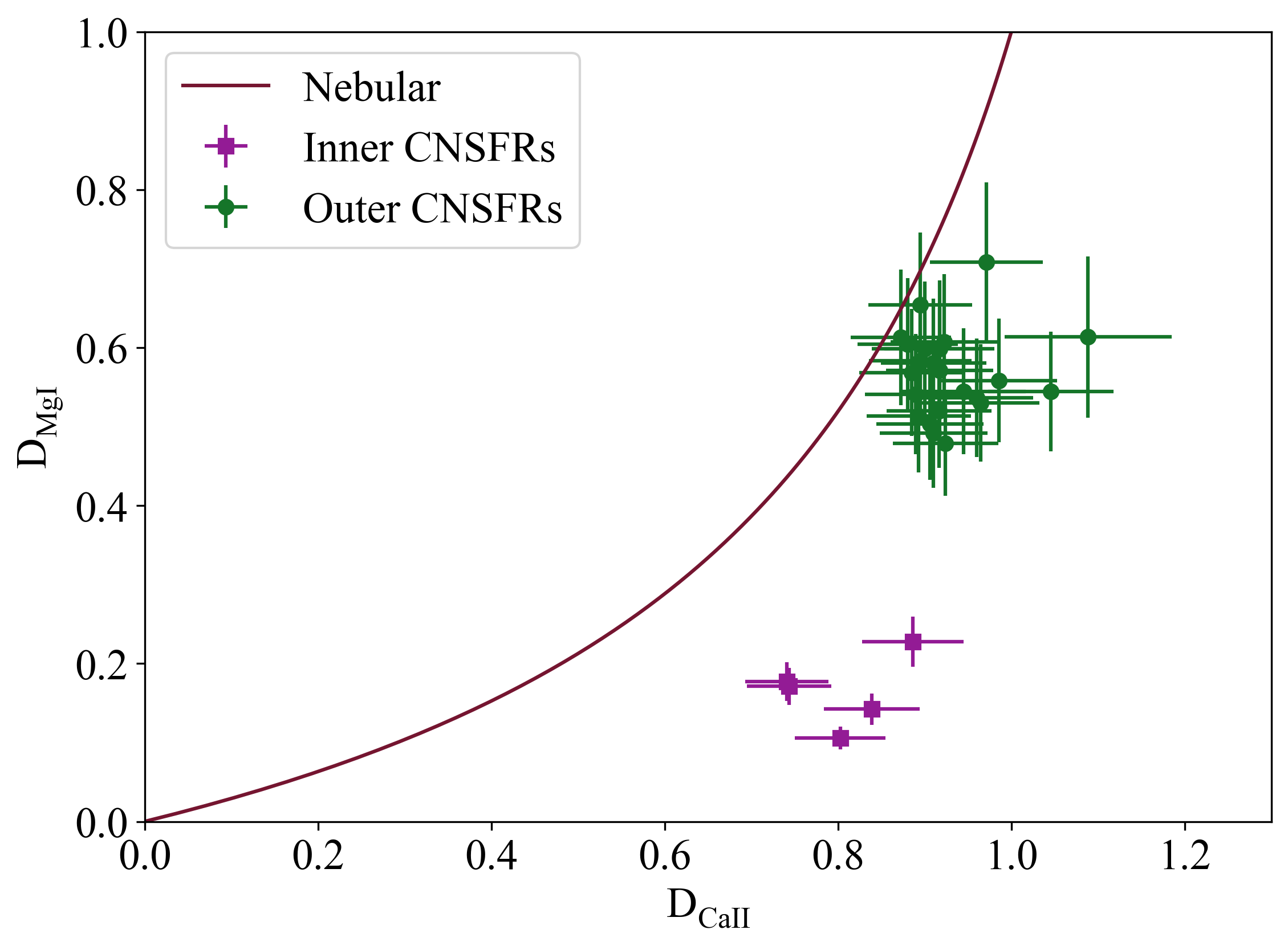}
\caption{Dilution of the MgI absorption line vs that of the CaII triplet lines. The red lines represent the dilution due to the nebular continuum in the optical and near IR.}
 \label{fig:Dilution_NGC7469}
\end{figure}

All of our outer ring regions show MgI dilutions consistent with the contribution by a nebular continuum (about 40\%) with the CaT lines looking almost undiluted. This would be expected from young clusters with red supergiant stars whose EW are larger than the reference value corresponding to a galaxy nucleus population dominated by red giants. In a few cases, the CaT EWs are even increased with respect to the reference value. In the case of the inner ring regions, larger dilutions are found in the MgI lines  (about 80\%) which could be due to an additional continuum originating in the AGN, but, again, the CaT lines show very little dilution pointing to a larger contribution by red supergiant stars. None of these effects can be ascribed to the additional presence of a metal rich stellar population since EW(MgI) increases with metal abundance faster than EW(CaII) in the high-metallicity regime \citep{1984ApJ...287..586B} and the MgI feature should then be larger than the CaT one. Thus, in order to explain the observations, the stellar population in our clusters should have stars with CaT features stronger than those in normal early spiral galaxies, like young red supergiants. Other authors have already suggested that the presence of strong CaT features in galactic nuclei is representative of relatively recent starbursts \citep[see for example][]{1990MNRAS.242..271T,1993Ap&SS.205...85G,1995A&A...301...55O}.

\subsubsection{Stellar velocity dispersions}
\label{sigma}

Stellar velocity dispersions can be obtained from the absorption stellar lines using the cross-correlation technique proposed by \citet{1979AJ.....84.1511T}. This method calculates the line-of-sight velocity dispersion by comparing a stellar template with the observed stellar population spectrum. This method is based on the assumption  that a galaxy spectrum is  represented by the sum of different stellar spectra with different velocity offsets convolved with a broadening function, 
\begin{equation}\label{eq:gn}
g(n)\sim \alpha [t(n)\ast b(n-\delta)],
\end{equation}
where $\ast$ means the convolution product, g(n) is the galaxy spectrum, $\alpha$ is the number of stars, t(n) is the template spectrum, b(n) is the broadening function and $\delta$ is the offset of the broadening function with respect to the template. The broadening function is assumed to be a Gaussian and the convolution product is applied assuming a periodic spectrum with discrete Fourier transforms. 

Determining young stellar cluster velocity dispersion at optical wavelengths is particularly complicated due to the shortage of prominent stellar absorption lines. There are some stellar lines from OB stars but there are weak and they coincide in wavelength with the nebular hydrogen and helium emission lines. However, as shown above, our clusters have red supergiant stars (see Sect. \ref{EWCaT}) which dominate the longer optical wavelengths where we can find the CaII$\lambda \lambda $ 8498, 8542, 8662 \AA\ triplet lines. The use of the CaT lines to measure the velocity dispersion of stars presents some advantages. From an observational perspective: (i) the red wavelengths are less affected by the presence of an AGN in the inner part of a galaxy since a lower dilution of stellar lines is observed; (ii) the measure of these lines is not affected by the presence of TiO bands and the nebular line contamination is small; and (iii) the velocity resolution at the near IR is higher than in the blue part of the spectrum for the same spectral dispersion. Also, although these lines show some dependence with metal abundance, at high metallicities like those found for our CNSFRs the surface gravity becomes the dominant parameter with the CaT strength increasing with decreasing gravity  \citep{1989MNRAS.239..325D}.
Hence we have used late-type red giants and supergiants as reference templates. They have been obtained from the MUSE stellar library presented by \citet{2019A&A...629A.100I}. 

A large sample of stellar types with different luminosities is present in an HII region, but we have selected only two stellar types as templates for our cross-correlation analysis. This fact can introduce errors in our velocity dispersion measures. We have done an effort to minimise these systematic offsets estimating previously the stars which dominate the CaII features. Furthermore, the use of the calcium triplet lines minimises the mismatch between the stellar template and the galaxy lines since there are very strong in most stars, with the exception of the hottest ones. For this reason, we finally use an average template to be correlated with our cluster spectrum. We have aligned all selected stellar spectra in velocity and then compute a direct average between all of them, verifying that no apparent broadening is introduced in the procedure.

The first step in the application of the cross-correlation method is the binning of each spectrum into logarithmic wavelengths in order to get a uniform velocity width. We have used 512 bins that correspond to a velocity resolution of $\sim$ 27.1 km/s in a wavelength range between 8450 \AA\ - 8850 \AA. Next, the emission lines present in each spectrum have been removed, the continuum has been subtracted and the spectra have been normalised. Finally, the high and low frequency variations associated with the noise component and continuum subtracted errors respectively have been filtered. This filtering is performed by applying a band-pass to the Fourier spectrum transform with minimum and maximum wave numbers, k$_{min}$ and k$_{max}$ respectively. We have used k$_{min}$ = 3, which corresponds to wavelength values lower than 10 \AA, and k$_{max}$ = 60, corresponding to the nominal MUSE spectral resolution (MUSE User Manual-ESO-261650). Further detail of this procedure can be found in \citep{CaT-proc}.

The internal error of the method can be estimated from the asymmetric part of the correlation peak and calculated its root mean square \citep[see][]{1978ApJ...221....1D}. However, this error is very small and not considered representative of the real one associated with the measurement. Hence we have calculated the velocity dispersion error as the semi-difference between the largest and smallest Gaussian width that can be fitted considering the asymmetries in the correlation peak, as suggested in \citet{CaT-proc}.

\begin{figure}
\centering
\includegraphics[width=\columnwidth]{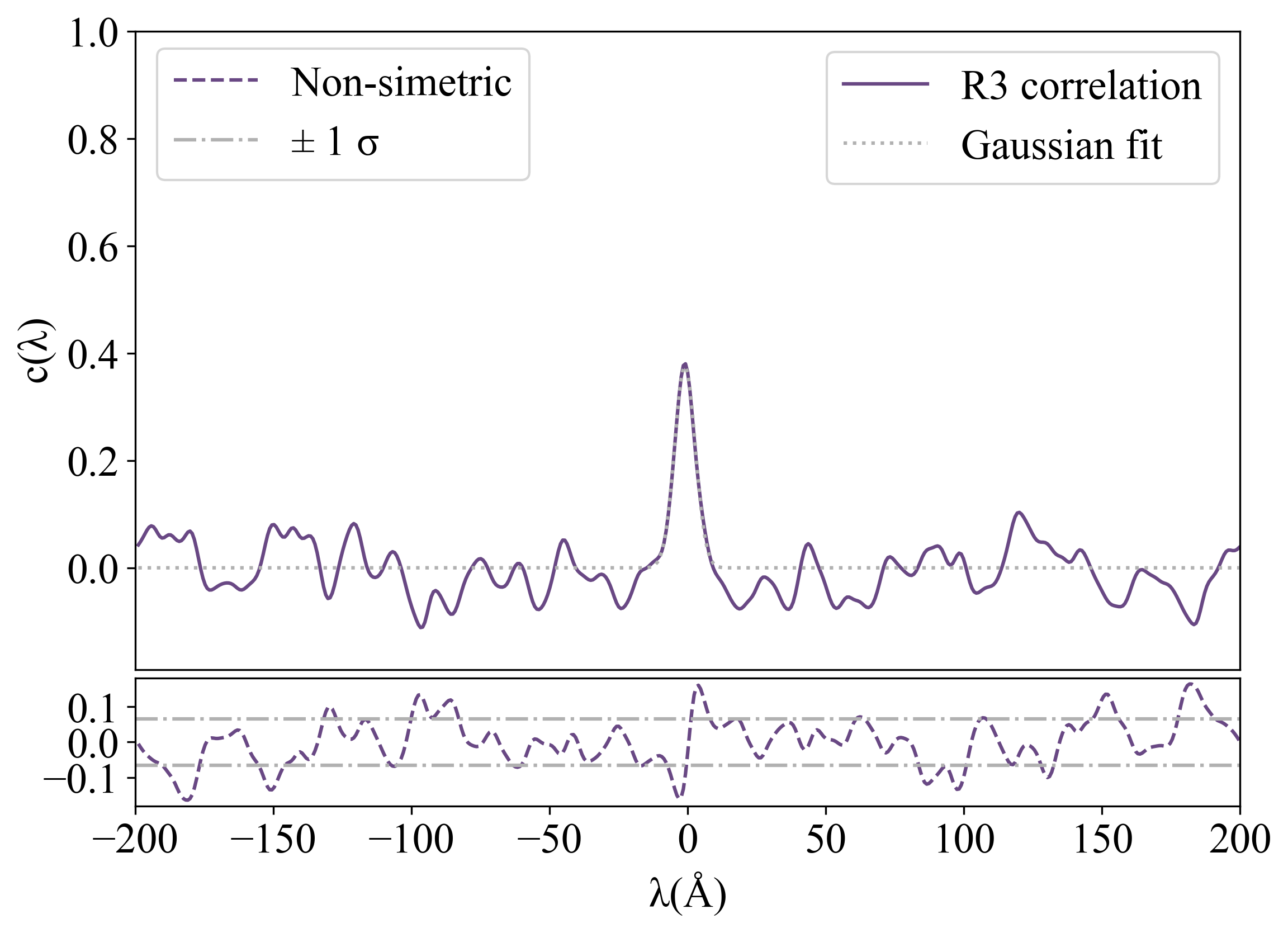}
\caption{Cross-correlation function example. The upper panel shows the cross-correlation function of region R3 with the stellar template used. The lower panel shows the asymmetric noise component of this function.}
 \label{fig:example_corr_NGC7469}
\end{figure}

Figure \ref{fig:example_corr_NGC7469} shows the correlation function between the R3 region and the stellar template as an example. We can see that the peak amplitude is lower than 0.5. This is due to an observational artefact that appears between the two strongest calcium lines mainly in the outer ring regions, which have lower S/N. However, the correlation peak is much greater than the asymmetric component and perfectly distinguishable. Additionally, we can see that the correlation peak has a Gaussian behaviour so the assumptions of the method  are justified.

Derived stellar velocity dispersions take median values of 97.38 km/s for the outer ring regions and 161 km/s for the inner ring ones.

\begin{table}
\centering
\setlength{\tabcolsep}{3.5pt}
\caption{Equivalent widths of stellar absorption lines and velocity dispersions of the observed CNSFRs.}
\label{tab:sigma}
\begin{tabular}{ccccc}
\hline
Region ID & \begin{tabular}[c]{@{}c@{}} EW(CaII)$^a$ \\ (\AA)\end{tabular} & \begin{tabular}[c]{@{}c@{}} EW(MgI) \\ (\AA)\end{tabular} & \begin{tabular}[c]{@{}c@{}} $\sigma_*$(CaT) \\ (km/s)\end{tabular}\\
\hline
R1 & 7.058 $\pm$ 0.091 & 2.692 $\pm$ 0.054 & 96.61 $\pm$ 0.02 & \\ 
R2 & 6.889 $\pm$ 0.087 & 3.022 $\pm$ 0.085 & 111.50 $\pm$ 0.05 & \\ 
R3 & 7.116 $\pm$ 0.086 & 2.481 $\pm$ 0.049 & 78.02 $\pm$ 0.03 & \\ 
R4 & 6.718 $\pm$ 0.082 & 3.175 $\pm$ 0.091 & 100.22 $\pm$ 0.15 & \\ 
R5 & 6.779 $\pm$ 0.079 & 3.128 $\pm$ 0.076 & 104.08 $\pm$ 0.07 & \\ 
R6 & 7.274 $\pm$ 0.292 & 2.821 $\pm$ 0.145 & 104.38 $\pm$ 0.25 & \\ 
R7 & 7.060 $\pm$ 0.175 & 3.100 $\pm$ 0.148 & 97.89 $\pm$ 0.06 & \\ 
R8 & 7.062 $\pm$ 0.132 & 2.958 $\pm$ 0.071 & 96.84 $\pm$ 0.11 & \\ 
R9 & 6.976 $\pm$ 0.151 & 2.607 $\pm$ 0.084 & 78.66 $\pm$ 0.00 & \\ 
R10 & 6.851 $\pm$ 0.085 & 2.802 $\pm$ 0.092 & 115.08 $\pm$ 0.14 & \\ 
R11 & 8.051 $\pm$ 0.194 & 2.820 $\pm$ 0.068 & 97.38 $\pm$ 0.01 & \\ 
R12 & 6.878 $\pm$ 0.118 & 2.661 $\pm$ 0.077 & 79.83 $\pm$ 0.07 & \\ 
R13 & 7.008 $\pm$ 0.114 & 3.008 $\pm$ 0.092 & 108.77 $\pm$ 0.08 & \\ 
R14 & 7.590 $\pm$ 0.151 & 2.892 $\pm$ 0.081 & 89.68 $\pm$ 0.04 & \\ 
R15 & 6.891 $\pm$ 0.106 & 3.387 $\pm$ 0.109 & 91.34 $\pm$ 0.20 & \\ 
R16 & 7.392 $\pm$ 0.141 & 2.778 $\pm$ 0.076 & 82.69 $\pm$ 0.13 & \\ 
R17 & 7.009 $\pm$ 0.148 & 2.548 $\pm$ 0.087 & 94.14 $\pm$ 0.05 & \\ 
R18 & 6.816 $\pm$ 0.146 & 2.945 $\pm$ 0.103 & 102.90 $\pm$ 0.21 & \\ 
R19 & 7.426 $\pm$ 0.204 & 2.745 $\pm$ 0.087 & 83.66 $\pm$ 0.06 & \\ 
R20 & 7.480 $\pm$ 0.125 & 3.668 $\pm$ 0.149 & 104.28 $\pm$ 0.09 & \\ 
R21 & 6.932 $\pm$ 0.135 & 3.102 $\pm$ 0.107 & 90.92 $\pm$ 0.14 & \\ 
R22 & 7.104 $\pm$ 0.114 & 3.144 $\pm$ 0.118 & 99.33 $\pm$ 0.07 & \\ 
R23 & 8.382 $\pm$ 0.502 & 3.177 $\pm$ 0.302 & 146.85 $\pm$ 0.35 & \\

\hline
Ra & 5.706 $\pm$ 0.033 & 0.918 $\pm$ 0.010 & 157.08 $\pm$ 0.25 & \\ 
Rb & 5.725 $\pm$ 0.055 & 0.886 $\pm$ 0.008 & 168.40 $\pm$ 0.30 & \\ 
Rc & 6.180 $\pm$ 0.025 & 0.547 $\pm$ 0.007 & 178.81 $\pm$ 0.25 & \\ 
Rd & 6.461 $\pm$ 0.074 & 0.737 $\pm$ 0.012 & 160.47 $\pm$ 0.20 & \\ 
Re & 6.826 $\pm$ 0.086 & 1.179 $\pm$ 0.026 & 161.01 $\pm$ 0.31 & \\ 
\hline
\end{tabular}
\begin{tablenotes}
\centering
\item $^a$ ($\lambda$8542\AA\ + $ \lambda$8662\AA + $ \lambda$8498\AA ). \\
\end{tablenotes}
\end{table}

Table \ref{tab:sigma} gives the identification of each line in column 1, the EW of CaII lines in column 2, the EW of MgI line in column 3 and the velocity dispersions of the observed CaII features in column 4.

\section{Discussion}
\label{discussion}

\subsection{Ionising clusters characteristics}

\subsubsection{Ionisation nature}
\label{sec:nature}

\begin{figure*}
\centering
\includegraphics[width=\textwidth]{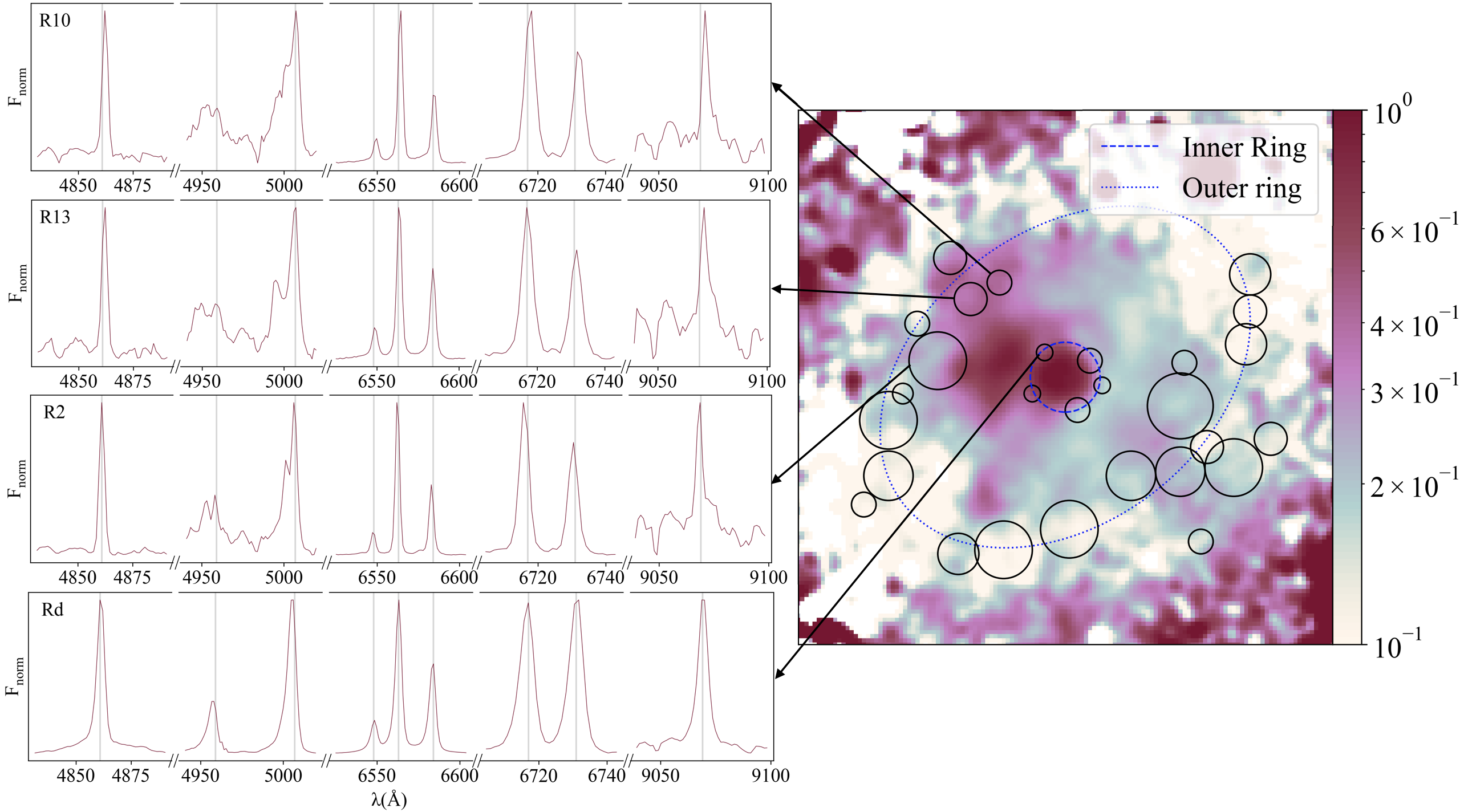}
\caption{Emission line spectra of regions with multiple kinematical components. Left panels, from top to bottom: Spectra of regions R10, R13, R2, and Rd. Shown from left to right are the emission lines of H$\beta$, [OIII]$\lambda \lambda$ 4069,5007 \AA, H$\alpha$ and [NII]$\lambda \lambda$ 6548,6584 \AA , [SII]$\lambda \lambda $ 6717,6731 \AA\ and [SIII]$\lambda$ 9069 \AA\ are  shown. Right panel: Map of the observed [OIII]$\lambda$ 5007 \AA\ / [NII]$\lambda$ 6583 \AA\ ratio smoothed with a Gaussian function of $\sigma$ = 1.5 pix.}
 \label{fig:OIII_vel_}
\end{figure*}

The left panels of Fig. \ref{fig:OIII_vel_} show 4 outer ring region spectra at different wavelengths. From left to right the H$\beta$, [OIII]$\lambda\lambda$ 4959,5007 \AA, [NII]$\lambda\lambda$ 6548,84 \AA\ and H$\alpha$,  HeI$\lambda$ 5875 \AA, [SII]$\lambda\lambda$ 6716,31 \AA and [SIII]$\lambda$ 9069 \AA. More than one kinematical component can be seen in the [OIII] emission lines as well as asymmetries with respect to a Gaussian behaviour also seen in some other emission lines, as already reported in previous works \citep{2021ApJ...906L...6R,2022ApJ...933..110X}, evidence of complex velocity flows. All emission lines with the required S/N (see Sect. \ref{sec:line measurements}) appear to have at least two components. In addition, we have found four regions: R2, R10, R13 and Rd, showing three components in the [OIII]$\lambda \lambda$ 4959,5007 \AA\ emission line. Neither of them corresponds to the previously identified star-forming complexes (see Fig. \ref{fig:HST-606}), suggesting that the third kinematical component might be associated with an outflow coming from the active galaxy nucleus. The right panel of Fig. \ref{fig:OIII_vel_} shows the [OIII]$\lambda$ 5007 \AA\ /[NII]$\lambda$ 6584 \AA\ map, spatially smoothed with a Gaussian function of $\sigma$ = 1.5 pix (0.3 arcsec). The four regions mentioned above are close to each other and located in the same area in which this emission line ratio shows an excess. 

\begin{figure}
\centering
\includegraphics[width=0.75\columnwidth]{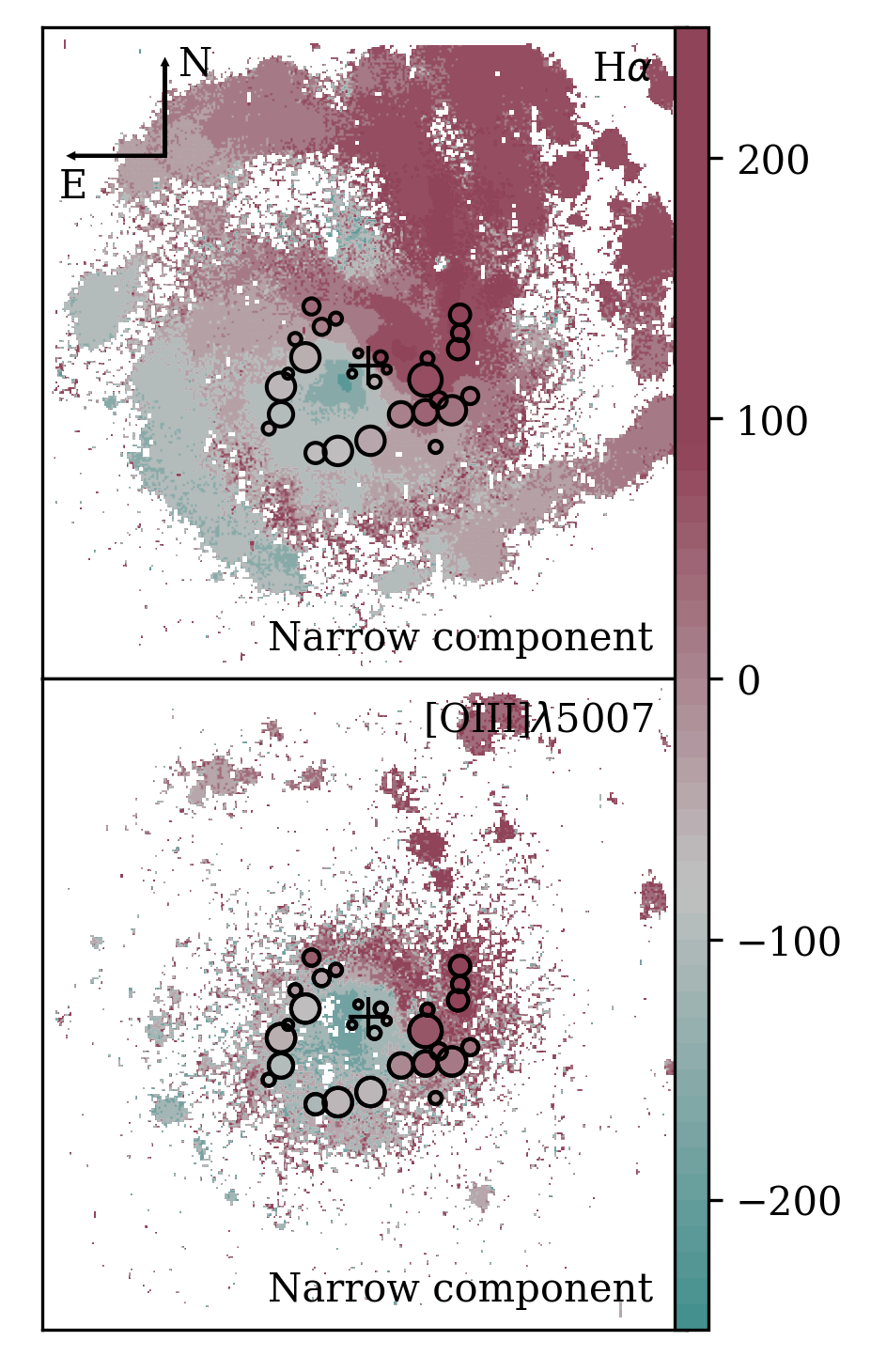}
\caption{Radial velocity maps. H$\alpha$ (upper panel) and [OIII] $\lambda$ 5007 \AA\ (lower panel) pixel-by-pixel radial velocity maps. The velocity of the narrow component associated with each cluster is superimposed.}
 \label{fig:Hb_vel}
\end{figure}

Figure \ref{fig:Hb_vel} shows H$\alpha$ (upper panel)  and [OIII]$\lambda$ 5007 \AA\ (lower panel) radial velocity maps showing that the selected young clusters follow the galaxy disc velocity distribution, hence assuring that the kinematical component associated with the observed HII regions is the correct one. 

\begin{figure}
\centering
\includegraphics[width=\columnwidth]{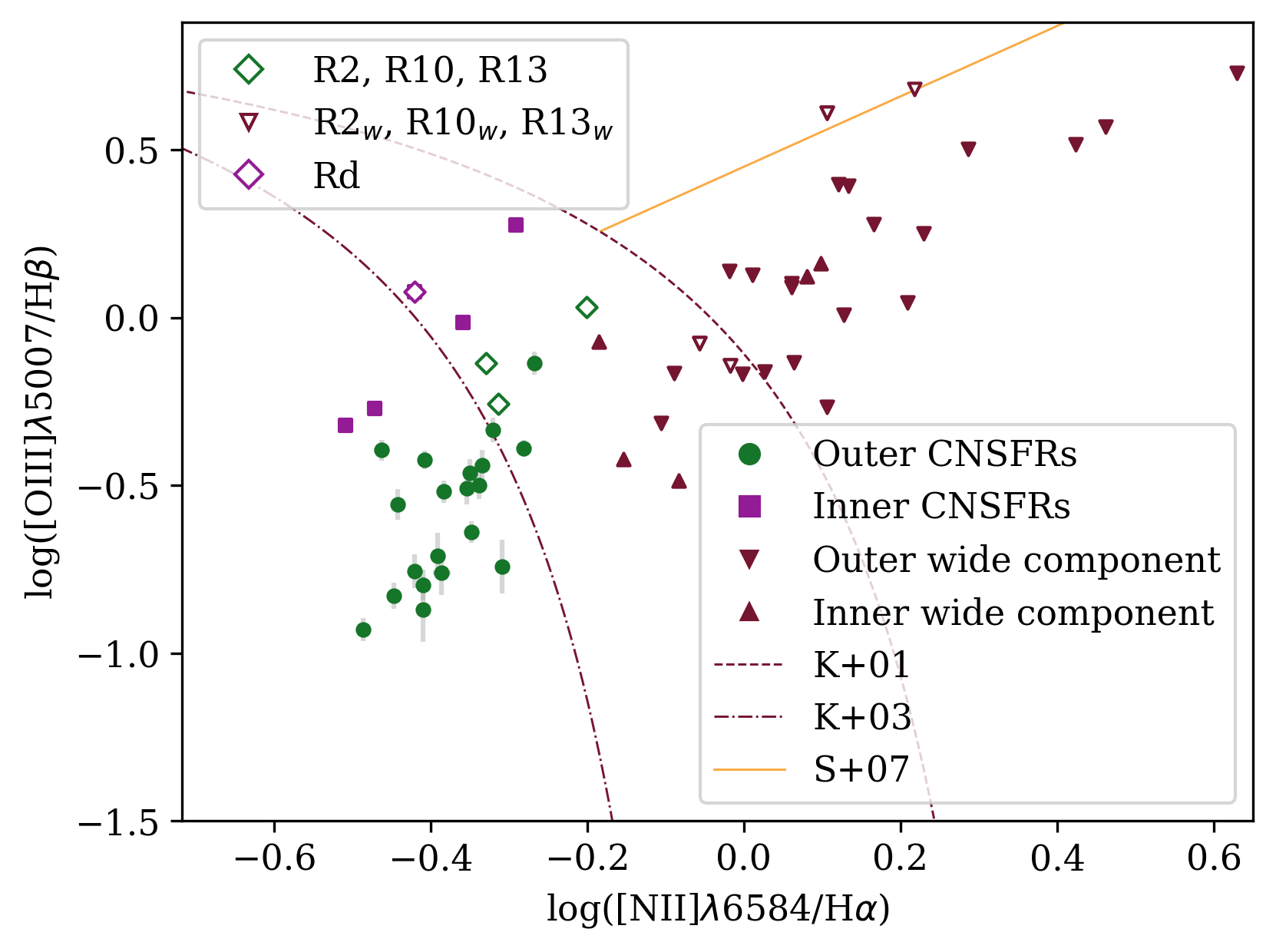}
\includegraphics[width=\columnwidth]{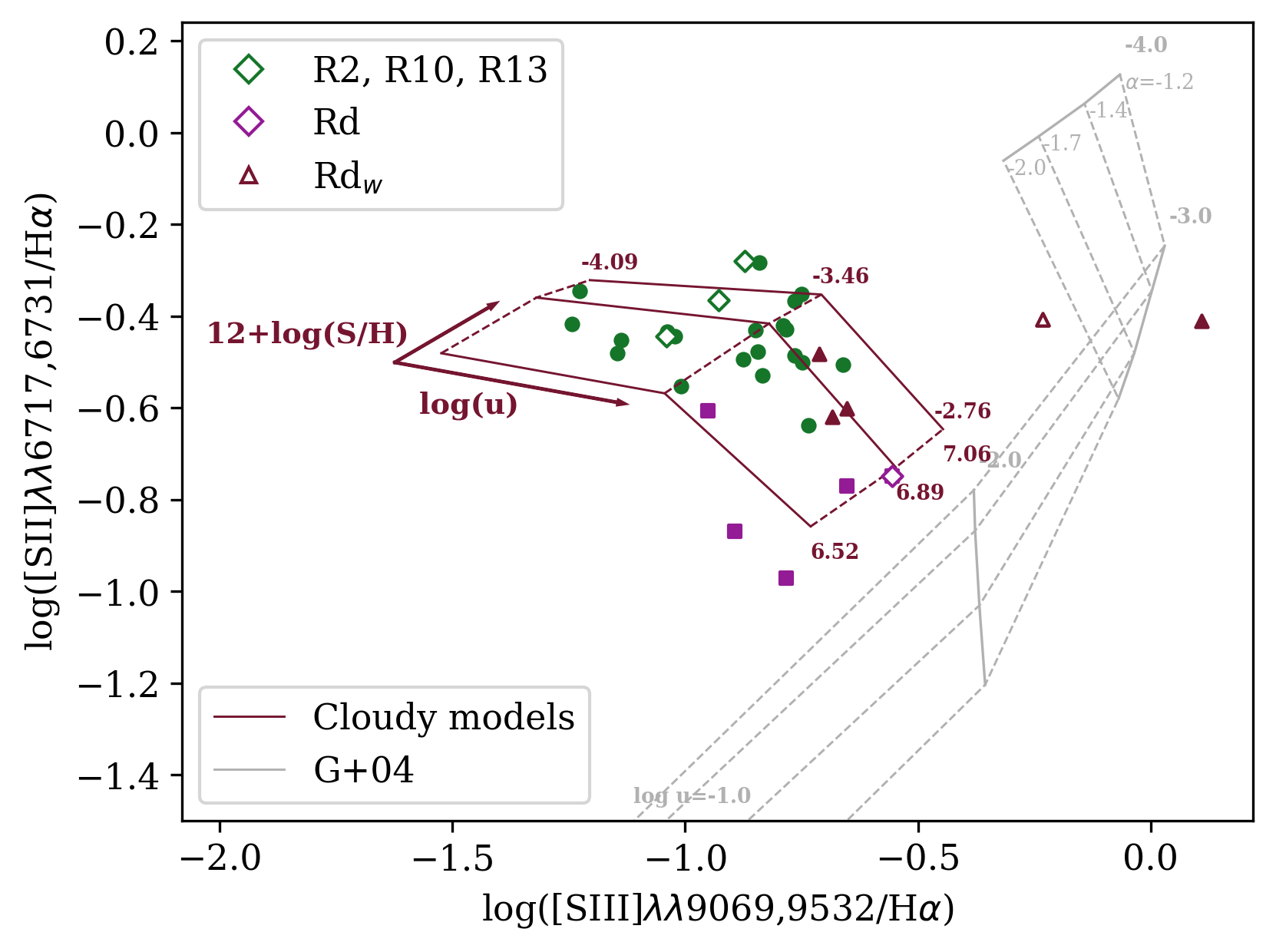}
\caption{Ionisation nature diagnostics. Upper panel: [OIII]/H$\beta$ vs [NII]/H$\alpha$ diagnostic diagram. Overplotted are the  derived separations between LINER/Seyfert \citep[S+07,][]{2007MNRAS.382.1415S} and HII regions \citep[K+01 and K+03,][]{2001ApJ...556..121K, 2003MNRAS.346.1055K}. Lower panel: [SII]/H$\alpha$ - [SIII]/H$\alpha$ diagnostic diagram. Overplotted are the dust-free AGN photo-ionisation models \citep[G+04,][]{2004ApJS..153....9G}.}
 \label{fig:bpt}
\end{figure}

Figure \ref{fig:bpt} shows in the upper panel the BPT diagram for the observed HII regions in both the outer and inner rings (narrow and broad components). This is the diagnostic more commonly used to distinguish between star-forming and shocked and/or non-thermal activity in ionised regions, although its sensitivity to the N/O ratio is a recognised caveat. Hence we better suggest the use of the near-infrared sulphur emission lines which constitute a powerful diagnostic to distinguish between shock and photo-ionisation mechanisms \citep[see][]{1985MNRAS.212..737D}, being independent of relative abundances and little sensitive to reddening. The lower panel of  Fig. \ref{fig:bpt} shows the location on this diagram of the same regions. Also shown are the shock models calculated by \citet{2004ApJS..153....9G} and the results of photo-ionisation models for nebulae ionised by young star clusters computed using the Cloudy \citep{cloudy} code with S/H abundances and ionisation parameter labelled \citep[see][for further information]{ngc7742paper}. Regions Rd, R2, R10 and R13 (regions with three kinematical components in [OIII]$\lambda$ 5007 \AA\ line) are shown with open diamond symbols for the narrow component and open triangles and inverted triangles for the broad ones. The examination of both figures shows the adequate selection of HII region components analysed.

\subsubsection{Characteristics of the observed CNSFRs}
\label{sec:CNSFR}

The H$\alpha$ luminosity, L(H$\alpha$), can be calculated from the extinction-corrected H$\alpha$ fluxes. For the outer ring HII regions this value is between 4.5 $\times$ 10$^{38}$ erg/s/cm$^2$ and 1.2 $\times$ 10$^{40}$ erg/s/cm$^2$ while for inner ring regions is from 2.4 $\times$ 10$^{40}$ erg/s/cm$^2$ to 2.0 $\times$ 10$^{41}$ erg/s/cm$^2$. These values are respectively on the central and higher side of the distribution found by \citet{2015MNRAS.451.3173A} for a large sample of CNSFRs. The H$\alpha$ luminosity of the star formation in the five inner regions dominates the emission at the galactic centre with 4.4 $\times$ 10$^{41}$ erg/s/cm$^2$ representing  77\% of the total H$\alpha$ luminosity within the central 2.5 arcsec. These high luminosities can be understood in the context of LIRGs which contain HII regions with H$\alpha$ luminosities about two orders of magnitude higher than the ones found in normal galaxies and consistent with the values found in our work \citep[see Fig. 4 ][]{2002AJ....124..166A}. The median value found for these galaxies is log(L(H$\alpha$)) $\sim$ 39.70 erg/s comparable to that of the giant H II region 30 Doradus in the Large Magellanic Cloud and can be a consequence of the regions forming in a high gas pressure and density environment in addition to the presence of interaction processes.

The corresponding number of hydrogen ionising photons per second has been derived using the recombination coefficient of the H$\alpha$ line assuming a constant value of electron density of 100 cm$^{-3}$, a temperature of 10$^4$ K and case B recombination. HII regions in the outer ring have H$\alpha$ luminosities, hence a number of ionising photons, between one and two orders of magnitude lower than inner ring regions, implying lower star formation rates (SFR). The dimensionless ionisation parameter, u, has been estimated from the [SII]/[SIII] ratio \citep[see][]{1991MNRAS.253..245D} and ranges from -4.47 to -3.15 in logarithmic units for the outer ring regions, with a median value of -3.78. Ionisation parameters are higher for the inner HII regions by factors between 5 and 10.

\begin{figure*}
\centering
 \includegraphics[width=0.66\textwidth]{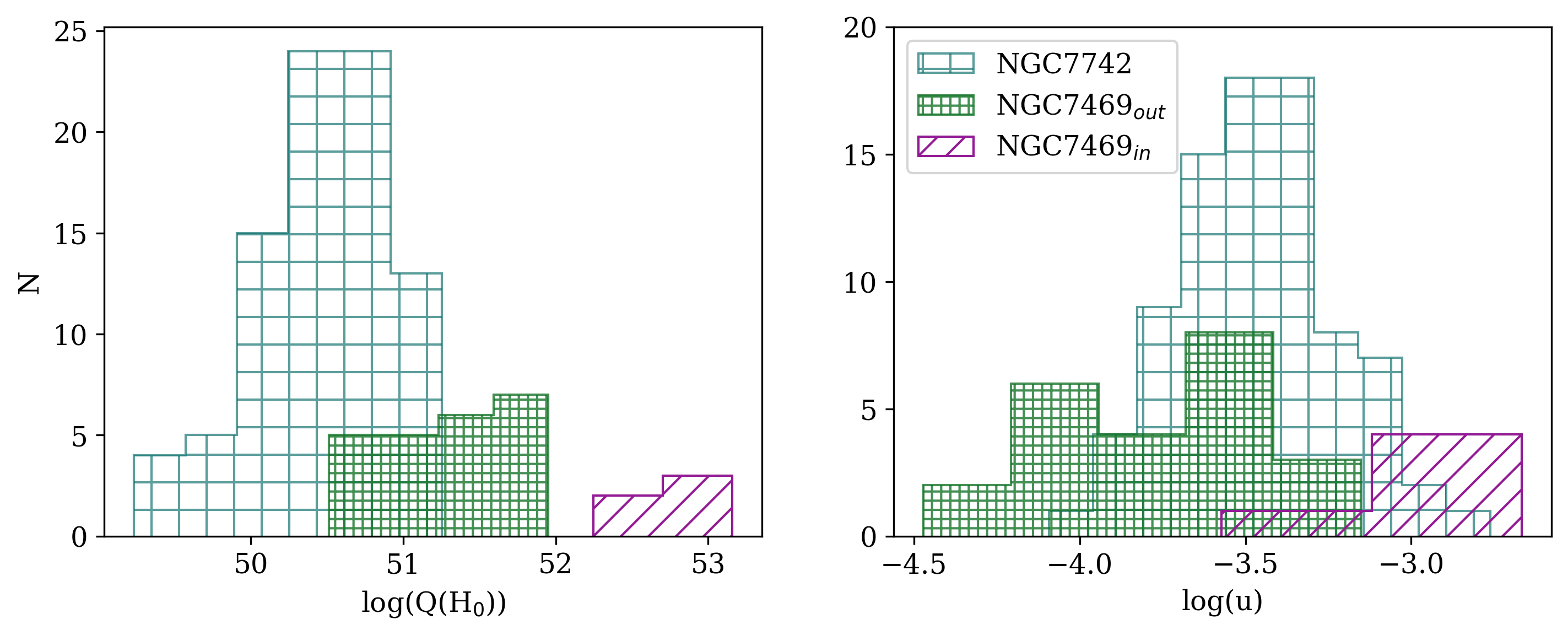}
 \includegraphics[width=\textwidth]{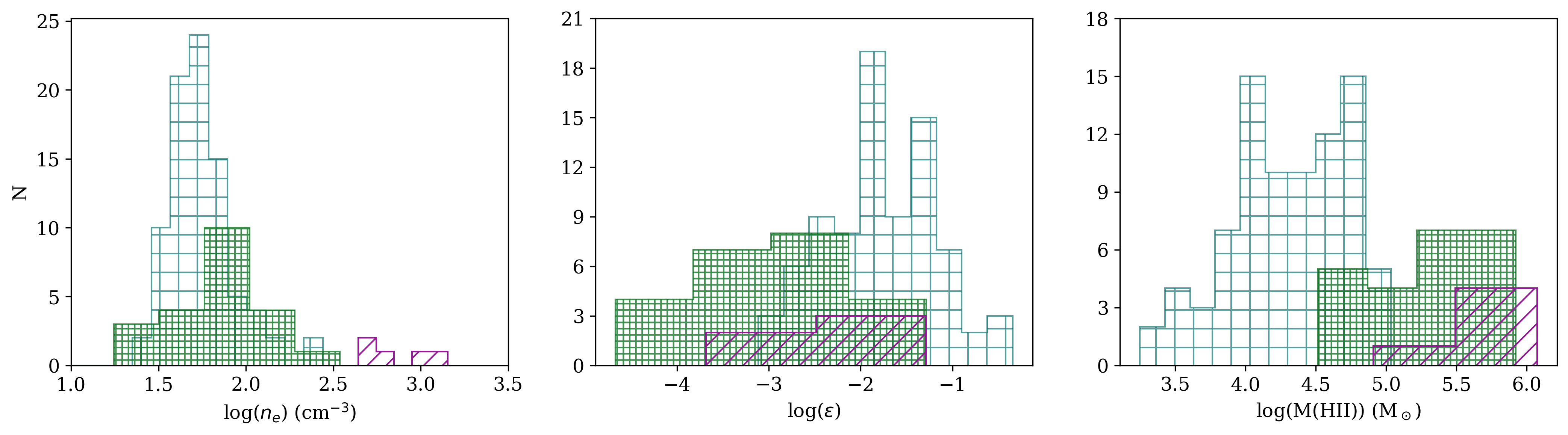}
 \caption{Physical properties of CNSFRs. The different histograms in the figure show for the ring HII regions (in green) and the regions outside (in purple) the distributions of the number of hydrogen ionising photons (upper left), the ionisation parameter (upper right), the electron density (bottom left), the filling factor (bottom centre), and the mass of ionised hydrogen (bottom right).}
 \label{fig:hist_r_ne_Q}
\end{figure*}

The distribution of both quantities, number of ionising photons Q(H$_0$), and ionisation parameter, log(u), are shown in the two upper panels of Fig. \ref{fig:hist_r_ne_Q} in comparison to the values found for the circumnuclear ring in Galaxy NGC~7742 \citep[][]{ngc7742paper}. On average, the outer ring regions in NGC~7469 are more luminous than those in NGC7742 by an order of magnitude. The lacking low-luminosity tail in the distribution of NGC~7469 might be due to lack of detection given the larger distance, by a factor of 3, to this galaxy. On the other hand, half of the regions show values of Q(H$_0$) larger than the maximum obtained in NGC~7742 pointing to larger structures being selected due to lower linear resolutions. In fact, all regions with L(H$\alpha$) larger than 10$^{39}$ erg/s have several ionising knots as revealed by HST images. For these regions, larger in extent, lower ionisation parameters are expected as found.

The bottom panels of Fig. \ref{fig:hist_r_ne_Q} show, from left to right, the distribution of electron density, filling factor and ionised hydrogen mass for the inner and outer ring HII regions as compared with the NGC 7742 ones.
The electron density can be calculated from the [SII]$\lambda$ 6717 \AA\ / [SII]$\lambda$ 6731 \AA\ ratio only for n$_e$ > 50 cm$^{-3}$. For the regions within the ring for which only upper limits could be estimated, the electron density has been derived from the observed region sizes, the ionisation parameter and the H$\alpha$ fluxes. The electron densities range from 50 cm$^{-3}$ to 345 cm$^{-3}$ for outer ring regions. Only six regions (R4, R5, R11, R16, R20 and R21) have density values larger than 100 cm$^{-3}$ and only one of them (R21) is significantly different from the median value (>3$\sigma$). These values are similar to those estimated for NGC 7742 regions. On the other hand, inner ring regions present higher values of n$_e$, between 439 cm$^{-3}$ and 1431 cm$^{-3}$ with a mean value of 848 cm$^{-3}$ which,  given their smaller sizes, could be due to their much closer to the galactic nucleus environment. 
Filling factors can be derived using the ionisation parameter and the measured angular radius of each observed HII region \citep[see][]{1991MNRAS.253..245D} with electron density larger than 50 cm$^{-3}$. Filling factors for the outer ring HII regions are low, ranging from 2.14 $\times$ 10$^{-5}$ to 5.25 $\times$ 10$^{-2}$, with a mean value of 1.26 $\times$ 10$^{-3}$. These values are lower than those estimated for high-metallicity disc HII regions \citep[between 0.008 and 0.52;][]{1991MNRAS.253..245D,diaz2000,2002MNRAS.337..540C}, CNSFRs \citep[from 0.0006 to 0.001;][]{2007MNRAS.382..251D} and for the case of NGC 7742 \citep[from 0.00077 to 0.45;][]{ngc7742paper}. This is consistent with the larger HII region sizes for similar electron densities. For the inner ring regions, filling factor range from 2.09 $\times$ 10$^{-4}$ to 5.13 $\times$ 10$^{-2}$, with a mean value of 4.17 $\times$ 10$^{-3}$, in the higher part of the distribution found in the outer ring regions. Finally, the mass of ionised hydrogen, in solar masses, have be derived using  the expression given in  \citet{1991MNRAS.253..245D}. Their mean value is 3.0 $\times$ 10$^5$ M$_\odot$ and 6.2 $\times$ 10$^5$ M$_\odot$ for the HII regions within the outer and inner ring respectively. Both rings show a similar distribution of this quantity, shifted to larger values with respect to the case of NGC~7742 galaxy. 

\begin{figure}
 \includegraphics[width=\columnwidth]{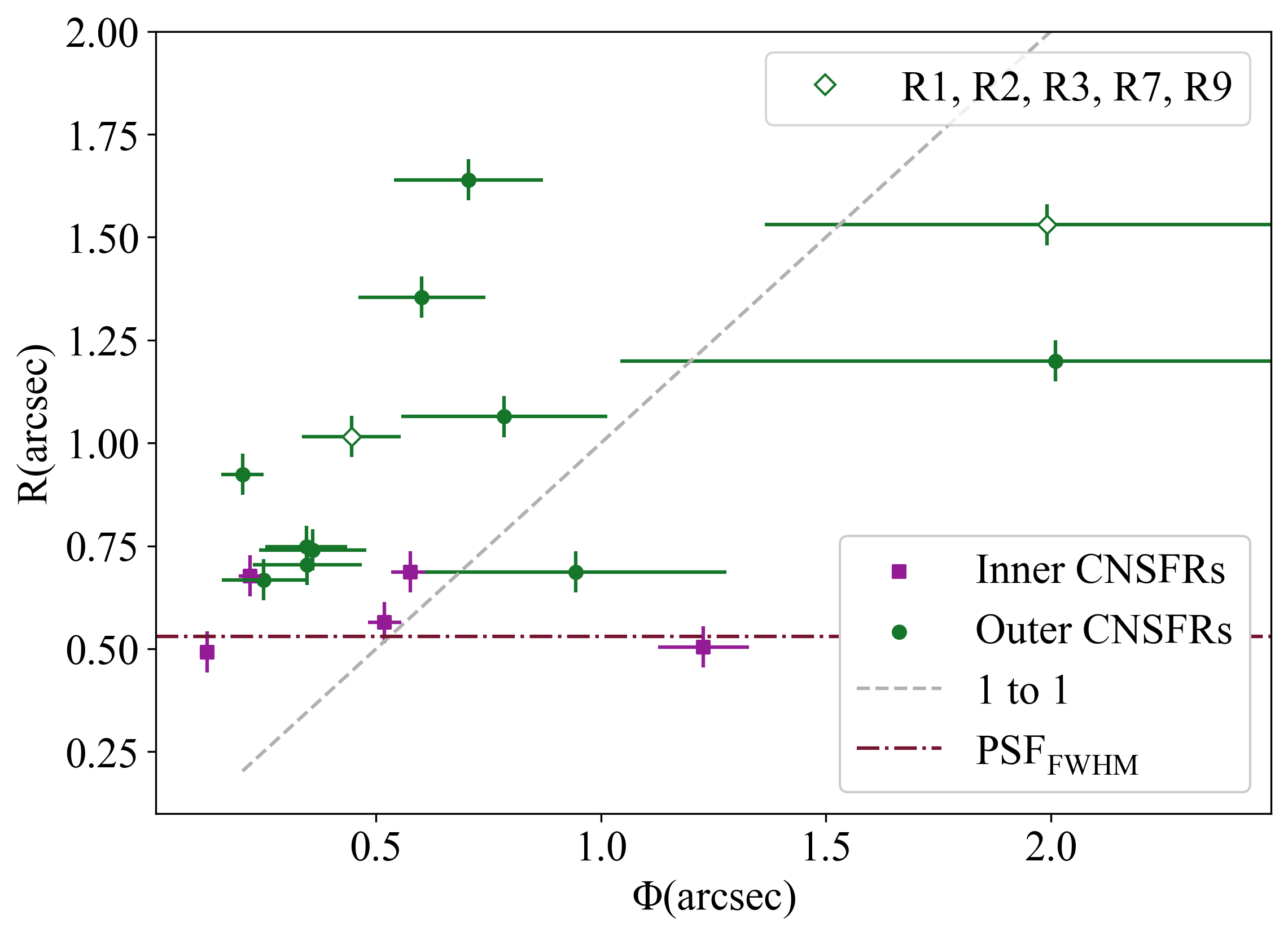}
 \caption{The ionisation derived angular radius against the angular radius measured from the HII region segmentation (see Sect. \ref{sec:segmentation}).}
 \label{fig:radios}
\end{figure}

We can compare the estimated angular radii of the observed ring HII regions, $\phi$, calculated using the definition of the ionisation parameter, with the actually measured ones (see Sect. \ref{sec:segmentation}). This has been done for regions with derived electron densities larger than 50 cm$^{-3}$ using the expressions given in \citet{2002MNRAS.337..540C}. Figure \ref{fig:radios} shows this comparison. In most cases, the predicted angular sizes are smaller than measured and only three cases with very large errors in the derivation of their predicted sizes, could correspond to radiation bounded ionised nebulae.  One of the inner ring regions, Re, shows the opposite behaviour. However, this region shows very little continuum emission at  either in 3360 \AA\ or 6600 \AA\ wavelengths (see Fig. \ref{fig:HST-606}) and hence the existence of an ionising young star cluster may be questioned.

\begin{table*}
\centering
\setlength{\tabcolsep}{1pt}
\caption{Characteristics of the observed CNSFRs (extract). The full table is available in Appendix \ref{ap:tables}.}
\label{tab:characteristics CNSFR}
\begin{tabular}{ccccccccc}
\hline
\begin{tabular}[c]{@{}c@{}}Region\\ID\end{tabular}&\begin{tabular}[c]{@{}c@{}}L(H$\alpha$) \\ (erg s$^{-1}$)\end{tabular}& \begin{tabular}[c]{@{}c@{}}Q(H$_0$) \\ (photons s$^{-1}$)\end{tabular}&\begin{tabular}[c]{@{}c@{}}log(u) \\ \end{tabular}&\begin{tabular}[c]{@{}c@{}}$\phi$ \\ (arcsec)\end{tabular} & \begin{tabular}[c]{@{}c@{}}R \\ (arcsec)\end{tabular} & \begin{tabular}[c]{@{}c@{}}n$_e$ \\ (cm$^{-3}$)\end{tabular}& \begin{tabular}[c]{@{}c@{}}log($\epsilon$) \\ \end{tabular} & \begin{tabular}[c]{@{}c@{}}M(HII) \\ (M$_\odot$)\end{tabular}\\ \hline
R1 & (114.3 $\pm$ 6.4) $\times$ 10$^{38}$ & (83.7 $\pm$ 4.7) $\times$ 10$^{50}$ & -3.756 $\pm$ 0.077 & - & 1.57 $\pm$ 0.05 & 35 $\pm$ 22 & -3.65 $\pm$ 0.16 & (51.2 $\pm$ 9.7) $\times$ 10$^{4}$\\ 
R2 & (12.2 $\pm$ 1.4) $\times$ 10$^{39}$ & (89.4 $\pm$ 7.5) $\times$ 10$^{50}$ & -3.992 $\pm$ 0.102 & 1.99 $\pm$ 0.63 & 1.53 $\pm$ 0.05 & 59 $\pm$ 34 & -4.14 $\pm$ 0.21 & (28.4 $\pm$ 6.9) $\times$ 10$^{4}$\\ 
R3 & (67.8 $\pm$ 7.8) $\times$ 10$^{38}$ & (49.6 $\pm$ 4.4) $\times$ 10$^{50}$ & -3.630 $\pm$ 0.082 & - & 1.36 $\pm$ 0.05 & 31 $\pm$ 7 & -3.11 $\pm$ 0.17 & (5.1 $\pm$ 1.0) $\times$ 10$^{5}$\\ 
R4 & (7.4 $\pm$ 1.0) $\times$ 10$^{39}$ & (54.4 $\pm$ 6.6) $\times$ 10$^{50}$ & -3.609 $\pm$ 0.091 & 0.71 $\pm$ 0.17 & 1.64 $\pm$ 0.05 & 119 $\pm$ 48 & -3.19 $\pm$ 0.19 & (7.9 $\pm$ 1.7) $\times$ 10$^{5}$\\ 
R5 & (41.8 $\pm$ 6.2) $\times$ 10$^{38}$ & (30.6 $\pm$ 4.0) $\times$ 10$^{50}$ & -3.657 $\pm$ 0.092 & 0.60 $\pm$ 0.14 & 1.35 $\pm$ 0.05 & 103 $\pm$ 41 & -2.95 $\pm$ 0.19 & (4.8 $\pm$ 1.1) $\times$ 10$^{5}$\\ 
R6 & (47.9 $\pm$ 5.0) $\times$ 10$^{38}$ & (35.1 $\pm$ 2.6) $\times$ 10$^{50}$ & -3.152 $\pm$ 0.060 & - & 0.94 $\pm$ 0.05 & 34 $\pm$ 30 & -1.84 $\pm$ 0.13 & (7.4 $\pm$ 1.3) $\times$ 10$^{5}$\\ 
R7 & (8.4 $\pm$ 1.0) $\times$ 10$^{39}$ & (61.1 $\pm$ 6.1) $\times$ 10$^{50}$ & -3.503 $\pm$ 0.090 & - & 1.50 $\pm$ 0.05 & 23 $\pm$ 6 & -2.98 $\pm$ 0.19 & (8.4 $\pm$ 1.8) $\times$ 10$^{5}$\\ 

\hline

\end{tabular}
\end{table*}

Tab. \ref{tab:characteristics CNSFR} shows the characteristics of each HII region within the ring and lists in column 1 to 9: (1) the region ID; (2) the extinction-corrected H$\alpha$ luminosity; (3) the number of hydrogen ionising photons; (4) the ionisation parameter; (5) the estimated angular radius; (6)  the measured linear radius; (7) the electron density; (8) the filling factor; and (9) the mass of ionised hydrogen.

\subsubsection{Chemical abundances}

\begin{figure}
\centering
 \includegraphics[width=0.87\columnwidth]{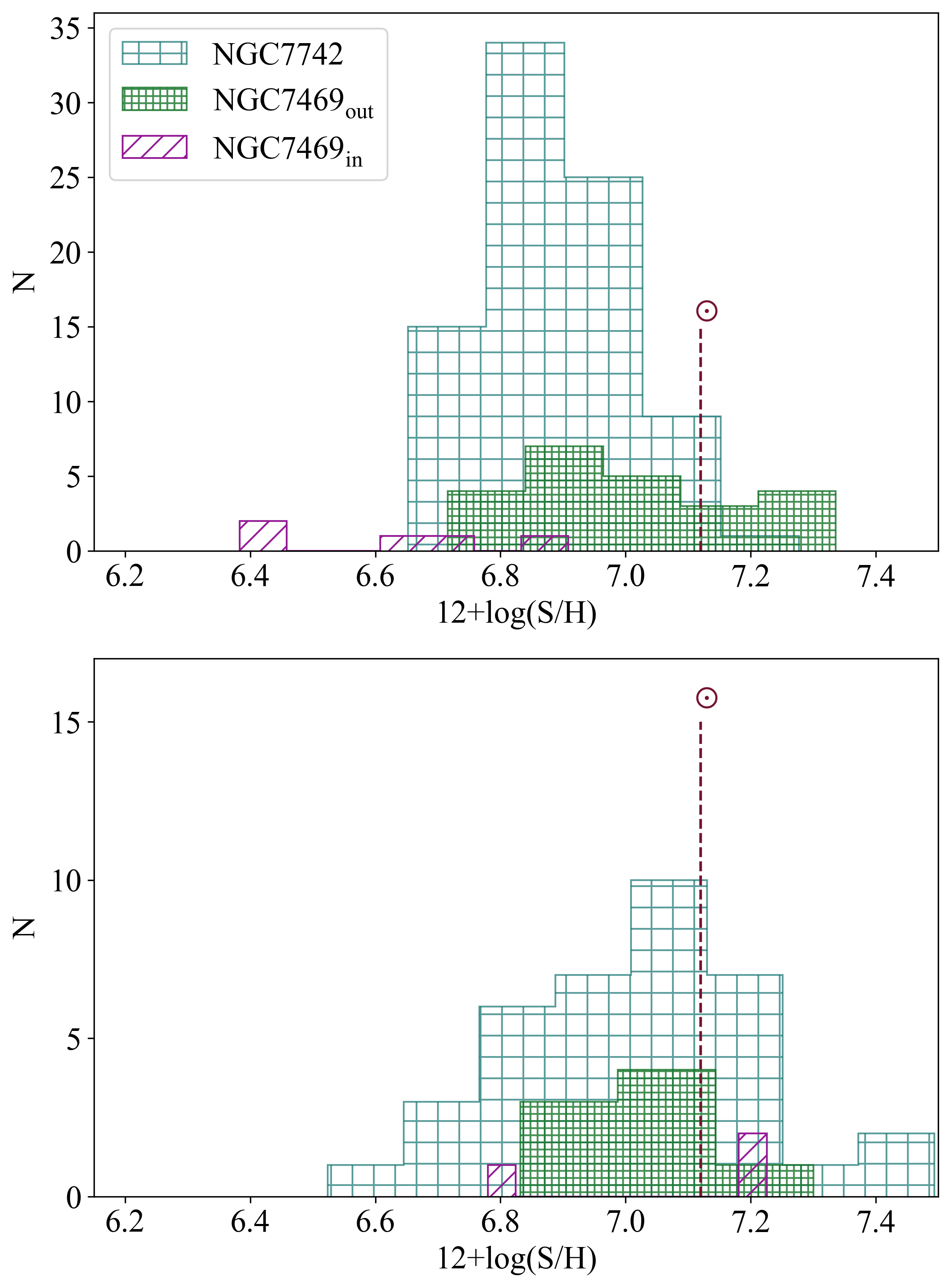}
 \caption{Distribution of the total empirically derived (upper panel) and directly derived (lower panel)  sulphur abundances for the CNSFRs. The dashed line corresponds to the solar value \citep[12+log(S/H)$_{\odot}$= 7.12,][]{2009ARA&A..47..481A}.} 
 \label{fig:hist_S}
\end{figure}

Regarding chemical abundances, the upper panel of Fig. \ref{fig:hist_S} shows the distribution of 12+log(S/H) derived from the empirical S$_{23}$ calibration. Outer ring regions show values ranging from 6.72 to 7.34 in units of 12+log(S/H) \citep[12+log(S/H)$_{\odot}$= 7.12,][]{2009ARA&A..47..481A}, with a mean of 7.00 and errors between 0.03 to 0.1 dex. Inner ring regions, on the other hand, show somewhat lower values (about a factor of two), ranging from 6.38 to 6.91 with a mean value of 6.62 and errors between 0.03 and 0.05 dex. The lower panel of the figure shows the directly derived S/H abundances. In this case, outer ring regions show logarithmic sulphur abundances between 6.832 $\pm$ 0.022 and 7.300 $\pm$ 0.016 in units of 12+log(S/H), i.e. between 0.52 and 1.51 times the solar value \citep[12+log(S/H)$_{\odot}$= 7.12,][]{2009ARA&A..47..481A}. The corresponding values for the inner ring regions are comparable within the errors, ranging from 6.780 $\pm$ 0.022 to 7.226 $\pm$ 0.013, between 0.46 and 1.28 times the solar value. In both cases a comparison is made with the CNSFR in NGC~7742.} 

\begin{figure}
 \includegraphics[width=\columnwidth]{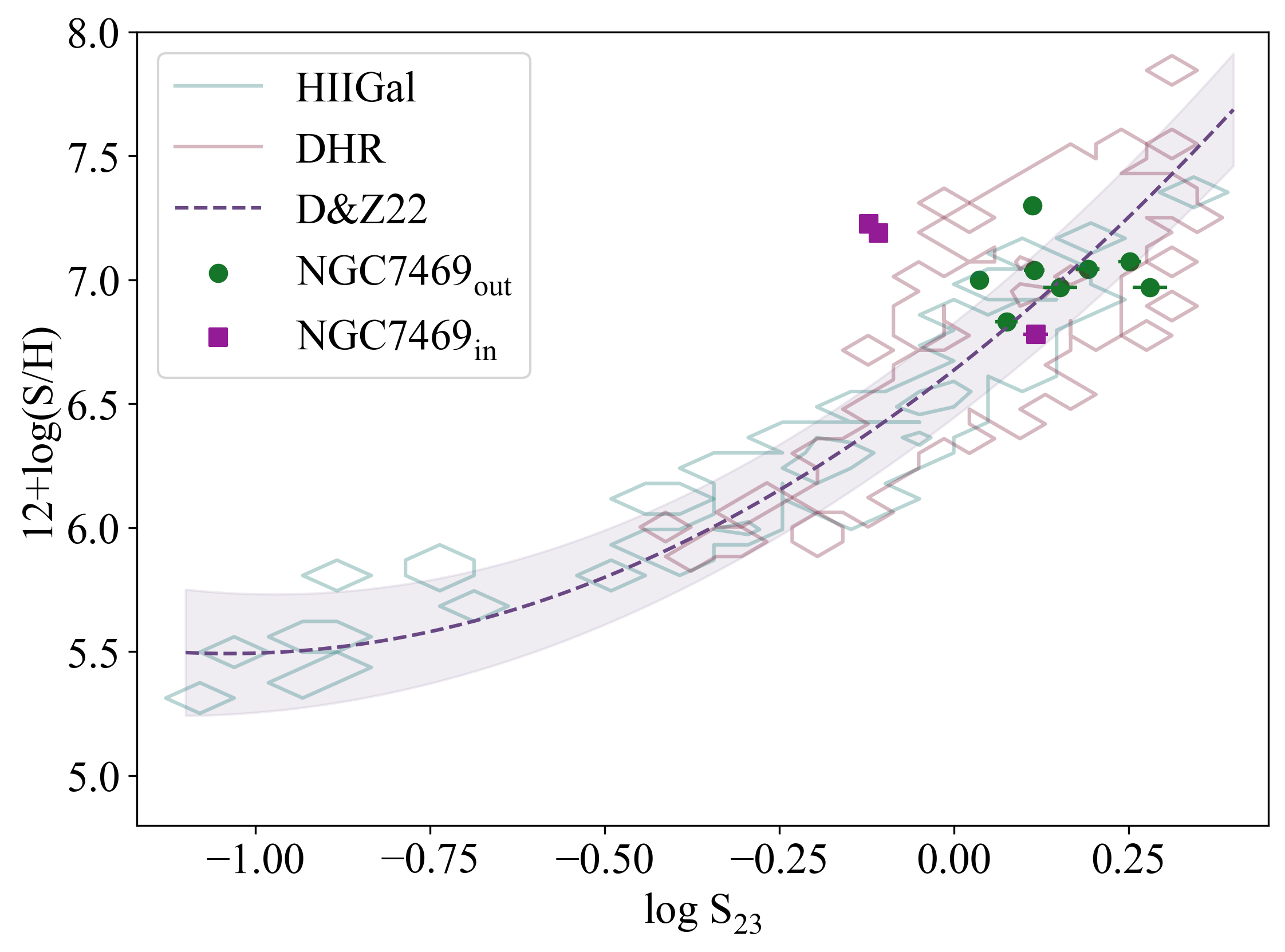}
 \caption{The S$_{23}$ abundance calibration from \citet{2022MNRAS.511.4377D}. The red contours correspond to disc HII regions, while the blue contours correspond to HII galaxies. The green dots and purple squares represent respectively the outer and inner ring regions analysed in this work. Observational errors for these data are inside the symbols in the graph.} 
 \label{fig:calibracion}
\end{figure}

We have represented in Fig. \ref{fig:calibracion} the empirical S$_{23}$ calibration together with red and blue contours corresponding to data for disc HII regions and HII galaxies respectively and individual symbols representing directly derived abundances for the 8 outer ring regions (green solid circles) and the 3 inner ring regions (purple squares).
In two inner HII regions, Ra and Rb, we have found empirically derived abundances somewhat lower that directly derived ones. However, the empirical S$_{23}$ calibration could be somewhat affected by the effective temperature of the ionising radiation. Low values of this temperature could move this calibration to higher abundances by up to 14\% in the log which would reconcile directly and empirically derived abundances. 

The spectral energy distribution of the ionising radiation can be estimated from the quotient between the number of helium and hydrogen ionising photons, Q(He$_0$)/Q(H$_0$) \citep[see][]{ngc7742paper}. This ratio can be used when there is no direct measurement of the ionic abundances of oxygen and sulphur and it is equivalent to the calculation made with the use of the $\eta$ parameter. We have calculated the number of ionising He$_0$ photons from the observed luminosity in the HeI$\lambda$ 6678 \AA\ emission line using its corresponding flux, the distance to NGC~7469 which has been taken as 66.47 Mpc (see Tab. \ref{tab:galaxy characteristics}) and the recombination coefficient of HeI$\lambda$ 6678 \AA\ emission line assuming a constant value of electron density of 100 cm$^{-3}$, a temperature of 10$^4$ K and  case B recombination \citep{Osterbrock2006}. 
We have detected and measured the HeI$\lambda$ 6678 \AA\ line in 10 outer ring regions and in all inner ring ones finding mean values of Q(He$_0$) of 2.5 $\times$ 10$^{48}$ photons s$^{-1}$ and 4.4 $\times$ 10$^{49}$ photons s$^{-1}$ for outer and inner ring regions respectively (see Table \ref{tab:cluster properties}).

\begin{figure}
\includegraphics[width=1\columnwidth]{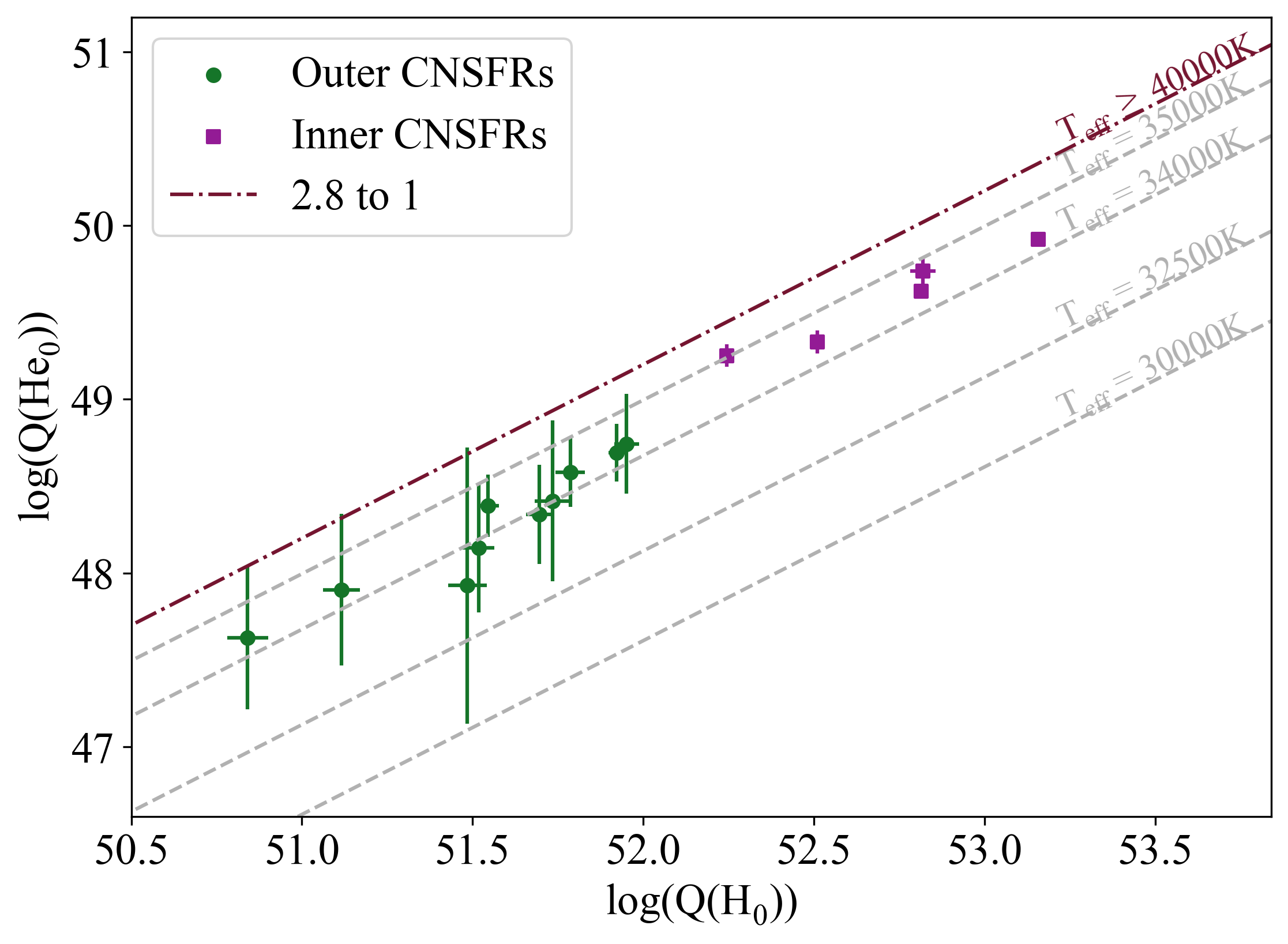}
\caption{Relation between the logarithmic numbers of HeI and HI ionising photons per second (see text for details).} 
 \label{fig:teff}
\end{figure}

Figure \ref{fig:teff} shows the relation between the logarithmic numbers of HeI and HI ionising photons. Superimposed are the lines corresponding to different Cloudy models with ionisation parameter values from -4.0 to -2.5, solar metallicity and a constant value of the electron density of 100 cm$^{-3}$. In these models the nebula is ionised by stellar atmospheres from \citet[][non-LTE models for B and O stars, log(g) = 4 and T$_{eff}$ from 30000 K to 55000 K]{1978stat.book.....M}. According to these models, we can deduce that the He$^+$ nebular zone is much smaller than that of H$^+$ in all the analysed ionising clusters.  All of them seem to have similar effective temperatures, around 34400 K, although the outer ring regions show much larger errors. 

\subsection{Cluster stellar populations}
\label{SC_pop}

\subsubsection{Ionising and photometric masses} \label{clusters}

We have estimated the mass of the ionising clusters powering the circumnuclear HII regions from the number of Lyman continuum photons and using single stellar population (SSP) models to obtain the number of ionising photons per unit solar mass, Q(H$_0$)/M$_{\odot}$, which decreases with the age of the cluster. We have used the equivalent width of the H$\beta$ emission line, EW(H$\beta$, to parametrise this age obtaining a linear relation between the ionising photon number and  EW(H$\beta$) \citep[see][]{ngc7742paper}. In order to do this, we have used PopStar models \citep{Popstar} for ages under 10 Ma and metallicities between 0.004 and 0.02. The slope of the initial mass function (IMF) and the lower mass limit affect this relation thus we have used the Salpeter IMF with $\phi (m) = m^{-\alpha}$, $\alpha = 2.35$, $m_{low}(M_\odot)$ = 0.85 and $m_{up}(M_\odot)$ = 120 that seems the most suitable for young regions.

EW(H$\beta$) values for the selected HII regions are between 2.2 to 12.24 \AA\ for outer ring regions and from 8.91 to 15.00 \AA\ for the inner ring ones, corresponding to regions of active star formation. The larger Balmer emission line luminosities and equivalent widths shown by inner ring regions could, in principle, imply an earlier evolutionary stage.

\begin{figure}
\centering
 \includegraphics[width=0.87\columnwidth]{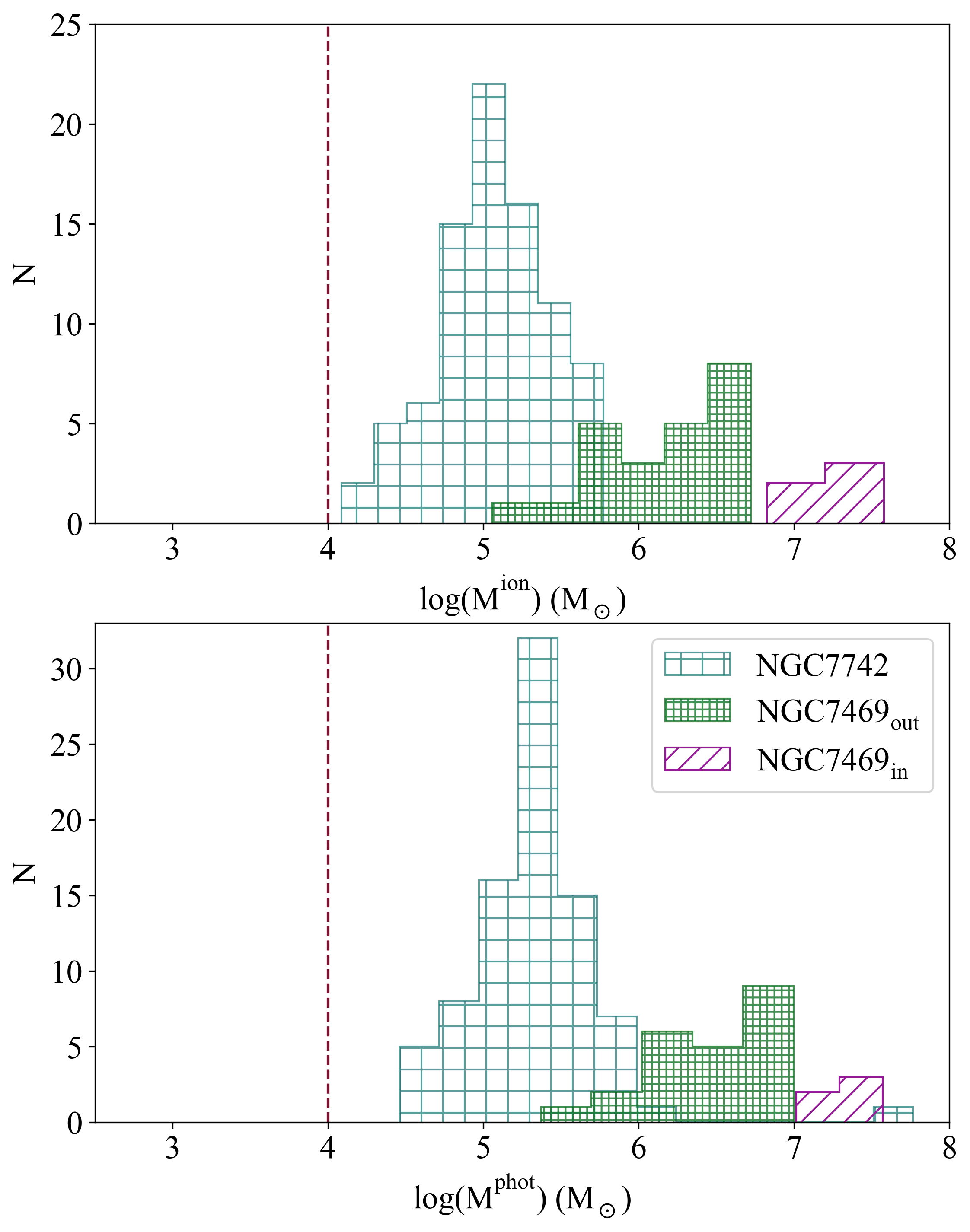}
 \caption{Histograms of the distributions of ionising (top panel) and  photometric (bottom panel) stellar masses for the outer and inner ring HII regions, in green and purple, respectively. The dashed line corresponds to 10$^4$ M$_\odot$ (see text for details).}
 \label{fig:ion_pho_masses}
\end{figure}

The upper panel of Fig. \ref{fig:ion_pho_masses} shows the distribution of ionising masses for the HII regions in the two rings of NGC~7469  compared with the NGC~7742 ones. Ionising cluster masses in the former galaxy are higher than in the latter, something to be expected given their larger sizes. Inner ring regions harbour the most  massive clusters with a mean mass value of 2.1 $\times$ 10$^7$ M$_\odot$ higher by an order of magnitude than the ionising clusters in the outer ring. All the studied clusters have masses higher than 10$^4$ M$_\odot$, which is the lower limit for a cluster to fully sample the IMF \citep{1994ApJS...91..553G,2010A&A...522A..49V}. Furthermore, our results are only lower limits to the ionising masses since we are assuming that: (i), there is no dust absorption and reemission at infrared wavelengths; and (ii), there is no photon escape from HII regions. 
Using the same SSP models, we have derived the photometric masses of our CNSFRs from their absolute r-magnitudes. The bottom panel of Fig. \ref{fig:ion_pho_masses} shows the distribution of these photometric masses. For the regions within the outer ring we have obtained values between 2.4 $\times$ 10$^5$ and 9.9 $\times$ 10$^6$ $M_\odot$ while inner ring regions are from 1.0 $\times$ 10$^7$ and 3.7 $\times$ 10$^7$ $M_\odot$. The photometric masses of outer ring regions follow ionising star masses in a constant proportion of about 3. However, this relation seems to be lost in the inner ring regions, where the most massive ionising clusters deviate being close to the 1:1 relation. Table \ref{tab:cluster properties} summarises these results listing 
in columns 1 to 5: (1) region ID; (2) Q(He$_0$); (3) EW(H$\beta$; (4) ionising cluster mass; (5) integrated photometric mass. 

\begin{table*}
\centering
\caption{Ionising cluster properties. This is a sample table consisting of the first seven rows of data. The full table is available in Appendix \ref{ap:tables}.}
\label{tab:cluster properties}
\begin{tabular}{ccccc}
\hline
Region ID & \begin{tabular}[c]{@{}c@{}}Q(He$_0$) \\ (photons s$^{-1}$)\end{tabular}& \begin{tabular}[c]{@{}c@{}}EW(H$\beta$)  \\ (\AA )\end{tabular} &  \begin{tabular}[c]{@{}c@{}}M$_{ion}$ \\ (M$_\odot$)\end{tabular} &  \begin{tabular}[c]{@{}c@{}}M$_{phot}$ \\ (M$_\odot$)\end{tabular} \\ \hline
R1 & (4.9 $\pm$ 1.9) $\times$ 10$^{48}$ & 10.54 $\pm$ 0.52 & (30.0 $\pm$ 2.9) $\times$ 10$^{5}$ & (54.4 $\pm$ 9.8) $\times$ 10$^{5}$ \\ 
R2 & (5.5 $\pm$ 3.7) $\times$ 10$^{48}$ & 6.69 $\pm$ 0.38 & (47.4 $\pm$ 5.4) $\times$ 10$^{5}$ & (8.9 $\pm$ 1.2) $\times$ 10$^{6}$ \\ 
R3 & (2.2 $\pm$ 1.4) $\times$ 10$^{48}$ & 6.24 $\pm$ 0.36 & (28.0 $\pm$ 3.3) $\times$ 10$^{5}$ & (53.8 $\pm$ 7.5) $\times$ 10$^{5}$ \\ 
R4 & (2.6 $\pm$ 2.8) $\times$ 10$^{48}$ & 4.98 $\pm$ 0.34 & (37.2 $\pm$ 5.5) $\times$ 10$^{5}$ & (9.1 $\pm$ 1.5) $\times$ 10$^{6}$ \\ 
R5 & (0.8 $\pm$ 1.6) $\times$ 10$^{48}$ & 5.10 $\pm$ 0.38 & (20.5 $\pm$ 3.2) $\times$ 10$^{5}$ & (50.3 $\pm$ 8.3) $\times$ 10$^{5}$ \\ 
R6 & (2.4 $\pm$ 1.0) $\times$ 10$^{48}$ & 12.24 $\pm$ 0.87 & (11.0 $\pm$ 1.3) $\times$ 10$^{5}$ & (14.5 $\pm$ 2.7) $\times$ 10$^{5}$ \\ 
R7 & (3.8 $\pm$ 1.8) $\times$ 10$^{48}$ & 9.09 $\pm$ 0.73 & (24.9 $\pm$ 3.4) $\times$ 10$^{5}$ & (38.4 $\pm$ 2.5) $\times$ 10$^{5}$ \\ 

\hline

\end{tabular}
\end{table*}

\subsubsection{CNSFR evolutionary stage}
\label{CNSFR-evol}

\begin{figure}
\includegraphics[width=\columnwidth]{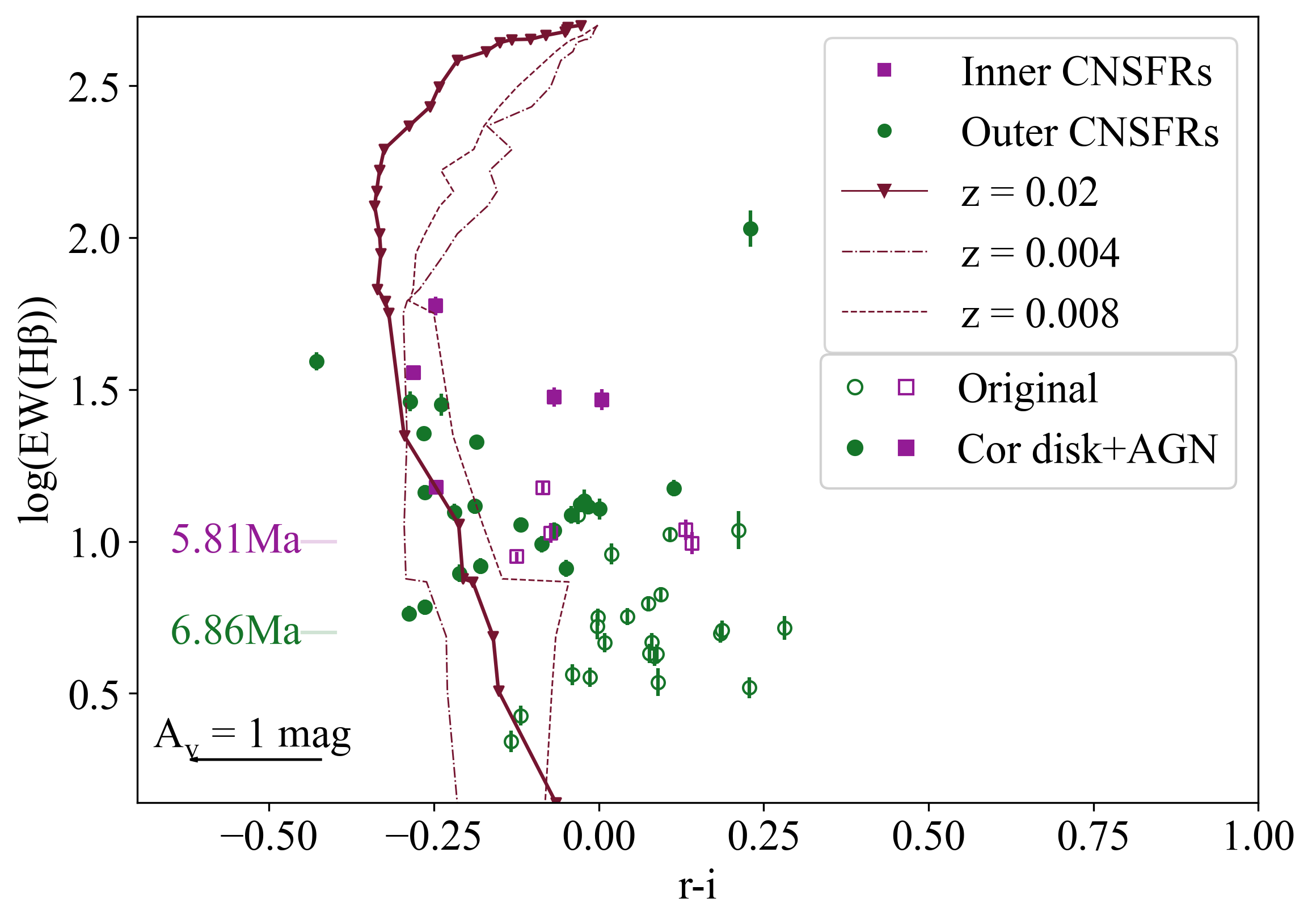}
\caption{Relation between the equivalent width of the H$\beta$ emission line and the r-i colour. The solid line was calculated using PopStar models \citep{Popstar}. The beginning and end of the line correspond to ages of 0.1 and 8.5 Ma. The observational errors are inside the symbols in the graph.}
\label{fig:EWHb_color}
\end{figure}

The evolutionary state of the analysed clusters can be interpreted with the help of evolutionary population synthesis models, which predict the EWs of Balmer lines as a function of the continuum colour of a SSP. Under this assumption, these EWs, mainly EW(H$\beta$), can be considered good estimators of the age of a given cluster \citep{1981Ap&SS..80..267D} since they provide the ratio of the present to past star formation rates. The former is related with the evolutionary time scale of the ionising star clusters and decreases with age up to 10 Ma, and the latter samples a longer time scale ( $\geq$ 300 Ma) and becomes redder with age \citep[see][]{diaz2000}. Figure \ref{fig:EWHb_color} shows the relation between logEW(H$\beta$), and the r-i colour together with evolutionary tracks calculated with the SSP PopStar models described above with IMF parameters more appropriate for the case of evolved star clusters (m$_{low}$ = 0.15 M$_\odot$, m$_{up}$ = 100 M$_\odot$). 

Our observed ring regions, taken at face value (green and purple open symbols for outer and inner regions respectively), lie to the right of the line defined by SSP  showing low values of logEW(H$\beta$) at mean ages of 6.9 Ma for the outer ring regions and 5.8 Ma for the inner ring ones. However, as shown in \citet[][]{2010MNRAS.403.2012M}, star clusters older than 5.2 Ma do not produce a detectable emission-line spectrum. After deprojection of the galaxy outer ring using observed inclination angle (45 $^{\circ}$) and position angle of the major axis of
(128 $^{\circ}$) from \citet{2004ApJ...602..148D}, we have calculated a radial profile for the r and i continuum bands and the continuum underlying the H$\beta$ emission line, fitting a Sérsic disc and subtracting it from the outer galaxy ring. We have repeated this procedure for the inner ring but using an additional component to fit the central emission and remeasured both EW(H$\beta$) and the r-i colour. Solid symbols (green for outer ring regions and purple for inner ring ones) show the locations of the isolated young ionising clusters at evolutionary track mean ages of 5.7 Ma for the outer ring regions and younger ages, between 5.1 and 5.7 Ma for the inner ring ones. These regions can be identified as the young (5-6 Ma), stellar population accounting for most of the IR luminosity in the central region of the galaxy \citep[][]{2007ApJ...661..149D}. 

\subsubsection{Wolf-Rayet stellar population}

We have detected carbon Wolf-Rayet (WRC) broad star features in the spectra of all analysed HII regions. The presence of these features would place the age of the regions between 3.2 and 5.25 Ma according to PopStar models.  Wolf-Rayet (WR) are massive stars (M > 25 M$_\odot$) which have left the main sequence having lost part of their hydrogen-rich envelope by means of powerful stellar winds. They can be classified as nitrogen stars (WN) and carbon stars (WC and WO) which are in the CNO and He burning evolutionary phases respectively. The Wolf-Rayet (WR) population of NGC 7469 was studied by \cite{2016A&A...592A.105M} as part of a systematic search of extragalactic regions including this kind of star. They used PMAS-PPAK IFU at the CAHA 3.5m telescope, with a field of view (FoV) of 74 arcsec × 64 arcsec providing a spatial sampling of 1 arcsec/pix and a spectral resolution R = 850. They observed the blue WR bump around He II$\lambda$ 4686 \AA\ which is associated with nitrogen WR stars (WN). However, the red feature is less prominent than the blue one and hence they did not detect the red WR bump around C IV $\lambda$ 5808 \AA , associated with carbon WR stars (WC).

\begin{figure}
\includegraphics[width=\columnwidth]{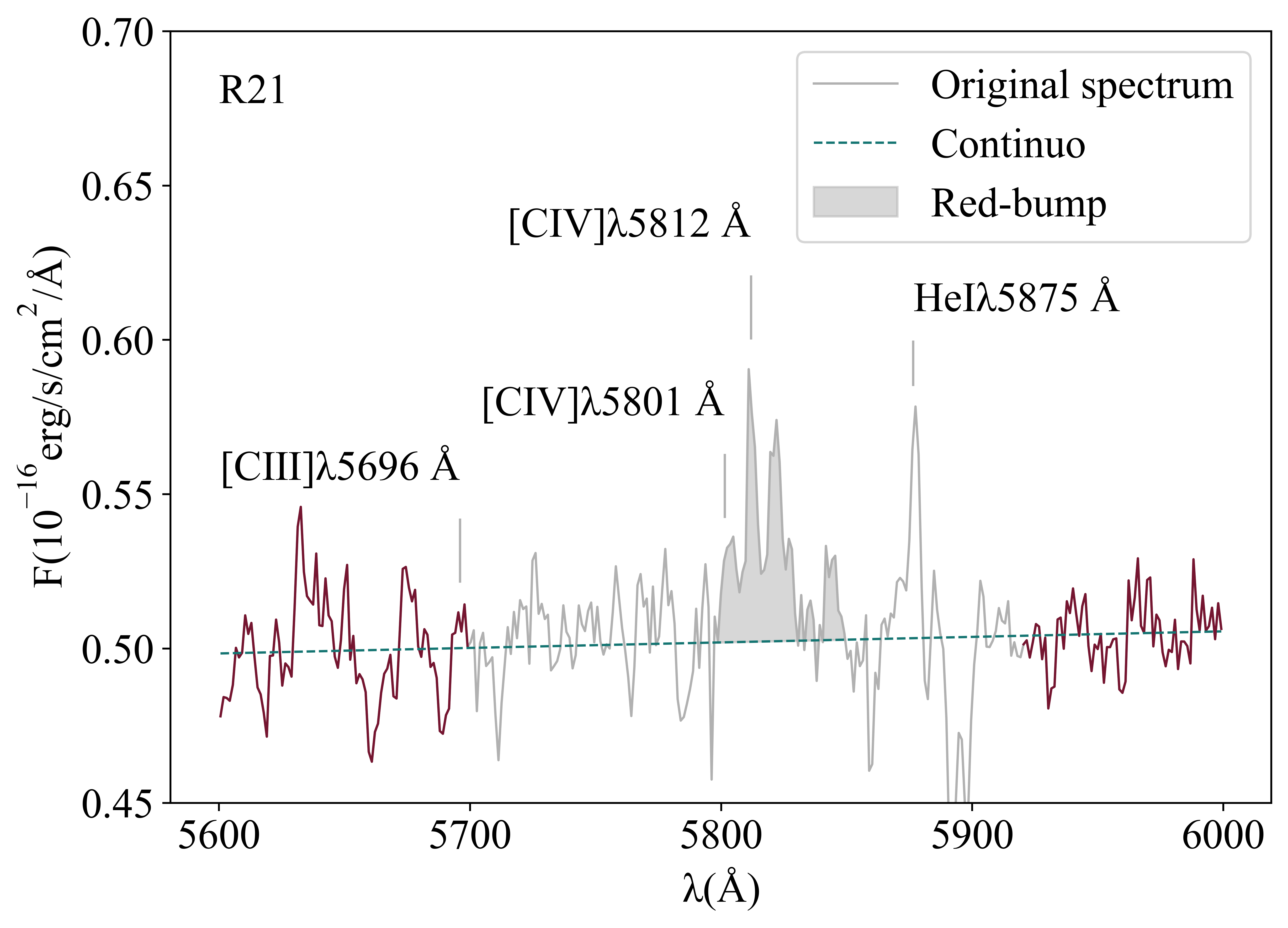}
\caption{Emission lines of WR stars in region R21. The red solid line shows the side-bands selected to calculate the continuum, shown with a blue dashed line. The integrated red bump is shown in grey.}
\label{fig:wr-bump}
\end{figure}

The larger sensitivity of the MUSE spectrograph in the red part of the spectrum has allowed the measurement of the fluxes in the called `red bump' centred at about 5808 \AA\ assuming a linear behaviour of the continuum and choosing the same side-bands as \cite{2016A&A...592A.105M} (5600 - 5700 \AA , 5920 - 6000 \AA ). We have integrated the bump flux from 5798 \AA\ to 5850 \AA. Figure \ref{fig:wr-bump} shows as an example the spectrum of region R21 showing the position of the CIV$\lambda\lambda$ 5801,12 \AA\ and CIII$\lambda$ 5696 \AA\ lines \citep[see][]{1992AJ....103.1159M} and the red WR bump. 
We can see, in the spectrum of this region, the two  CIV lines although we cannot detect the  CIII line, hence we assume that the red bump emission originates in early WC stars (WCE). Assuming the average luminosity of WCE stars as L$_{WCE}$(CIV$\lambda$ 5808 \AA) $\sim$ 3.3 $\times$ 10$^{36}$ erg/s \citep{1992ApJ...401..543V}, their number can be calculated as
\begin{equation}
 N_{WCE} = 36.26\cdot 10^{-3} \left(\frac{F(Red-bump)}{10^{-17}}\right) \left(\frac{D}{10}\right)^2 ,
\end{equation}
where F(Red-bump) is expressed in erg s$^{-1}$ cm$^{-2}$ and D is the distance to NGC~7469 which has been taken as 66.47 Mpc (see Tab. \ref{tab:galaxy characteristics}). Outer ring 
regions contain between 12 and 415 stars, while these numbers are increased to 185 and 1477 in the inner ring regions. This is something to be expected since the inner regions are larger, more massive and their metallicity is super-solar.

\cite{2016A&A...592A.105M} reported 21772 late-type WN stars (WNL) inside a radius of 1306 pc (4.146 arcsec; see Tab. \ref{tab:galaxy characteristics}) from the galactic centre. This area includes our five inner regions, which we have found to include a total of 3672 WCE. This gives a WCE/WNL ratio of 0.17 consistent with the range calculated by these authors (15 - 25 \%). 

The large number of WR in these CNSFRs could affect their properties. In fact, the low filling factors and large radii measured in the HII regions of this galaxy can be explained by the stellar winds produced by these WR stars. 

The radius associated with the kinetic energy produced by the combined wind of a population of WR stars can be found using the expression in \citet{1975ApJ...200L.107C} as 
\begin{equation}
 R_{WR}= 1.6 \left(\epsilon /n\right )^{1/5} \cdot t^{3/5} ,
\end{equation}
\noindent where $\epsilon$ is the total ejected energy in units of 10$^{36}$ erg/s, n is the interstellar medium density in cm$^{-3}$, t is the age of the expanding shell in units of 10$^4$ yr, and R$_{WR}$ is expressed in parsec.

Using this information PopStar models provide this radius, R$_{PopStar}$, for the number of WR present in ionising clusters of a given age, metallicity, and IMF.  Hence, we estimate R$_{WR}$  of our CNSFRs as

\begin{equation}
 R_{WR}= R_{PopStar} \left[ \frac{n_{WR}}{n_{WR,PopStar}} \cdot \frac{\epsilon_{WR,PopStar}}{\epsilon_{total,PopStar}}\right]^{1/5} ,
\end{equation}

\noindent where n$_{WR}$ is the number of WR in each observed cluster, and $n_{WR,PopStar}$, $\epsilon_{WR,PopStar}$, and $n_{WR,PopStar}$ are the  number of WR stars, the total energy ejected by WRs, and the total energy ejected by WR and supernova winds in the used model. In our case we have assumed an age of 5.5 Ma, solar metallicity and the IMF parameters given in Sect. \ref{SC_pop}.

\begin{figure}
\includegraphics[width=\columnwidth]{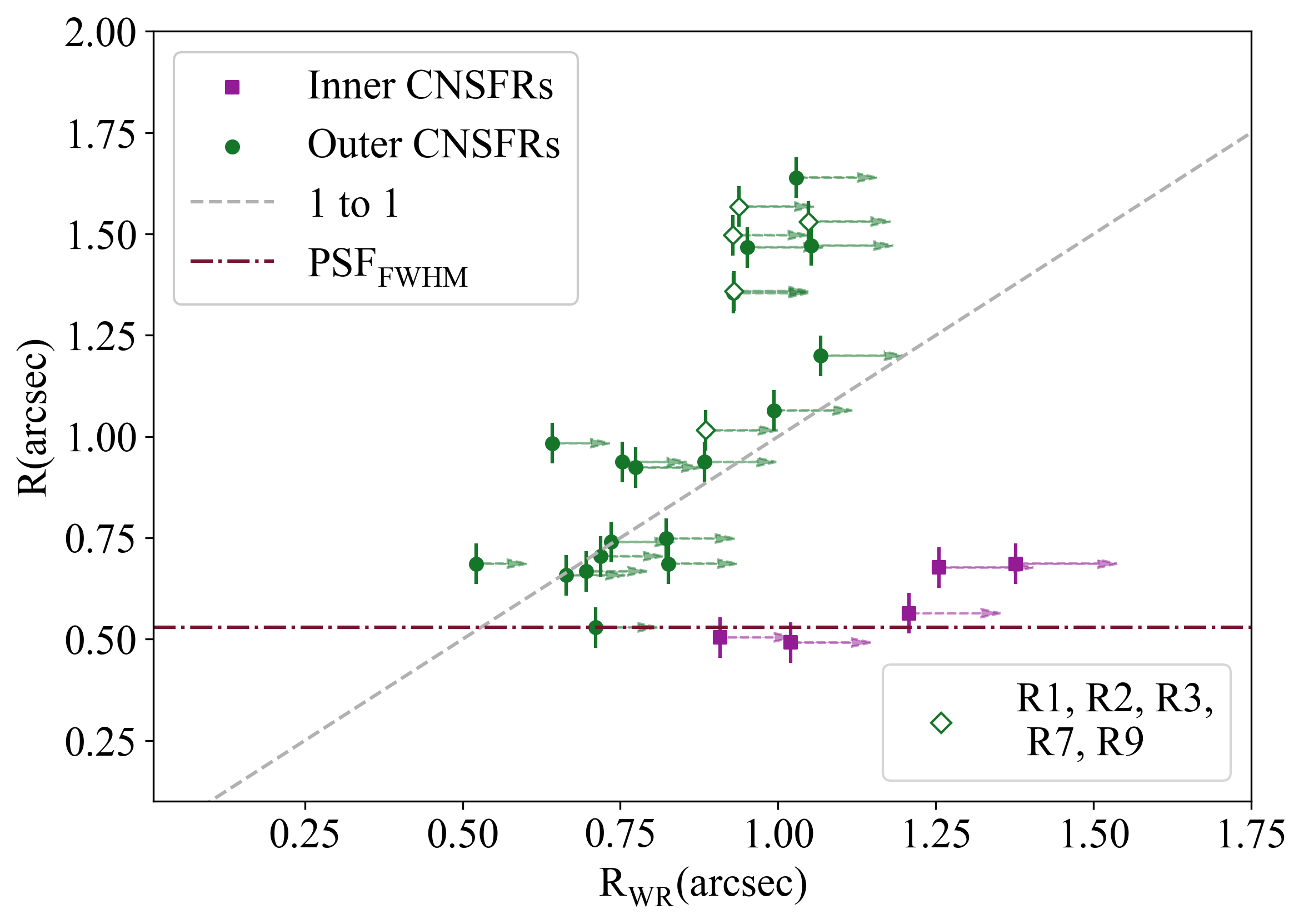}
 \caption{Derived angular radii associated with the kinetic energy produced by the carbon WR stars against the angular radii measured from the HII region segmentation. The arrows show these values with the addition of the energy produced by the nitrogen WR stars assuming the ratio proposed by the  PopStar models (63 \% WC and 37 \% WN). Open markers show the HII complexes identified in H$\alpha$.}
 \label{fig:radio_WR}
\end{figure}

Figure \ref{fig:radio_WR} shows the measured radii of the selected HII regions (see Sect. \ref{sec:segmentation}) vs  those blown out by WR winds. We can see that both radii are in good agreement contrary to the case of the ionisation derived angular radii ($\phi$; see Fig. \ref{fig:radios} in Sect. \ref{sec:CNSFR}) and all regions look spatially resolved, as from the H$\alpha$ map. Most regions predicted angular radii fully compatible with those measured, so we can assume that their ionised bubbles are inflated by the winds from WR stars. There are still 7 outer ring regions whose position in the plot cannot be explained by only by the model computed WR winds. They are located in the upper right area of the graph and have been represented by green solid dots. However, we should keep in mind that PopStar models only include single WR stars and the inclusion of binaries in the evolutionary codes would increase the production of WRs also increasing the duration of the WR evolutionary phase \citep{2008MNRAS.384.1109E}. On the other hand, the inner ring regions have the greatest number of WRs, but the radii predicted by the model winds are too large. In these regions, the pressure is probably balanced by the more crowded central environment of the galaxy. We have found a relation between the number of WR and the ionisation and the photometric masses of each region. From this result, we can conclude that the IMF is similar among clusters. Additionally, we have verified that the constant ratio between these two masses (M$^{phot}$ / M$^{ion}$ $\sim$ 3) is lost for regions with the highest number of this kind of stars (see Sect. \ref{clusters}).

\begin{table}
\centering
\caption{WR features and number for observed CNSFRs.}
\label{tab:WR}
\begin{tabular}{cccc}
\hline
Region ID &\begin{tabular}[c]{@{}c@{}}Red-bump \\ (erg s$^{-1}$ cm$^{2}$)\end{tabular}& N$_{WCE}$ & \begin{tabular}[c]{@{}c@{}}R$_{WR}$ \\ (arcsec)\end{tabular} \\ \hline
R1 & (135.4 $\pm$ 7.5) $\times$ 10$^{-17}$ & 217 $\pm$ 2 & 0.94\\ 
R2 & (23.6 $\pm$ 1.9) $\times$ 10$^{-16}$ & 378 $\pm$ 4 & 1.05\\ 
R3 & (13.0 $\pm$ 1.1) $\times$ 10$^{-16}$ & 208 $\pm$ 3 & 0.93\\ 
R4 & (21.5 $\pm$ 2.5) $\times$ 10$^{-16}$ & 345 $\pm$ 4 & 1.03\\ 
R5 & (12.9 $\pm$ 1.6) $\times$ 10$^{-16}$ & 207 $\pm$ 2 & 0.93\\ 
R6 & (45.2 $\pm$ 3.3) $\times$ 10$^{-17}$ & 72 $\pm$ 1 & 0.75\\ 
R7 & (12.9 $\pm$ 1.2) $\times$ 10$^{-16}$ & 206 $\pm$ 3 & 0.93\\ 
R8 & (14.5 $\pm$ 1.5) $\times$ 10$^{-16}$ & 232 $\pm$ 3 & 0.95\\ 
R9 & (10.1 $\pm$ 1.1) $\times$ 10$^{-16}$ & 162 $\pm$ 2 & 0.89\\ 
R10 & (7.2 $\pm$ 1.2) $\times$ 10$^{-16}$ & 115 $\pm$ 2 & 0.83\\ 
R11 & (24.1 $\pm$ 3.6) $\times$ 10$^{-16}$ & 386 $\pm$ 4 & 1.05\\ 
R12 & (18.0 $\pm$ 3.2) $\times$ 10$^{-16}$ & 289 $\pm$ 4 & 0.99\\ 
R13 & (10.0 $\pm$ 2.2) $\times$ 10$^{-16}$ & 161 $\pm$ 2 & 0.88\\ 
R14 & (20.4 $\pm$ 3.1) $\times$ 10$^{-17}$ & 33 $\pm$ 0 & 0.64\\ 
R15 & (40.2 $\pm$ 5.5) $\times$ 10$^{-17}$ & 64 $\pm$ 1 & 0.74\\ 
R16 & (7.0 $\pm$ 1.1) $\times$ 10$^{-16}$ & 113 $\pm$ 1 & 0.82\\ 
R17 & (34.0 $\pm$ 4.4) $\times$ 10$^{-17}$ & 54 $\pm$ 1 & 0.71\\ 
R18 & (30.5 $\pm$ 4.1) $\times$ 10$^{-17}$ & 49 $\pm$ 1 & 0.70\\ 
R19 & (24.2 $\pm$ 3.4) $\times$ 10$^{-17}$ & 39 $\pm$ 1 & 0.66\\ 
R20 & (35.8 $\pm$ 6.2) $\times$ 10$^{-17}$ & 57 $\pm$ 1 & 0.72\\ 
R21 & (51.9 $\pm$ 6.4) $\times$ 10$^{-17}$ & 83 $\pm$ 1 & 0.77\\ 
R22 & (25.9 $\pm$ 5.4) $\times$ 10$^{-16}$ & 415 $\pm$ 5 & 1.07\\ 
R23 & (7.2 $\pm$ 1.1) $\times$ 10$^{-17}$ & 12 $\pm$ 0 & 0.52\\ 
\hline
Ra & (92.2 $\pm$ 2.0) $\times$ 10$^{-16}$ & 1477 $\pm$ 20 & 1.38\\ 
Rb & (47.9 $\pm$ 4.0) $\times$ 10$^{-16}$ & 767 $\pm$ 14 & 1.21\\ 
Rc & (58.1 $\pm$ 5.1) $\times$ 10$^{-16}$ & 931 $\pm$ 10 & 1.26\\ 
Rd & (20.6 $\pm$ 2.4) $\times$ 10$^{-16}$ & 330 $\pm$ 3 & 1.02\\ 
Re & (11.5 $\pm$ 1.5) $\times$ 10$^{-16}$ & 185 $\pm$ 2 & 0.91\\ 
\hline
\end{tabular}
\end{table}

Table \ref{tab:WR} summarises our results listing for each region in columns 1 to 4: (1) the region ID; (2) the integrated red bump flux; (3) the number of early WC stars; and (4) the radius of the WR wind blown region.

\subsubsection{Dynamical masses}

 We have calculated the dynamical mass for each observed star-forming complex using their measured stellar velocity dispersion (see \ref{sigma}) and size assuming the system to be virialised and that: (i) it has spherical symmetry; (ii) it is gravitationally bounded; and (iii) it has an isotropic velocity distribution. Then, the dynamical mass is given by \citep{1996ApJ...466L..83H, 1996ApJ...472..600H} as
\begin{equation}
M^{dyn}=3\cdot \sigma^2 \cdot \frac{R}{G}   , 
\end{equation}
where $\sigma$ is the velocity dispersion of the system, R is its radius, and G is the gravitational constant.

\begin{figure}
\centering
\includegraphics[width=\columnwidth]{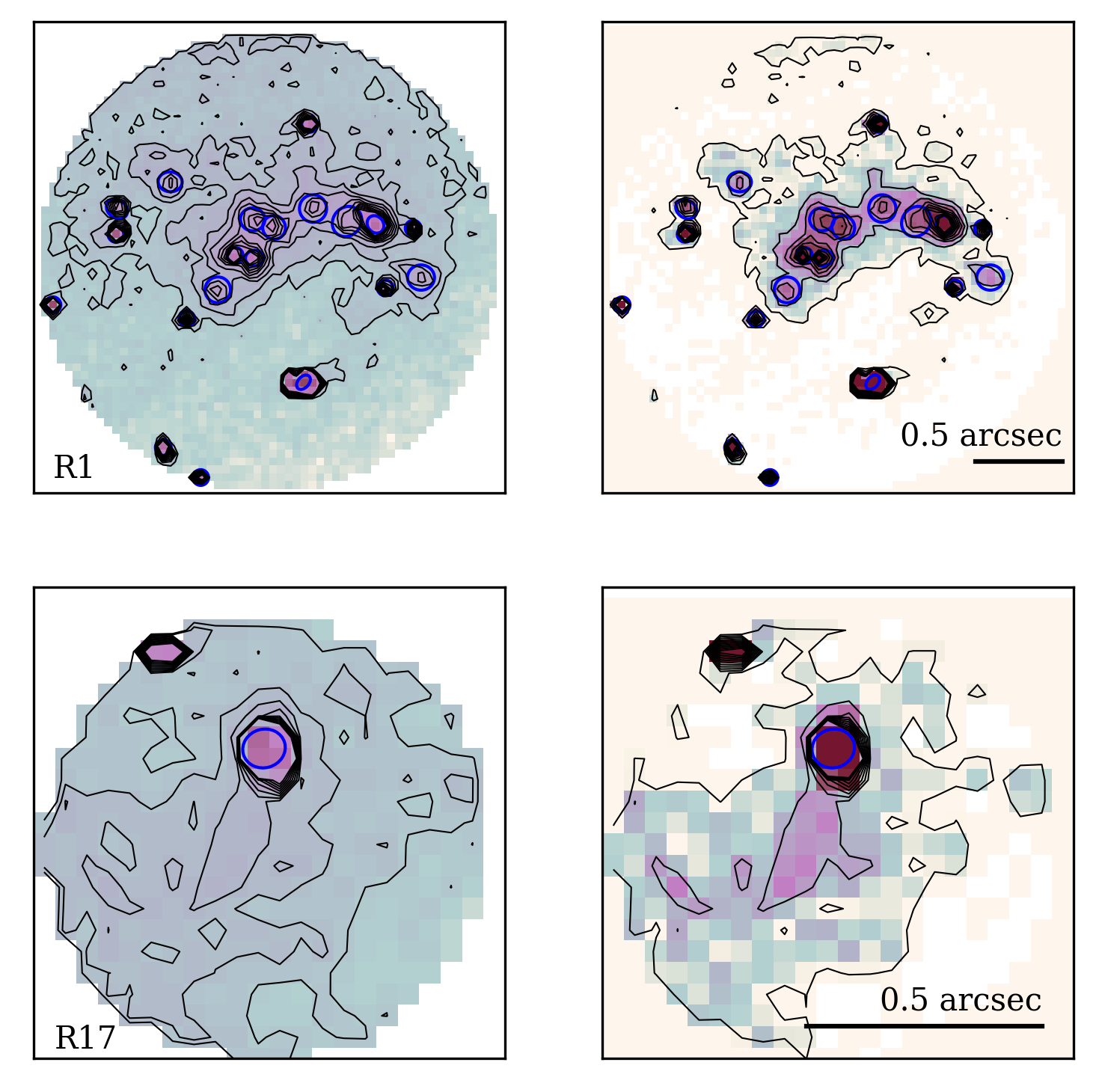}
\caption{Results of the star cluster radius measurement procedure for regions R1 and R17 (upper and lower panels, respectively). The left panels show the F606W WFPC2-HST image for each selected region and the right panels show the same image after the background subtraction. Selected clusters are shown with blue circles, and the angular scale is shown in the corner of each panel.}
 \label{fig:plot_radios_stars_NGC7469}
\end{figure}

In Sect. \ref{sec:segmentation} we measured the radii of our CNSFRs from the selected HII regions on the observed H$\alpha$ image thus including the total gas emission. However, in order to better identify the stellar clusters, we have now calculated the star cluster sizes from the F606W WFPC2-HST image. First, we have calculated and subtracted the region background by fitting a third-order polynomial. Next, we  fitted each knot present in the observed clusters assuming a two-dimensional Gaussian profile. The radius of each knot has been taken as 1/2$\cdot$ FWHM. Figure \ref{fig:plot_radios_stars_NGC7469} shows two examples of the described procedure. 

All the observed  regions are composed by more than one knot. Only regions R17, R19, R20 and Rb seem to host single clusters. There are two large complexes R1 and R13, with 20 and 19 knots respectively.  The radii of the single knots vary between 7.3 pc and 58.2 pc for outer ring regions. Inner ring regions have knots with very similar sizes, with their radii taking values from 11.0 to 27.5 pc.
No knot is appreciable in the continuum at 6060 \AA\ in the position of region Re and hence we have not measured any size for Re region.  Also, no cluster can be seen either in UV-HST images \citep[see][]{2007ApJ...661..149D}. However, this region shows CaII triplet lines with the same equivalent width as the rest of regions and is identified in H$\alpha$, [ArII]$ \lambda$ 6.99 $\mu$m, PAH$\lambda$ 6.2 $\mu$m and in 11.7 $\mu$m imaging \citep{2022ApJ...940L...5U,2022A&A...666L...5G,1994ApJ...425L..37M}. A possible explanation could come from its high  interstellar extinction value, since photons at short wavelengths could be absorbed.

\begin{figure}
\centering
\includegraphics[width=\columnwidth]{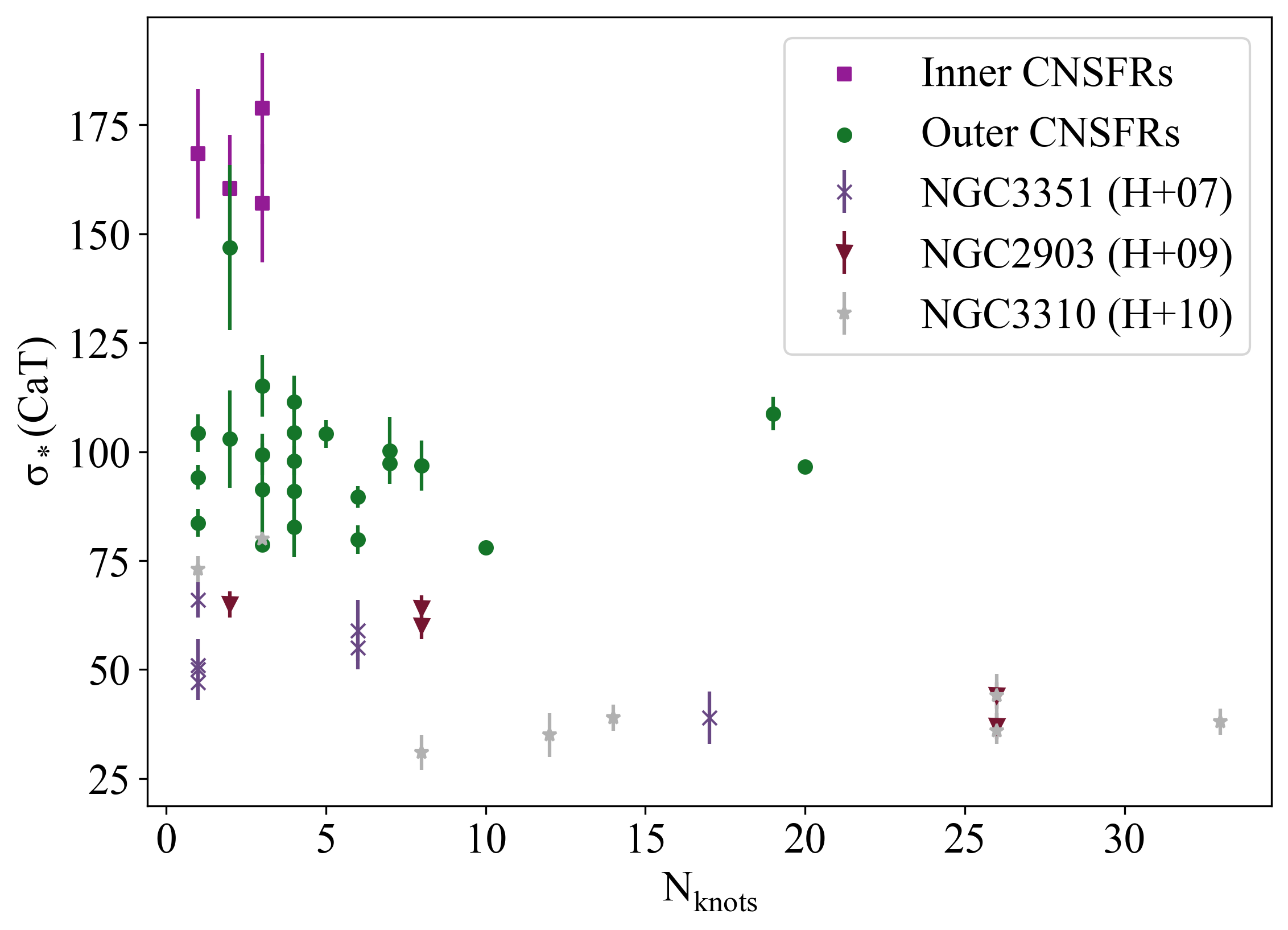}
\caption{CaT velocity dispersion as a function of the number of knots (subclusters) in the segmented regions compared to those from the literature.}
 \label{fig:referee}
\end{figure}

We have used the total velocity dispersion corresponding to the entire selected region to infer the mass of each knot inside it. Then, we have added all masses to calculate the mass of each complete CNSFR. It must be kept in mind that, as in the case of our previous works \citep[][]{hagele2007, hagele2009, hagele2010}, we have measurements for the sizes of each knot, as measured on HST images, but we do not have direct access to the stellar velocity dispersion of each individual cluster, since our spectroscopic measurements encompass  wider circular areas  with typical radii of 1 arcsec which, at the distance of NGC7469 corresponds to 360 pc, that is very large structures.
This method can overestimate the mass of the individual SFC but MUSE spatial resolution is insufficient to allow the integration of each individual cluster. However, the velocity dispersions derived from the gas emission lines associated with the ionising SFC of the outer ring regions is of the same order as those derived from the CaT lines and in most cases somewhat larger than these.
{Moreover, regarding the issue of the existence of several clusters in the segmented regions, Fig. \ref{fig:referee} shows the CaT velocity dispersion as a function of the number of knots in each SFC in this work, together with the ones obtained from the literature \citep[][(H+07), (H+10), and  (H+10) respectively]{hagele2007,hagele2009,hagele2010}. No trend is evident either for outer or inner regions, showing that there is no correlation between these two quantities and therefore no artificial broadening effect due to sub-clustering. This is reasonable since the rotation among the different unresolved subclusters (knots) must also respond to the local gravity and it is also clear from the figure that the inner regions show larger velocity dispersions than those in the outer ring.

Another factor to take into account is the possible contribution of binaries \citep[see][]{hagele2010} among the red supergiant and red giant populations from which we have derived the stellar velocity dispersions. The single-star assumption introduces a systematic error that depends on the properties of the star cluster and the binary population \citep{2008A&A...480..103K}. From numerical simulations these authors conclude that the importance of this effect decreases with increasing cluster mass and therefore the dynamical masses of massive star clusters are only mildly affected by the presence of binaries.

\begin{figure}
\centering
\includegraphics[width=\columnwidth]{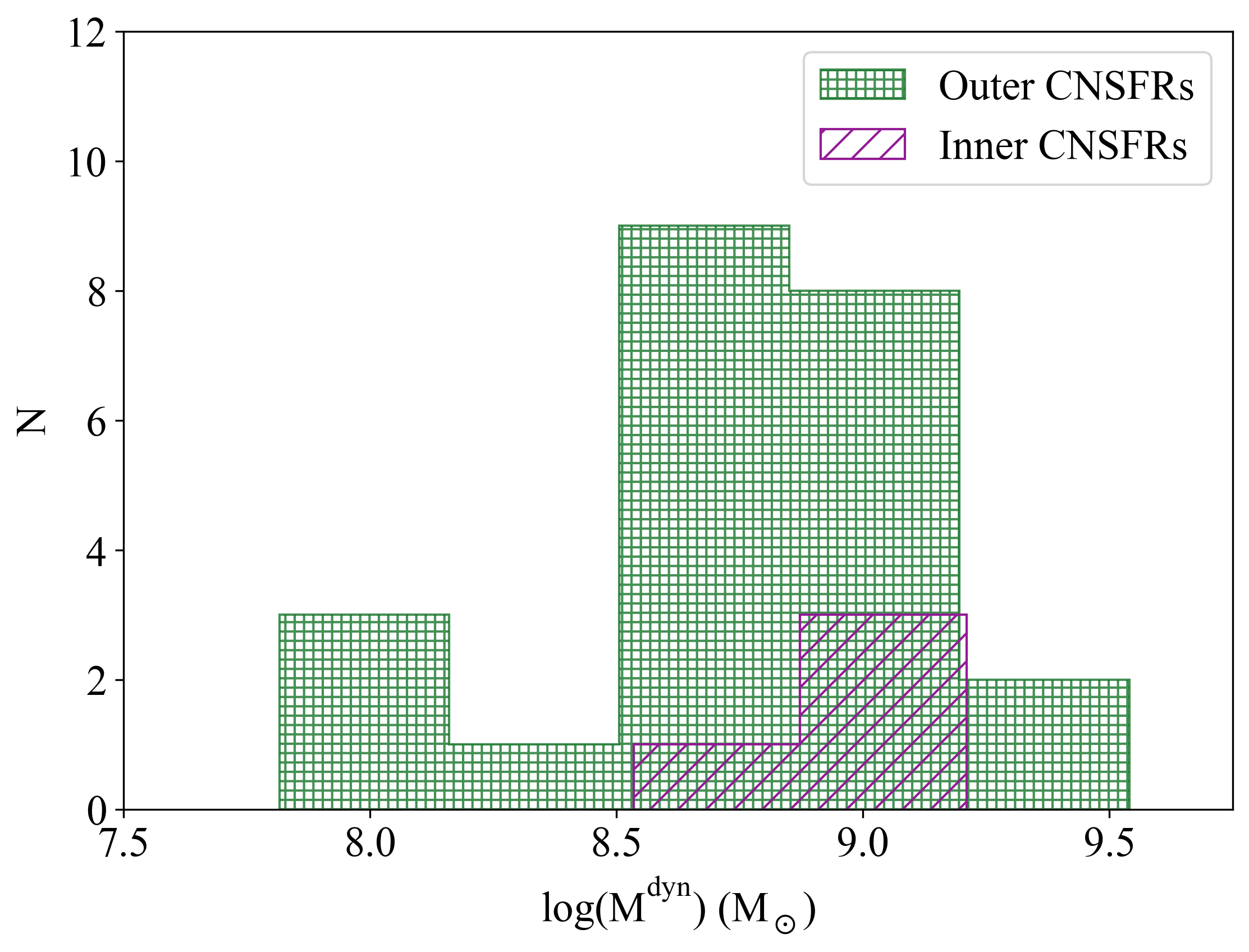}
\caption{Histograms of the distribution of dynamical masses for the  outer and inner ring HII regions, in green and purple, respectively.}
 \label{fig:hist_masa_din_NGC7469}
\end{figure}

\begin{table}
\centering
\setlength{\tabcolsep}{3pt}
\caption{Star-forming complexes masses and cluster sizes for observed CNSFRs.}
\label{tab:mdyn}
\begin{tabular}{ccccc}
\hline
\begin{tabular}[c]{@{}c@{}}Region \\ ID\end{tabular} & N$_{knots}$ & \begin{tabular}[c]{@{}c@{}}R$^* _{mean}$ \\ (pc/knot)\end{tabular} 
& \begin{tabular}[c]{@{}c@{}} M$_{dyn}$ \\ (M$_\odot$)\end{tabular} & \begin{tabular}[c]{@{}c@{}} M$_{ion}$/M$_{dyn}$ \\ (per cent)\end{tabular} \\
\hline
R1 & 20 & 21.0 & (2734.8 $\pm$ 8.1) $\times$ 10$^{6}$ & 0.11\\ 
R2 & 4 & 28.0 & (98.3 $\pm$ 2.5) $\times$ 10$^{7}$ & 0.48\\ 
R3 & 10 & 20.3 & (88.1 $\pm$ 2.6) $\times$ 10$^{7}$ & 0.32\\ 
R4 & 7 & 16.5 & (80.6 $\pm$ 5.7) $\times$ 10$^{7}$ & 0.46\\ 
R5 & 5 & 19.5 & (74.1 $\pm$ 2.8) $\times$ 10$^{7}$ & 0.28\\ 
R6 & 4 & 20.9 & (57.7 $\pm$ 6.0) $\times$ 10$^{7}$ & 0.19\\ 
R7 & 4 & 36.4 & (99.6 $\pm$ 6.7) $\times$ 10$^{7}$ & 0.25\\ 
R8 & 8 & 24.6 & (13.1 $\pm$ 1.3) $\times$ 10$^{8}$ & 0.23\\ 
R9 & 3 & 28.2 & (3681.5 $\pm$ 5.0) $\times$ 10$^{5}$ & 0.55\\ 
R10 & 3 & 23.3 & (62.3 $\pm$ 4.4) $\times$ 10$^{7}$ & 0.24\\ 
R11 & 7 & 25.9 & (1229.2 $\pm$ 6.9) $\times$ 10$^{6}$ & 0.40\\ 
R12 & 6 & 32.9 & (88.6 $\pm$ 6.1) $\times$ 10$^{7}$ & 0.50\\ 
R13 & 19 & 21.7 & (34.2 $\pm$ 1.4) $\times$ 10$^{8}$ & 0.05\\ 
R14 & 6 & 17.7 & (568.3 $\pm$ 6.6) $\times$ 10$^{6}$ & 0.06\\ 
R15 & 3 & 27.6 & (44.6 $\pm$ 4.7) $\times$ 10$^{7}$ & 0.15\\ 
R16 & 4 & 22.4 & (39.6 $\pm$ 3.0) $\times$ 10$^{7}$ & 0.33\\ 
R17 & 1 & 16.1 & (102.9 $\pm$ 2.8) $\times$ 10$^{6}$ & 0.61\\ 
R18 & 2 & 18.6 & (25.6 $\pm$ 2.4) $\times$ 10$^{7}$ & 0.21\\ 
R19 & 1 & 24.3 & (114.4 $\pm$ 2.9) $\times$ 10$^{6}$ & 0.40\\ 
R20 & 1 & 8.5 & (64.1 $\pm$ 3.8) $\times$ 10$^{6}$ & 0.91\\ 
R21 & 4 & 22.6 & (48.9 $\pm$ 2.3) $\times$ 10$^{7}$ & 0.19\\ 
R22 & 3 & 31.8 & (628.1 $\pm$ 4.5) $\times$ 10$^{6}$ & 0.84\\ 
R23 & 2 & 19.6 & (54.0 $\pm$ 5.3) $\times$ 10$^{7}$ & 0.02\\ 
\hline
Ra & 3 & 21.3 & (78.1 $\pm$ 5.7) $\times$ 10$^{7}$ & 4.86\\ 
Rb & 1 & 16.9 & (30.8 $\pm$ 4.0) $\times$ 10$^{7}$ & 8.77\\ 
Rc & 3 & 23.8 & (14.8 $\pm$ 1.1) $\times$ 10$^{8}$ & 0.76\\ 
Rd & 2 & 26.5 & (72.4 $\pm$ 1.4) $\times$ 10$^{7}$ & 0.92\\ 
\hline
\end{tabular}
\end{table}

Table \ref{tab:mdyn} shows our results listing for each region in columns 1 to 5: (1) the region ID; (2) the number of stellar knots present in each region, (3) the mean radius of them, (4) the dynamical masses and (5) the ratio of ionising to dynamical masses.

Figure \ref{fig:hist_masa_din_NGC7469} shows the distribution of the calculated dynamical masses for each CNSFR. Contrary to the case of ionising and photometric masses, the star-forming complexes in both rings show similar masses, with median values 1.07 $\times$ 10$^9$ and  6.58 $\times$ 10$^8$ for inner and outer ring regions respectively. These values are larger than the ones found in CNSFR reported for other galaxies which range from 5.0 $\times$ 10$^6$ and 2.0 $\times$ 10$^8$ \citep[][]{hagele2007,hagele2009,hagele2010}. The total masses  of the outer and inner rings are (197.3 $\pm$ 3.1) $\times$ 10$^8$ M$_\odot$ and (41.1 $\pm$ 2.3) $\times$ 10$^8$ M$_\odot$. From gas stellar dynamics, \citet{1995ApJ...444..129G} inferred a dynamical mass
for the inner ring of $4.5 \times 10^9 $ M$_\odot$, very similar to the value found here, whereas \citet{2004ApJ...602..148D} estimated a dynamical mass $6.5 \times 10^9\,$ M$_\odot$, within a radius of 2.5 arcsec including $\sim 15\%$ corresponding to the galaxy nucleus.
A more significant insight into the characteristics of star formation in CNSFRs comes from the ratio of ionising to dynamical masses. This ratio ranges between 0.02 and 0.91 \% for the outer ring regions and between 0.76 and 4.86 \% for the inner ring ones implying a larger contribution by recent star formation for the regions closer to the galactic nucleus. For comparison, this percentage ranges from 1\% to 11\% for the CNSFR analysed in the galaxies  NGC~3351, NGC~2903 and NGC~3310 \citep[see respectively][]{hagele2007, hagele2009, hagele2010}, whose average distances to their host galaxy nuclei are about 300 pc, comparable to the inner regions of NGC~7469.

\section{Summary and conclusions}

In this second paper in the series, we analysed the circumnuclear environment of the almost face-on galaxy NGC 7469, an early spiral (SABa) hosting a Seyfert 1 nucleus. The galaxy shows two prominent star-forming rings, one of them very close to their active galactic nucleus, within 1.5 arcsec from the galaxy centre, and a second incomplete ring with elliptical appearance and with dimensions of 21 and 13.2 arcsec respectively for its major and minor axes. We  used publicly available observations obtained by the IFS MUSE spectrograph as part of the first Science Verification run. These data were analysed following the methodology already described in the first paper in the series. 

We  constructed 2D flux maps of different emission lines and two continuum bands. The [OIII]$\lambda $ 5007 \AA\ line is predominant in the emission from the active nucleus and it seems to blur along the galaxy disc. A map of the EW(H$\alpha$) emission shows the circumnuclear regions within the rings, the object of this study, having EW(H$\alpha$) > 50 \AA, consistent with the presence of recent star formation.

This map was used to select the ionised outer ring regions. For the inner ring regions we   used the observed HeI$\lambda $ 6678 \AA\ flux map, due to the presence of saturation effects in the central parts of the galaxy. At the end of the entire procedure, we   selected a total of 23 HII regions in the outer ring and 5 in the inner ring. In the same way, extinction correction were estimated using the ratios of H$\alpha$ and H$\beta$ and HeI$\lambda$ 5875 \AA/HeI$\lambda$ 6678 \AA\ respectively for the   outer and inner ring regions. 

All emission lines appear to have at least two kinematical components and four of them show three components in the [OIII]$\lambda \lambda$ 4959,5007 \AA\ emission lines, which might be associated with an outflow coming from the active galaxy nucleus. We   ascribed the most intense and narrow component to the emission lines originated by the ionising cluster since they follow the radial velocity of the galaxy disc and their emission, and the emission line ratios of the HII regions within the ring are consistent with the predictions of star-forming models, thus assuring that the kinematical component associated with the observed HII regions is the appropriate one.

The first part of this work concerns the properties of the ionising gas of the CNSFRs from the measured emission lines in their spectra: H$\alpha$ and H$\beta$ Balmer lines; [OIII]$\lambda \lambda$ 4959,5007 \AA, [NII]$\lambda \lambda$ 6548,84 \AA, [SII]$\lambda \lambda$ 6716,31 \AA, [ArIII]$\lambda$ 7136 \AA, and [SIII]$\lambda$ 9069 \AA\ forbidden lines; and also the weaker lines of [SIII]$\lambda$ 6312 \AA, HeI$\lambda$ 6678 \AA, and [OII]$\lambda \lambda$ 7320,30 \AA. 
For each observed ring HII region, we  derived (1) the number of hydrogen ionising photons per second, Q(H$_0$); (2) the electron density of the emitting gas per cubic centimetre, n$_e$; (3) the ionisation parameter, u; (4) the corresponding angular radius, $\phi$; (5) the filling factor, $\epsilon$; and (6) the mass of ionised hydrogen in solar masses, M(HII). The inner ring regions seem to be more compact, showing smaller sizes, higher filling factors, and higher electron densities than those in the outer ring. 
The outer ring regions in NGC~7469 are more luminous than those in NGC~7742, the first galaxy studied in this series. They also have higher electron densities, lower filling factors, lower ionisation parameters, and higher masses of ionised hydrogen. This is consistent with its larger HII region sizes. Although their model-predicted angular sizes, $\phi$, are found to be smaller than the measured ones, implying that their behaviour does not correspond to a radiation or matter bounded nebula,  the larger radii measured in the HII regions of this galaxy can be explained by the stellar winds produced by WR stars present in all the observed HII regions in numbers of up to a few hundred for the outer ring and as large as $\sim$ 1500 for the inner ring. 

We used sulphur as a tracer for chemical abundances of the selected HII regions with the temperature-sensitive [SIII]$\lambda$ 6312 \AA\ emission line having been measured in $\sim$48 \% of the outer ring regions and in three out of five regions within the inner ring regions, which allowed the abundances to be derived by the direct method. For the rest of the regions the empirical calibration S$_{23}$ was used. The outer ring regions show sulphur abundances, 12+log(S/H), ranging from 6.72 to 7.34;  the inner ring regions show  similar values within the errors. 

Ionising and photometric cluster masses were estimated from the number of Lyman continuum photons and absolute r-magnitudes, respectively, using SSP models. The inner ring regions show masses more than an order of magnitude higher than the outer ring regions. Additionally, the photometric masses of the outer ring regions follow the ionising star masses in a constant proportion of about 3, although this relation seems to be lost in the inner ring regions. 

In order to infer stellar population properties from the stellar absorption lines, we  calculated a dilution factor for the CaII and MgI lines. The outer ring regions show MgI dilutions consistent with the contribution by a nebular continuum, but the CaT lines look almost undiluted, pointing to a larger contribution by red supergiant stars. In the inner regions, the behaviour of the CaT is similar;  we  found larger dilutions in the MgI lines, which could be due to an additional continuum originating in the AGN.

Stellar velocity dispersions were derived from the measured CaT absorption lines using the cross-correlation technique proposed by Tonry \& Davis \citep{1979AJ.....84.1511T}. Subsequently, the dynamical masses were derived from the measured CaT velocity dispersion and the stellar sizes of each cluster, assuming virialisation. Contrary to the case of ionising and photometric masses, the SFC in both rings show similar masses, with median values 1.07 $\times$ 10$^9$ M$_\odot$ and 6.58 $\times$ 10$^8$ M$_\odot$ for the inner and the outer ring regions, respectively. The ratio of the ionising to the dynamical masses takes mean values of 3.8 \%  for the inner ring regions and 0.3 \% for the outer ring regions. 

The evolutionary state of the analysed clusters can be inferred from SSP evolutionary population synthesis models using the EWs of Balmer lines as a function of the continuum colour. Our observed ring regions show mean ages of 5.7 Ma for the outer ring regions and younger ages, between 5.1 and 5.7 Ma, for the inner ring regions, according to the SSP models. These regions can be identified as the young (5-6 Ma) stellar population accounting for most of the IR luminosity in the central region of the galaxy \citep{2007ApJ...661..149D}. The detection of carbon Wolf-Rayet star features places the age of the regions between 3.2 and 5.25 Ma, which is compatible with the ages found using SSP models. 

Finally, the comparison between the characteristics of inner and outer ring ionising clusters, together with their derived dynamical masses, point to circumnuclear regions close to the active galactic nucleus being more compact and having higher gas density. This suggests that they may be survivors in the hostile environment powered by nuclear activity.

\begin{acknowledgements}
This research has made use of the services of the ESO Science Archive Facility and NASA’s Astrophysics Data System Abstract Service. It is based on observations collected at the European Organisation for Astronomical Research in the Southern Hemisphere under ESO programme 60.A-9301(A) and data products created thereof. Also we have used observations obtained with the NASA/ESA HST and obtained from the \textit{Hubble} Legacy Archive, which is a collaboration between the Space Telescope Science Institute (STScI/NASA), the Space Telescope European Coordinating Facility (ST-ECF/ESA), and the Canadian Astronomy Data Centre (CADC/NRC/CSA). 
This work has been supported by Spanish grants from the  former Ministry of Economy, Industry and Competitiveness through the MINECO-FEDER research grants AYA2016-79724-C4-1-P, and PID2019-107408GB-C42 and the present Ministry of Science and Innovation through the research grant PID2022-136598NB-C33 by MCIN/AEI/10.13039/501100011033 and 
by “ERDF A way of making Europe”.
S.Z. acknowledges the support from contract: BES-2017-080509 associated with the first of these grants. 
\end{acknowledgements}

%
\bibliographystyle{aa} 
\bibliography{bibliografia} 
%

\appendix
\onecolumn
\section{Tables} \label{ap:tables}
\begin{table*}[h!]
\centering
\setlength{\tabcolsep}{2pt}
\caption{Reddening-corrected emission line intensities.}
\begin{tabular}{ccccccccccc}
\cline{2-11}
& Line & Hb & [OIII] & [OIII] &  [NII] & H$\alpha$ & [NII] & [SII] & [SII] & [SIII]\\
& $\lambda$ & 4861 & 4959 & 5007  & 6548 & 6563 & 6584 & 6717 & 6731 & 9069 \\
& f($\lambda$) & 0.000 & -0.024 & -0.035 & -0.311 & -0.313 & -0.316 & -0.334 & -0.336 & -0.561  \\ \hline
\multicolumn{1}{c}{Region ID} & \multicolumn{1}{c}{c(H$\beta$)} & \multicolumn{1}{c}{I(H$\beta$)$^a$}& \multicolumn{8}{c}{I($\lambda $)$^b$} \\ \hline
R1 & 0.50 $\pm$ 0.02 & 7.53 $\pm$ 0.42 & 42 $\pm$ 10 & 118 $\pm$ 9 & 304 $\pm$ 6 & 2870 $\pm$ 68 & 936 $\pm$ 9 & 469 $\pm$ 7 & 335 $\pm$ 7 & 82 $\pm$ 9\\ 
R2 & 0.78 $\pm$ 0.03 & 8.05 $\pm$ 0.67 & 206 $\pm$ 18 & 553 $\pm$ 18 & 457 $\pm$ 9 & 2870 $\pm$ 101 & 1394 $\pm$ 13 & 595 $\pm$ 13 & 438 $\pm$ 12 & 76 $\pm$ 11\\ 
R3 & 0.57 $\pm$ 0.04 & 4.47 $\pm$ 0.40 & 83 $\pm$ 17 & 229 $\pm$ 17 & 387 $\pm$ 9 & 2870 $\pm$ 108 & 1290 $\pm$ 13 & 551 $\pm$ 9 & 368 $\pm$ 8 & 111 $\pm$ 12\\ 
R4 & 0.54 $\pm$ 0.05 & 4.89 $\pm$ 0.59 & 109 $\pm$ 23 & 303 $\pm$ 23 & 415 $\pm$ 13 & 2870 $\pm$ 146 & 1188 $\pm$ 17 & 610 $\pm$ 15 & 478 $\pm$ 14 & 135 $\pm$ 17\\ 
R5 & 0.46 $\pm$ 0.05 & 2.75 $\pm$ 0.36 & 145 $\pm$ 24 & 407 $\pm$ 24 & 498 $\pm$ 13 & 2870 $\pm$ 158 & 1504 $\pm$ 19 & 720 $\pm$ 16 & 555 $\pm$ 15 & 149 $\pm$ 19\\ 
R6 & 0.88 $\pm$ 0.03 & 3.16 $\pm$ 0.23 & 56 $\pm$ 14 & 148 $\pm$ 13 & 315 $\pm$ 8 & 2870 $\pm$ 90 & 1025 $\pm$ 13 & 385 $\pm$ 10 & 275 $\pm$ 9 & 154 $\pm$ 12\\ 
R7 & 0.81 $\pm$ 0.04 & 5.51 $\pm$ 0.54 & 60 $\pm$ 18 & 159 $\pm$ 17 & 335 $\pm$ 11 & 2870 $\pm$ 120 & 1118 $\pm$ 16 & 521 $\pm$ 12 & 327 $\pm$ 11 & 122 $\pm$ 15\\ 
R8 & 0.72 $\pm$ 0.04 & 4.38 $\pm$ 0.46 & 65 $\pm$ 22 & 175 $\pm$ 21 & 313 $\pm$ 11 & 2870 $\pm$ 128 & 1089 $\pm$ 16 & 575 $\pm$ 13 & 372 $\pm$ 11 & 60 $\pm$ 14\\ 
R9 & 0.85 $\pm$ 0.04 & 2.96 $\pm$ 0.32 & 142 $\pm$ 25 & 376 $\pm$ 23 & 433 $\pm$ 11 & 2870 $\pm$ 131 & 1122 $\pm$ 15 & 512 $\pm$ 11 & 385 $\pm$ 10 & 182 $\pm$ 15\\ 
R10 & 1.10 $\pm$ 0.07 & 2.09 $\pm$ 0.34 & 285 $\pm$ 32 & 732 $\pm$ 28 & 465 $\pm$ 11 & 2870 $\pm$ 198 & 1346 $\pm$ 16 & 713 $\pm$ 18 & 526 $\pm$ 17 & 99 $\pm$ 12\\ 
R11 & 0.90 $\pm$ 0.06 & 4.86 $\pm$ 0.74 & 131 $\pm$ 34 & 345 $\pm$ 32 & 406 $\pm$ 15 & 2870 $\pm$ 184 & 1283 $\pm$ 20 & 653 $\pm$ 15 & 444 $\pm$ 14 & 48 $\pm$ 16\\ 
R12 & 1.11 $\pm$ 0.08 & 3.41 $\pm$ 0.62 & 181 $\pm$ 44 & 463 $\pm$ 39 & 476 $\pm$ 16 & 2870 $\pm$ 220 & 1373 $\pm$ 21 & 705 $\pm$ 16 & 527 $\pm$ 15 & 143 $\pm$ 17\\ 
R13 & 0.85 $\pm$ 0.09 & 1.50 $\pm$ 0.34 & 405 $\pm$ 49 & 1076 $\pm$ 41 & 647 $\pm$ 20 & 2870 $\pm$ 271 & 1809 $\pm$ 28 & 887 $\pm$ 27 & 619 $\pm$ 25 & 112 $\pm$ 22\\ 
R14 & 0.12 $\pm$ 0.06 & 0.46 $\pm$ 0.07 & 46 $\pm$ 30 & 135 $\pm$ 30 & 264 $\pm$ 18 & 2870 $\pm$ 186 & 1118 $\pm$ 24 & 662 $\pm$ 20 & 350 $\pm$ 18 & 61 $\pm$ 31\\ 
R15 & 0.66 $\pm$ 0.06 & 0.78 $\pm$ 0.11 & 102 $\pm$ 31 & 278 $\pm$ 30 & 393 $\pm$ 17 & 2870 $\pm$ 170 & 1038 $\pm$ 22 & 606 $\pm$ 18 & 460 $\pm$ 17 & 118 $\pm$ 20\\ 
R16 & 0.91 $\pm$ 0.07 & 1.33 $\pm$ 0.22 & 118 $\pm$ 39 & 310 $\pm$ 36 & 420 $\pm$ 17 & 2870 $\pm$ 200 & 1271 $\pm$ 23 & 598 $\pm$ 17 & 473 $\pm$ 16 & 137 $\pm$ 22\\ 
R17 & 0.89 $\pm$ 0.06 & 0.77 $\pm$ 0.10 & 74 $\pm$ 33 & 195 $\pm$ 30 & 376 $\pm$ 14 & 2870 $\pm$ 162 & 1167 $\pm$ 20 & 641 $\pm$ 15 & 415 $\pm$ 13 & 76 $\pm$ 13\\ 
R18 & 0.68 $\pm$ 0.06 & 0.62 $\pm$ 0.09 & 66 $\pm$ 35 & 181 $\pm$ 33 & 482 $\pm$ 16 & 2870 $\pm$ 167 & 1411 $\pm$ 22 & 518 $\pm$ 16 & 388 $\pm$ 16 & 149 $\pm$ 22\\ 
R19 & 0.59 $\pm$ 0.06 & 0.53 $\pm$ 0.07 & 115 $\pm$ 31 & 316 $\pm$ 30 & 436 $\pm$ 18 & 2870 $\pm$ 172 & 1318 $\pm$ 25 & 634 $\pm$ 19 & 397 $\pm$ 17 & 79 $\pm$ 21\\ 
R20 & 0.57 $\pm$ 0.07 & 0.54 $\pm$ 0.09 & 131 $\pm$ 41 & 362 $\pm$ 39 & 476 $\pm$ 22 & 2870 $\pm$ 212 & 1329 $\pm$ 28 & 839 $\pm$ 28 & 653 $\pm$ 27 & 120 $\pm$ 29\\ 
R21 & 0.72 $\pm$ 0.05 & 1.17 $\pm$ 0.15 & 149 $\pm$ 31 & 402 $\pm$ 30 & 428 $\pm$ 17 & 2870 $\pm$ 151 & 990 $\pm$ 21 & 493 $\pm$ 18 & 463 $\pm$ 17 & 119 $\pm$ 22\\ 
R22 & 1.18 $\pm$ 0.09 & 3.42 $\pm$ 0.73 & 288 $\pm$ 65 & 731 $\pm$ 57 & 565 $\pm$ 23 & 2870 $\pm$ 260 & 1552 $\pm$ 31 & 743 $\pm$ 23 & 556 $\pm$ 21 & 50 $\pm$ 19\\ 
R23 & 0.41 $\pm$ 0.07 & 0.29 $\pm$ 0.05 & 62 $\pm$ 26 & 173 $\pm$ 27 & 235 $\pm$ 19 & 2870 $\pm$ 194 & 1181 $\pm$ 31 & 571 $\pm$ 26 & 367 $\pm$ 23 & 144 $\pm$ 36\\ 
\hline
Ra & 1.081 $\pm$ 0.007 & 138.50 $\pm$ 2.87 & 164 $\pm$ 10 & 476 $\pm$ 12 & 312 $\pm$ 6 & 2870 $\pm$ 56 & 891 $\pm$ 10 & 152 $\pm$ 12 & 155 $\pm$ 12 & 138 $\pm$ 3\\ 
Rb & 0.9506 $\pm$ 0.0004 & 57.87 $\pm$ 0.87 & 211 $\pm$ 11 & 537 $\pm$ 12 & 479 $\pm$ 14 & 2870 $\pm$ 71 & 970 $\pm$ 19 & 186 $\pm$ 12 & 202 $\pm$ 11 & 107 $\pm$ 4\\ 
Rc & 0.4102 $\pm$ 0.0001 & 28.92 $\pm$ 0.78 & 301 $\pm$ 24 & 968 $\pm$ 29 & 565 $\pm$ 10 & 2870 $\pm$ 112 & 1256 $\pm$ 14 & 217 $\pm$ 38 & 270 $\pm$ 37 & 186 $\pm$ 8\\ 
Rd & 0.661 $\pm$ 0.001 & 16.30 $\pm$ 0.52 & 432 $\pm$ 24 & 1197 $\pm$ 25 & 596 $\pm$ 12 & 2870 $\pm$ 132 & 1091 $\pm$ 15 & 219 $\pm$ 28 & 293 $\pm$ 28 & 232 $\pm$ 10\\ 
Re & 1.46 $\pm$ 0.04 & 68.98 $\pm$ 5.97 & 686 $\pm$ 27 & 1894 $\pm$ 25 & 504 $\pm$ 8 & 2870 $\pm$ 112 & 1467 $\pm$ 13 & 363 $\pm$ 18 & 349 $\pm$ 18 & 93 $\pm$ 4\\ 
\hline
\end{tabular}
\begin{tablenotes}
\centering
\item $^a$ In units of 10$^{-15}$ erg/s/cm$^2$.\
\item $^b$ Values normalised to I(H$\beta$) 10$^{-3}$. 
\end{tablenotes}
\end{table*}

\begin{table*}[h!]
\centering
\setlength{\tabcolsep}{2pt}
\caption{Colours and magnitudes results.}
\begin{tabular}{cccccc}
\hline
Region ID & m$_i$ (mag) & m$_r$ (mag) & M$_i$ (mag) & M$_r$ (mag) & r-i ( mag)\\ \hline
R1 & 17.48 $\pm$ (0.35 $ \times$ 10$^{-4}$) & 17.81 $\pm$ (0.30 $ \times$ 10$^{-4}$) & -14.25 $\pm$ (0.35 $ \times$ 10$^{-4}$) & -13.92 $\pm$ (0.46 $ \times$ 10$^{-4}$) & 0.324 $\pm$ (0.460 $ \times$ 10$^{-4}$)\\ 
R2 & 17.36 $\pm$ (0.31 $ \times$ 10$^{-4}$) & 17.78 $\pm$ (0.30 $ \times$ 10$^{-4}$) & -14.37 $\pm$ (0.31 $ \times$ 10$^{-4}$) & -13.95 $\pm$ (0.43 $ \times$ 10$^{-4}$) & 0.427 $\pm$ (0.431 $ \times$ 10$^{-4}$)\\ 
R3 & 17.66 $\pm$ (0.33 $ \times$ 10$^{-4}$) & 17.99 $\pm$ (0.28 $ \times$ 10$^{-4}$) & -14.07 $\pm$ (0.33 $ \times$ 10$^{-4}$) & -13.75 $\pm$ (0.43 $ \times$ 10$^{-4}$) & 0.322 $\pm$ (0.429 $ \times$ 10$^{-4}$)\\ 
R4 & 16.99 $\pm$ (0.27 $ \times$ 10$^{-4}$) & 17.40 $\pm$ (0.25 $ \times$ 10$^{-4}$) & -14.74 $\pm$ (0.27 $ \times$ 10$^{-4}$) & -14.33 $\pm$ (0.37 $ \times$ 10$^{-4}$) & 0.414 $\pm$ (0.367 $ \times$ 10$^{-4}$)\\ 
R5 & 17.52 $\pm$ (0.29 $ \times$ 10$^{-4}$) & 17.90 $\pm$ (0.27 $ \times$ 10$^{-4}$) & -14.21 $\pm$ (0.29 $ \times$ 10$^{-4}$) & -13.83 $\pm$ (0.40 $ \times$ 10$^{-4}$) & 0.385 $\pm$ (0.396 $ \times$ 10$^{-4}$)\\ 
R6 & 19.50 $\pm$ (0.76 $ \times$ 10$^{-4}$) & 19.85 $\pm$ (0.67 $ \times$ 10$^{-4}$) & -12.23 $\pm$ (0.76 $ \times$ 10$^{-4}$) & -11.88 $\pm$ (1.02 $ \times$ 10$^{-4}$) & 0.347 $\pm$ (1.017 $ \times$ 10$^{-4}$)\\ 
R7 & 18.31 $\pm$ (0.64 $ \times$ 10$^{-4}$) & 18.68 $\pm$ (0.57 $ \times$ 10$^{-4}$) & -13.42 $\pm$ (0.64 $ \times$ 10$^{-4}$) & -13.05 $\pm$ (0.86 $ \times$ 10$^{-4}$) & 0.370 $\pm$ (0.856 $ \times$ 10$^{-4}$)\\ 
R8 & 17.86 $\pm$ (0.42 $ \times$ 10$^{-4}$) & 18.22 $\pm$ (0.37 $ \times$ 10$^{-4}$) & -13.87 $\pm$ (0.42 $ \times$ 10$^{-4}$) & -13.52 $\pm$ (0.56 $ \times$ 10$^{-4}$) & 0.354 $\pm$ (0.556 $ \times$ 10$^{-4}$)\\ 
R9 & 18.55 $\pm$ (0.42 $ \times$ 10$^{-4}$) & 18.92 $\pm$ (0.37 $ \times$ 10$^{-4}$) & -13.18 $\pm$ (0.42 $ \times$ 10$^{-4}$) & -12.82 $\pm$ (0.56 $ \times$ 10$^{-4}$) & 0.366 $\pm$ (0.560 $ \times$ 10$^{-4}$)\\ 
R10 & 19.25 $\pm$ (0.35 $ \times$ 10$^{-4}$) & 19.72 $\pm$ (0.34 $ \times$ 10$^{-4}$) & -12.48 $\pm$ (0.35 $ \times$ 10$^{-4}$) & -12.01 $\pm$ (0.49 $ \times$ 10$^{-4}$) & 0.470 $\pm$ (0.492 $ \times$ 10$^{-4}$)\\ 
R11 & 17.63 $\pm$ (0.36 $ \times$ 10$^{-4}$) & 18.01 $\pm$ (0.32 $ \times$ 10$^{-4}$) & -14.10 $\pm$ (0.36 $ \times$ 10$^{-4}$) & -13.73 $\pm$ (0.48 $ \times$ 10$^{-4}$) & 0.373 $\pm$ (0.479 $ \times$ 10$^{-4}$)\\ 
R12 & 18.31 $\pm$ (0.37 $ \times$ 10$^{-4}$) & 18.68 $\pm$ (0.33 $ \times$ 10$^{-4}$) & -13.42 $\pm$ (0.37 $ \times$ 10$^{-4}$) & -13.06 $\pm$ (0.50 $ \times$ 10$^{-4}$) & 0.364 $\pm$ (0.497 $ \times$ 10$^{-4}$)\\ 
R13 & 18.56 $\pm$ (0.34 $ \times$ 10$^{-4}$) & 19.02 $\pm$ (0.33 $ \times$ 10$^{-4}$) & -13.17 $\pm$ (0.34 $ \times$ 10$^{-4}$) & -12.71 $\pm$ (0.47 $ \times$ 10$^{-4}$) & 0.454 $\pm$ (0.470 $ \times$ 10$^{-4}$)\\ 
R14 & 18.80 $\pm$ (0.46 $ \times$ 10$^{-4}$) & 19.13 $\pm$ (0.39 $ \times$ 10$^{-4}$) & -12.93 $\pm$ (0.46 $ \times$ 10$^{-4}$) & -12.60 $\pm$ (0.60 $ \times$ 10$^{-4}$) & 0.331 $\pm$ (0.604 $ \times$ 10$^{-4}$)\\ 
R15 & 19.21 $\pm$ (0.38 $ \times$ 10$^{-4}$) & 19.58 $\pm$ (0.34 $ \times$ 10$^{-4}$) & -12.52 $\pm$ (0.38 $ \times$ 10$^{-4}$) & -12.15 $\pm$ (0.51 $ \times$ 10$^{-4}$) & 0.368 $\pm$ (0.515 $ \times$ 10$^{-4}$)\\ 
R16 & 19.13 $\pm$ (0.41 $ \times$ 10$^{-4}$) & 19.48 $\pm$ (0.36 $ \times$ 10$^{-4}$) & -12.61 $\pm$ (0.41 $ \times$ 10$^{-4}$) & -12.25 $\pm$ (0.54 $ \times$ 10$^{-4}$) & 0.352 $\pm$ (0.542 $ \times$ 10$^{-4}$)\\ 
R17 & 19.85 $\pm$ (0.36 $ \times$ 10$^{-4}$) & 20.24 $\pm$ (0.33 $ \times$ 10$^{-4}$) & -11.88 $\pm$ (0.36 $ \times$ 10$^{-4}$) & -11.49 $\pm$ (0.49 $ \times$ 10$^{-4}$) & 0.391 $\pm$ (0.491 $ \times$ 10$^{-4}$)\\ 
R18 & 19.49 $\pm$ (0.39 $ \times$ 10$^{-4}$) & 19.87 $\pm$ (0.35 $ \times$ 10$^{-4}$) & -12.24 $\pm$ (0.39 $ \times$ 10$^{-4}$) & -11.86 $\pm$ (0.52 $ \times$ 10$^{-4}$) & 0.380 $\pm$ (0.523 $ \times$ 10$^{-4}$)\\ 
R19 & 19.58 $\pm$ (0.48 $ \times$ 10$^{-4}$) & 19.91 $\pm$ (0.41 $ \times$ 10$^{-4}$) & -12.15 $\pm$ (0.48 $ \times$ 10$^{-4}$) & -11.82 $\pm$ (0.63 $ \times$ 10$^{-4}$) & 0.331 $\pm$ (0.632 $ \times$ 10$^{-4}$)\\ 
R20 & 18.89 $\pm$ (0.28 $ \times$ 10$^{-4}$) & 19.36 $\pm$ (0.28 $ \times$ 10$^{-4}$) & -12.84 $\pm$ (0.28 $ \times$ 10$^{-4}$) & -12.37 $\pm$ (0.40 $ \times$ 10$^{-4}$) & 0.473 $\pm$ (0.398 $ \times$ 10$^{-4}$)\\ 
R21 & 19.01 $\pm$ (0.48 $ \times$ 10$^{-4}$) & 19.40 $\pm$ (0.44 $ \times$ 10$^{-4}$) & -12.72 $\pm$ (0.48 $ \times$ 10$^{-4}$) & -12.33 $\pm$ (0.65 $ \times$ 10$^{-4}$) & 0.389 $\pm$ (0.649 $ \times$ 10$^{-4}$)\\ 
R22 & 18.20 $\pm$ (0.39 $ \times$ 10$^{-4}$) & 18.57 $\pm$ (0.35 $ \times$ 10$^{-4}$) & -13.53 $\pm$ (0.39 $ \times$ 10$^{-4}$) & -13.16 $\pm$ (0.53 $ \times$ 10$^{-4}$) & 0.374 $\pm$ (0.530 $ \times$ 10$^{-4}$)\\ 
R23 & 20.66 $\pm$ (1.16 $ \times$ 10$^{-4}$) & 21.05 $\pm$ (1.06 $ \times$ 10$^{-4}$) & -11.07 $\pm$ (1.16 $ \times$ 10$^{-4}$) & -10.68 $\pm$ (1.57 $ \times$ 10$^{-4}$) & 0.388 $\pm$ (1.568 $ \times$ 10$^{-4}$)\\ 
\hline
Ra & 16.20 $\pm$ (0.19 $ \times$ 10$^{-4}$) & 16.54 $\pm$ (0.17 $ \times$ 10$^{-4}$) & -15.53 $\pm$ (0.19 $ \times$ 10$^{-4}$) & -15.20 $\pm$ (0.25 $ \times$ 10$^{-4}$) & 0.332 $\pm$ (0.253 $ \times$ 10$^{-4}$)\\ 
Rb & 16.41 $\pm$ (0.19 $ \times$ 10$^{-4}$) & 16.70 $\pm$ (0.16 $ \times$ 10$^{-4}$) & -15.32 $\pm$ (0.19 $ \times$ 10$^{-4}$) & -15.03 $\pm$ (0.25 $ \times$ 10$^{-4}$) & 0.291 $\pm$ (0.254 $ \times$ 10$^{-4}$)\\ 
Rc & 16.00 $\pm$ (0.23 $ \times$ 10$^{-4}$) & 16.30 $\pm$ (0.19 $ \times$ 10$^{-4}$) & -15.73 $\pm$ (0.23 $ \times$ 10$^{-4}$) & -15.43 $\pm$ (0.30 $ \times$ 10$^{-4}$) & 0.295 $\pm$ (0.304 $ \times$ 10$^{-4}$)\\ 
Rd & 16.91 $\pm$ (0.20 $ \times$ 10$^{-4}$) & 17.30 $\pm$ (0.18 $ \times$ 10$^{-4}$) & -14.83 $\pm$ (0.20 $ \times$ 10$^{-4}$) & -14.43 $\pm$ (0.27 $ \times$ 10$^{-4}$) & 0.396 $\pm$ (0.272 $ \times$ 10$^{-4}$)\\ 
Re & 17.22 $\pm$ (0.18 $ \times$ 10$^{-4}$) & 17.68 $\pm$ (0.18 $ \times$ 10$^{-4}$) & -14.51 $\pm$ (0.18 $ \times$ 10$^{-4}$) & -14.06 $\pm$ (0.26 $ \times$ 10$^{-4}$) & 0.454 $\pm$ (0.257 $ \times$ 10$^{-4}$)\\ 
\hline
\end{tabular}
\end{table*}

\begin{table*}[h!]
\centering
\setlength{\tabcolsep}{1pt}
\caption{Characteristics of the observed CNSFRs.}
\begin{tabular}{ccccccccc}
\hline
\begin{tabular}[c]{@{}c@{}}Region\\ID\end{tabular}&\begin{tabular}[c]{@{}c@{}}L(H$\alpha$) \\ (erg s$^{-1}$)\end{tabular}& \begin{tabular}[c]{@{}c@{}}Q(H$_0$) \\ (photons s$^{-1}$)\end{tabular}&\begin{tabular}[c]{@{}c@{}}log(u) \\ \end{tabular}&\begin{tabular}[c]{@{}c@{}}$\phi$ \\ (arcsec)\end{tabular} & \begin{tabular}[c]{@{}c@{}}R \\ (arcsec)\end{tabular} & \begin{tabular}[c]{@{}c@{}}n$_e$ \\ (cm$^{-3}$)\end{tabular}& \begin{tabular}[c]{@{}c@{}}log($\epsilon$) \\ \end{tabular} & \begin{tabular}[c]{@{}c@{}}M(HII) \\ (M$_\odot$)\end{tabular}\\ \hline
R1 & (114.3 $\pm$ 6.4) $\times$ 10$^{38}$ & (83.7 $\pm$ 4.7) $\times$ 10$^{50}$ & -3.756 $\pm$ 0.077 & - & 1.57 $\pm$ 0.05 & 35 $\pm$ 22 & -3.65 $\pm$ 0.16 & (51.2 $\pm$ 9.7) $\times$ 10$^{4}$\\ 
R2 & (12.2 $\pm$ 1.4) $\times$ 10$^{39}$ & (89.4 $\pm$ 7.5) $\times$ 10$^{50}$ & -3.992 $\pm$ 0.102 & 1.99 $\pm$ 0.63 & 1.53 $\pm$ 0.05 & 59 $\pm$ 34 & -4.14 $\pm$ 0.21 & (28.4 $\pm$ 6.9) $\times$ 10$^{4}$\\ 
R3 & (67.8 $\pm$ 7.8) $\times$ 10$^{38}$ & (49.6 $\pm$ 4.4) $\times$ 10$^{50}$ & -3.630 $\pm$ 0.082 & - & 1.36 $\pm$ 0.05 & 31 $\pm$ 7 & -3.11 $\pm$ 0.17 & (5.1 $\pm$ 1.0) $\times$ 10$^{5}$\\ 
R4 & (7.4 $\pm$ 1.0) $\times$ 10$^{39}$ & (54.4 $\pm$ 6.6) $\times$ 10$^{50}$ & -3.609 $\pm$ 0.091 & 0.71 $\pm$ 0.17 & 1.64 $\pm$ 0.05 & 119 $\pm$ 48 & -3.19 $\pm$ 0.19 & (7.9 $\pm$ 1.7) $\times$ 10$^{5}$\\ 
R5 & (41.8 $\pm$ 6.2) $\times$ 10$^{38}$ & (30.6 $\pm$ 4.0) $\times$ 10$^{50}$ & -3.657 $\pm$ 0.092 & 0.60 $\pm$ 0.14 & 1.35 $\pm$ 0.05 & 103 $\pm$ 41 & -2.95 $\pm$ 0.19 & (4.8 $\pm$ 1.1) $\times$ 10$^{5}$\\ 
R6 & (47.9 $\pm$ 5.0) $\times$ 10$^{38}$ & (35.1 $\pm$ 2.6) $\times$ 10$^{50}$ & -3.152 $\pm$ 0.060 & - & 0.94 $\pm$ 0.05 & 34 $\pm$ 30 & -1.84 $\pm$ 0.13 & (7.4 $\pm$ 1.3) $\times$ 10$^{5}$\\ 
R7 & (8.4 $\pm$ 1.0) $\times$ 10$^{39}$ & (61.1 $\pm$ 6.1) $\times$ 10$^{50}$ & -3.503 $\pm$ 0.090 & - & 1.50 $\pm$ 0.05 & 23 $\pm$ 6 & -2.98 $\pm$ 0.19 & (8.4 $\pm$ 1.8) $\times$ 10$^{5}$\\ 
R8 & (66.5 $\pm$ 8.5) $\times$ 10$^{38}$ & (48.7 $\pm$ 5.1) $\times$ 10$^{50}$ & -4.106 $\pm$ 0.177 & - & 1.47 $\pm$ 0.05 & 78 $\pm$ 33 & -4.08 $\pm$ 0.36 & (20.0 $\pm$ 8.3) $\times$ 10$^{4}$\\ 
R9 & (45.0 $\pm$ 5.9) $\times$ 10$^{38}$ & (32.9 $\pm$ 3.6) $\times$ 10$^{50}$ & -3.252 $\pm$ 0.062 & 0.44 $\pm$ 0.11 & 1.02 $\pm$ 0.05 & 80 $\pm$ 37 & -2.04 $\pm$ 0.13 & (6.9 $\pm$ 1.2) $\times$ 10$^{5}$\\ 
R10 & (31.7 $\pm$ 5.7) $\times$ 10$^{38}$ & (23.2 $\pm$ 3.8) $\times$ 10$^{50}$ & -3.935 $\pm$ 0.090 & 0.94 $\pm$ 0.34 & 0.69 $\pm$ 0.05 & 60 $\pm$ 40 & -3.09 $\pm$ 0.20 & (6.5 $\pm$ 1.6) $\times$ 10$^{4}$\\ 
R11 & (7.4 $\pm$ 1.2) $\times$ 10$^{39}$ & (54.0 $\pm$ 8.2) $\times$ 10$^{50}$ & -4.377 $\pm$ 0.243 & - & 1.47 $\pm$ 0.05 & 160 $\pm$ 93 & -4.67 $\pm$ 0.49 & (10.8 $\pm$ 6.1) $\times$ 10$^{4}$\\ 
R12 & (5.2 $\pm$ 1.0) $\times$ 10$^{39}$ & (37.9 $\pm$ 6.9) $\times$ 10$^{50}$ & -3.657 $\pm$ 0.089 & 0.78 $\pm$ 0.23 & 1.06 $\pm$ 0.05 & 75 $\pm$ 39 & -2.94 $\pm$ 0.20 & (29.6 $\pm$ 6.7) $\times$ 10$^{4}$\\ 
R13 & (22.8 $\pm$ 5.4) $\times$ 10$^{38}$ & (16.7 $\pm$ 3.7) $\times$ 10$^{50}$ & -3.985 $\pm$ 0.143 & - & 0.94 $\pm$ 0.05 & 49 $\pm$ 20 & -3.18 $\pm$ 0.30 & (10.8 $\pm$ 3.7) $\times$ 10$^{4}$\\ 
R14 & (7.0 $\pm$ 1.2) $\times$ 10$^{38}$ & (51.3 $\pm$ 7.9) $\times$ 10$^{49}$ & -4.141 $\pm$ 0.370 & - & 0.98 $\pm$ 0.05 & 20 $\pm$ 17 & -3.00 $\pm$ 0.74 & (8.3 $\pm$ 7.1) $\times$ 10$^{4}$\\ 
R15 & (11.8 $\pm$ 1.9) $\times$ 10$^{38}$ & (8.6 $\pm$ 1.2) $\times$ 10$^{50}$ & -3.693 $\pm$ 0.123 & 0.36 $\pm$ 0.12 & 0.74 $\pm$ 0.05 & 89 $\pm$ 52 & -2.21 $\pm$ 0.26 & (13.2 $\pm$ 4.1) $\times$ 10$^{4}$\\ 
R16 & (20.2 $\pm$ 3.6) $\times$ 10$^{38}$ & (14.8 $\pm$ 2.4) $\times$ 10$^{50}$ & -3.587 $\pm$ 0.118 & 0.34 $\pm$ 0.09 & 0.75 $\pm$ 0.05 & 129 $\pm$ 54 & -2.24 $\pm$ 0.25 & (17.2 $\pm$ 5.2) $\times$ 10$^{4}$\\ 
R17 & (11.7 $\pm$ 1.8) $\times$ 10$^{38}$ & (8.6 $\pm$ 1.2) $\times$ 10$^{50}$ & -4.006 $\pm$ 0.129 & - & 0.53 $\pm$ 0.05 & 84 $\pm$ 31 & -2.69 $\pm$ 0.27 & (3.3 $\pm$ 1.2) $\times$ 10$^{4}$\\ 
R18 & (9.5 $\pm$ 1.5) $\times$ 10$^{38}$ & (69.4 $\pm$ 9.6) $\times$ 10$^{49}$ & -3.405 $\pm$ 0.109 & 0.25 $\pm$ 0.09 & 0.67 $\pm$ 0.05 & 76 $\pm$ 52 & -1.49 $\pm$ 0.23 & (20.9 $\pm$ 6.1) $\times$ 10$^{4}$\\ 
R19 & (8.0 $\pm$ 1.3) $\times$ 10$^{38}$ & (58.4 $\pm$ 8.3) $\times$ 10$^{49}$ & -3.961 $\pm$ 0.199 & - & 0.66 $\pm$ 0.05 & 33 $\pm$ 17 & -2.52 $\pm$ 0.40 & (5.6 $\pm$ 2.7) $\times$ 10$^{4}$\\ 
R20 & (8.2 $\pm$ 1.6) $\times$ 10$^{38}$ & (6.0 $\pm$ 1.0) $\times$ 10$^{50}$ & -3.927 $\pm$ 0.178 & 0.35 $\pm$ 0.12 & 0.70 $\pm$ 0.05 & 113 $\pm$ 61 & -2.49 $\pm$ 0.36 & (7.0 $\pm$ 3.0) $\times$ 10$^{4}$\\ 
R21 & (17.8 $\pm$ 2.6) $\times$ 10$^{38}$ & (13.0 $\pm$ 1.6) $\times$ 10$^{50}$ & -3.607 $\pm$ 0.135 & 0.20 $\pm$ 0.05 & 0.92 $\pm$ 0.05 & 345 $\pm$ 109 & -2.31 $\pm$ 0.28 & (25.1 $\pm$ 8.2) $\times$ 10$^{4}$\\ 
R22 & (5.2 $\pm$ 1.2) $\times$ 10$^{39}$ & (38.0 $\pm$ 8.1) $\times$ 10$^{50}$ & -4.471 $\pm$ 0.283 & 2.01 $\pm$ 0.97 & 1.20 $\pm$ 0.05 & 75 $\pm$ 50 & -4.62 $\pm$ 0.57 & (5.8 $\pm$ 3.8) $\times$ 10$^{4}$\\ 
R23 & (44.6 $\pm$ 7.8) $\times$ 10$^{37}$ & (32.6 $\pm$ 5.2) $\times$ 10$^{49}$ & -3.458 $\pm$ 0.185 & - & 0.69 $\pm$ 0.05 & - & -1.28 $\pm$ 0.38 & (19.5 $\pm$ 8.8) $\times$ 10$^{4}$\\ 
\hline

Ra & (196.5 $\pm$ 3.5) $\times$ 10$^{39}$ & (143.8 $\pm$ 2.6) $\times$ 10$^{51}$ & -2.675 $\pm$ 0.045 & 0.58 $\pm$ 0.04 & 0.69 $\pm$ 0.05 & 551 $\pm$ 56 & -2.36 $\pm$ 0.10 & (11.8 $\pm$ 2.1) $\times$ 10$^{5}$\\ 
Rb & (89.0 $\pm$ 6.5) $\times$ 10$^{39}$ & (651.2 $\pm$ 5.9) $\times$ 10$^{50}$ & -3.032 $\pm$ 0.042 & 0.52 $\pm$ 0.04 & 0.56 $\pm$ 0.05 & 701 $\pm$ 73 & -2.65 $\pm$ 0.09 & (35.1 $\pm$ 7.1) $\times$ 10$^{4}$\\ 
Rc & (44.2 $\pm$ 3.2) $\times$ 10$^{39}$ & (323.2 $\pm$ 1.9) $\times$ 10$^{50}$ & -2.793 $\pm$ 0.085 & 0.22 $\pm$ 0.02 & 0.68 $\pm$ 0.05 & 1119 $\pm$ 118 & -1.94 $\pm$ 0.17 & (8.8 $\pm$ 2.2) $\times$ 10$^{5}$\\ 
Rd & (24.1 $\pm$ 1.8) $\times$ 10$^{39}$ & (176.1 $\pm$ 1.2) $\times$ 10$^{50}$ & -2.667 $\pm$ 0.065 & 0.12 $\pm$ 0.01 & 0.49 $\pm$ 0.05 & 1431 $\pm$ 150 & -1.29 $\pm$ 0.14 & (6.2 $\pm$ 1.6) $\times$ 10$^{5}$\\ 
Re & (9.0 $\pm$ 1.0) $\times$ 10$^{40}$ & (66.0 $\pm$ 5.7) $\times$ 10$^{51}$ & -3.572 $\pm$ 0.043 & 1.23 $\pm$ 0.10 & 0.50 $\pm$ 0.05 & 439 $\pm$ 43 & -3.68 $\pm$ 0.10 & (8.1 $\pm$ 1.8) $\times$ 10$^{4}$\\ 
\hline

\end{tabular}
\end{table*}

\begin{table*}[h!]
\centering
\caption{Ionising cluster properties.}
\begin{tabular}{ccccc}
\hline
Region ID & \begin{tabular}[c]{@{}c@{}}Q(He$_0$) \\ (photons s$^{-1}$)\end{tabular}& \begin{tabular}[c]{@{}c@{}}EW(H$\beta$)  \\ (\AA )\end{tabular} &  \begin{tabular}[c]{@{}c@{}}M$_{ion}$ \\ (M$_\odot$)\end{tabular} &  \begin{tabular}[c]{@{}c@{}}M$_{phot}$ \\ (M$_\odot$)\end{tabular} \\ \hline
R1 & (4.9 $\pm$ 1.9) $\times$ 10$^{48}$ & 10.54 $\pm$ 0.52 & (30.0 $\pm$ 2.9) $\times$ 10$^{5}$ & (54.4 $\pm$ 9.8) $\times$ 10$^{5}$ \\ 
R2 & (5.5 $\pm$ 3.7) $\times$ 10$^{48}$ & 6.69 $\pm$ 0.38 & (47.4 $\pm$ 5.4) $\times$ 10$^{5}$ & (8.9 $\pm$ 1.2) $\times$ 10$^{6}$ \\ 
R3 & (2.2 $\pm$ 1.4) $\times$ 10$^{48}$ & 6.24 $\pm$ 0.36 & (28.0 $\pm$ 3.3) $\times$ 10$^{5}$ & (53.8 $\pm$ 7.5) $\times$ 10$^{5}$ \\ 
R4 & (2.6 $\pm$ 2.8) $\times$ 10$^{48}$ & 4.98 $\pm$ 0.34 & (37.2 $\pm$ 5.5) $\times$ 10$^{5}$ & (9.1 $\pm$ 1.5) $\times$ 10$^{6}$ \\ 
R5 & (0.8 $\pm$ 1.6) $\times$ 10$^{48}$ & 5.10 $\pm$ 0.38 & (20.5 $\pm$ 3.2) $\times$ 10$^{5}$ & (50.3 $\pm$ 8.3) $\times$ 10$^{5}$ \\ 
R6 & (2.4 $\pm$ 1.0) $\times$ 10$^{48}$ & 12.24 $\pm$ 0.87 & (11.0 $\pm$ 1.3) $\times$ 10$^{5}$ & (14.5 $\pm$ 2.7) $\times$ 10$^{5}$ \\ 
R7 & (3.8 $\pm$ 1.8) $\times$ 10$^{48}$ & 9.09 $\pm$ 0.73 & (24.9 $\pm$ 3.4) $\times$ 10$^{5}$ & (38.4 $\pm$ 2.5) $\times$ 10$^{5}$ \\ 
R8 & - & 5.67 $\pm$ 0.36 & (29.8 $\pm$ 4.0) $\times$ 10$^{5}$ & (56.8 $\pm$ 8.7) $\times$ 10$^{5}$ \\ 
R9 & (1.4 $\pm$ 1.2) $\times$ 10$^{48}$ & 5.64 $\pm$ 0.37 & (20.3 $\pm$ 2.8) $\times$ 10$^{5}$ & (37.3 $\pm$ 5.7) $\times$ 10$^{5}$ \\ 
R10 & - & 5.25 $\pm$ 0.51 & (15.2 $\pm$ 2.9) $\times$ 10$^{5}$ & (27.6 $\pm$ 4.4) $\times$ 10$^{5}$ \\ 
R11 & - & 3.58 $\pm$ 0.26 & (49.3 $\pm$ 8.6) $\times$ 10$^{5}$ & (9.9 $\pm$ 1.7) $\times$ 10$^{6}$ \\ 
R12 & - & 2.67 $\pm$ 0.20 & (44.5 $\pm$ 9.0) $\times$ 10$^{5}$ & (8.0 $\pm$ 1.4) $\times$ 10$^{6}$ \\ 
R13 & - & 3.44 $\pm$ 0.36 & (15.8 $\pm$ 3.9) $\times$ 10$^{5}$ & (35.8 $\pm$ 6.3) $\times$ 10$^{5}$ \\ 
R14 & - & 5.19 $\pm$ 0.46 & (33.9 $\pm$ 6.2) $\times$ 10$^{4}$ & (9.0 $\pm$ 1.5) $\times$ 10$^{5}$ \\ 
R15 & - & 4.19 $\pm$ 0.30 & (6.9 $\pm$ 1.1) $\times$ 10$^{5}$ & (15.7 $\pm$ 2.6) $\times$ 10$^{5}$ \\ 
R16 & - & 3.65 $\pm$ 0.29 & (13.3 $\pm$ 2.5) $\times$ 10$^{5}$ & (26.1 $\pm$ 4.5) $\times$ 10$^{5}$ \\ 
R17 & - & 4.64 $\pm$ 0.34 & (6.3 $\pm$ 1.0) $\times$ 10$^{5}$ & (12.5 $\pm$ 2.1) $\times$ 10$^{5}$ \\ 
R18 & (4.2 $\pm$ 4.0) $\times$ 10$^{47}$ & 4.26 $\pm$ 0.31 & (54.4 $\pm$ 8.9) $\times$ 10$^{4}$ & (12.4 $\pm$ 2.1) $\times$ 10$^{5}$\\ 
R19 & - & 4.27 $\pm$ 0.31 & (45.7 $\pm$ 7.6) $\times$ 10$^{4}$ & (10.3 $\pm$ 1.7) $\times$ 10$^{5}$ \\ 
R20 & - & 3.30 $\pm$ 0.26 & (5.8 $\pm$ 1.2) $\times$ 10$^{5}$ & (16.3 $\pm$ 2.9) $\times$ 10$^{5}$ \\ 
R21 & (8.0 $\pm$ 8.0) $\times$ 10$^{47}$ & 4.67 $\pm$ 0.32 & (9.4 $\pm$ 1.4) $\times$ 10$^{5}$ & (20.4 $\pm$ 3.4) $\times$ 10$^{5}$ \\ 
R22 & - & 2.20 $\pm$ 0.18 & (5.3 $\pm$ 1.2) $\times$ 10$^{6}$ & (9.6 $\pm$ 1.7) $\times$ 10$^{6}$ \\ 
R23 & - & 10.90 $\pm$ 1.56 & (11.4 $\pm$ 2.4) $\times$ 10$^{4}$ & (23.5 $\pm$ 4.2) $\times$ 10$^{4}$ \\ 

\hline
Ra & (83.2 $\pm$ 2.4) $\times$ 10$^{48}$ & 15.00 $\pm$ 0.68 & (38.0 $\pm$ 3.0) $\times$ 10$^{6}$ & (37.2 $\pm$ 6.3) $\times$ 10$^{6}$ \\
Rb & (41.9 $\pm$ 3.8) $\times$ 10$^{48}$ & 8.91 $\pm$ 0.33 & (27.0 $\pm$ 1.9) $\times$ 10$^{6}$ & (30.8 $\pm$ 6.2) $\times$ 10$^{6}$ \\ 
Rc & (21.4 $\pm$ 3.3) $\times$ 10$^{48}$ & 10.93 $\pm$ 0.84 & (11.2 $\pm$ 1.0) $\times$ 10$^{6}$ & (18.8 $\pm$ 3.3) $\times$ 10$^{6}$ \\ 
Rd & (17.8 $\pm$ 2.6) $\times$ 10$^{48}$ & 9.87 $\pm$ 0.84 & (66.8 $\pm$ 6.5) $\times$ 10$^{5}$ & (10.4 $\pm$ 1.9) $\times$ 10$^{6}$ \\ 
Re & (5.5 $\pm$ 1.0) $\times$ 10$^{49}$ & 10.67 $\pm$ 0.81 & (23.4 $\pm$ 3.0) $\times$ 10$^{6}$ & (21.8 $\pm$ 3.9) $\times$ 10$^{6}$ \\ 
\hline

\end{tabular}
\end{table*}

\end{document}